\pdfoutput=1
\documentclass[11pt,twoside,a4paper,cmspaper,final,collab]{cms-tdr}
\def\svnVersion{1ef28b8-D}\def\svnDate{2021/07/21}\def\cmsCernNoTag{CERN-EP-2020-171}\def\cmsCernDate{\today}\def\cmsMessage{"Published in the European Physical Journal C as \href{http://dx.doi.org/10.1140/epjc/s10052-021-09236-z}{\doi{10.1140/epjc/s10052-021-09236-z}.}"}
\begin{document}\cmsNoteHeader{EXO-19-008}

\providecommand{\cmsTable}[1]{\resizebox*{!}{\textheight}{#1}}
\newlength\cmsTabSkip\setlength\cmsTabSkip{1ex}
\newlength\cmsFigWidth
\ifthenelse{\boolean{cms@external}}{\setlength\cmsFigWidth{0.49\textwidth}}{\setlength\cmsFigWidth{0.65\textwidth}}
\newlength\cmsFigWidthApp
\ifthenelse{\boolean{cms@external}}{\setlength\cmsFigWidthApp{0.65\textwidth}}{\setlength\cmsFigWidthApp{0.75\textwidth}}
\ifthenelse{\boolean{cms@external}}{\providecommand{\cmsLeft}{upper\xspace}}{\providecommand{\cmsLeft}{left\xspace}}
\ifthenelse{\boolean{cms@external}}{\providecommand{\cmsRight}{lower\xspace}}{\providecommand{\cmsRight}{right\xspace}}
\ifthenelse{\boolean{cms@external}}{\providecommand{\cmsMiddle}{middle\xspace}}{\providecommand{\cmsMiddle}{lower left\xspace}}
\newcommand{\music}{\mbox{MUSiC}\xspace}
\newcommand{\minv}{\ensuremath{M}\xspace}
\newcommand{\MT}{\ensuremath{M_{\mathrm{T}}}\xspace}
\newcommand{\ST}{\ensuremath{S_{\mathrm{T}}}\xspace}
\newcommand{\pvalue}{\ensuremath{p\text{-value}}\xspace}
\newcommand{\pvalues}{\ensuremath{p\text{-values}}\xspace}
\newcommand{\ptilde}{\ensuremath{\tilde{p}}\xspace}
\newcommand{\ptildevalue}{\ensuremath{\tilde{p}\text{-value}}\xspace}
\newcommand{\ptildevalues}{\ensuremath{\tilde{p}\text{-values}}\xspace}

\cmsNoteHeader{EXO-19-008} 
\title{MUSiC: a model-unspecific search for new physics in proton-proton collisions at \texorpdfstring{$\sqrt{s} = 13\TeV$}{sqrt(s) = 13 TeV}}

\date{\today}

\abstract{
  \tolerance = 1600
  Results of the Model Unspecific Search in CMS (MUSiC), using proton-proton collision data recorded at the LHC at a centre-of-mass energy of 13\TeV, corresponding to an integrated luminosity of 35.9\fbinv, are presented. The MUSiC analysis searches for anomalies that could be signatures of physics beyond the standard model. The analysis is based on the comparison of observed data with the standard model prediction, as determined from simulation, in several hundred final states and multiple kinematic distributions. Events containing at least one electron or muon are classified based on their final state topology, and an automated search algorithm surveys the observed data for deviations from the prediction. The sensitivity of the search is validated using multiple methods. No significant deviations from the predictions have been observed. For a wide range of final state topologies, agreement is found between the data and the standard model simulation. This analysis complements dedicated search analyses by significantly expanding the range of final states covered using a model independent approach with the largest data set to date to probe phase space regions beyond the reach of previous general searches.
  \par
  }

\hypersetup{
pdfauthor={CMS Collaboration},%
pdftitle={MUSiC: a model unspecific search for new physics in proton-proton collisions at sqrt(s)=13 TeV},%
pdfsubject={CMS},
pdfkeywords={CMS, MUSiC, model independent, new physics, BSM}}

\maketitle 

\section{Introduction}
\label{sec:intro}

The CERN LHC  has produced proton-proton ($\Pp\Pp$) collisions at an unprecedented centre-of-mass energy of 13\TeV since 2015, providing an excellent opportunity to search for new phenomena in regions that were previously inaccessible to collider experiments. While the standard model (SM) of particle physics is well established as the theory that describes the fundamental particles and their interactions, it cannot explain certain phenomena such as dark matter, neutrino oscillations, and the matter-antimatter asymmetry in the universe. Several theories of physics beyond the standard model (BSM) have been developed to address the inadequacies of the SM, and a wide range of parameter and phase space regions of such theoretical models is accessible for a direct search for the first time at the LHC. A large number of searches for a range of BSM signatures have been conducted by the experiments at the LHC, including the CMS experiment \cite{Chatrchyan:2008zzk}, but no direct evidence for BSM physics has been found to date. Thus, it becomes imperative to expand the scope of searches so that signs of new physics that are in principle detectable by the CMS experiment are not missed.

Dedicated searches targeting specific BSM theories are often restricted in their scope to a few final states that are sensitive to the particular models probed. Practical constraints on the number of such analyses mean that there are models and final states that remain unexplored, where BSM signatures could possibly be hidden. Furthermore, new phenomena may exist that are not described by any of the existing models. Hence, complementary to the existing searches for specific BSM scenarios, a generalised model-independent approach is employed in the analysis reported here: \emph{Model Unspecific Search in CMS} (\music{}). The \music analysis uses an automated approach to quantify deviations between a Monte Carlo (MC) simulation of SM processes, as seen in the CMS detector, and the observed data in a wide variety of final states, in order to detect anomalies and identify discrepancies that could be hints of BSM physics or other neglected or unknown phenomena. Following the \music approach, events from data and SM simulation are classified based on the so-called final-state objects in an event, i.e. electrons (\Pe), muons (\PGm), photons (\PGg), jets originating from light-flavour quarks or gluons, jets originating from \PQb quarks (\PQb jets), and missing transverse momentum (\ptmiss), resulting in several hundred different event classes. Then, an automated statistical method is used to scan the different event classes and multiple kinematic distributions in each event class for deviations between the data and simulation, identifying either excesses or deficits. A deviation is considered significant if the measured significance of the deviation is beyond the expectation of the SM-only hypothesis. The discovery of significant deviations by \music would lead to a detailed investigation of both data and simulation in the final states of interest. Such deviations or anomalies could result from a possible insufficient description of the SM or detector effects in the simulation, from systematic effects that are unknown or incorrectly modelled, or they could be the first hints of BSM phenomena. This last interpretation cannot be the result of the general search algorithm itself, since the algorithm uses a simplified approach in order to probe a wide variety of diverse final states.  Rather, it would require additional study in the form of a dedicated analysis of the final states of interest, ideally performed on statistically independent data sets.

Since the analysis relies on the simulation to estimate the SM expectation, only final-state objects that are well modelled in the simulation are incorporated. In particular, \PGt leptons are not considered separately because of challenges in modelling the effects of misidentification of hadronic jets as \PGt leptons in the simulation. However, \PGt leptons enter the analysis in the form of electrons or muons from leptonic \PGt decays, or as jets from hadronic \PGt lepton decays. Other more complex objects, e.g.~hadronic decays of highly boosted \PW or \PZ bosons, are not considered in the current analysis. 
Furthermore, because beyond the leading order (LO) MC simulations of the quantum chromodymanics (QCD) multijet and $\PGg$+jets processes of the SM are not available to this analysis, the analysis is restricted to those final states that contain at least one isolated lepton (electron or muon), since the contributions of these processes are expected to be low in such final states. Finally, the electric charges of the final-state objects are not considered in the analysis.

Dedicated analyses in specific final states with search strategies optimised for particular signatures are expected to have greater sensitivity than the present, more general approach. Moreover, further final-state objects, kinematic distributions, and phase space regions remain to be explored.

General model-unspecific searches have been performed in the past by the D0 \cite{PhysRevD.62.092004, PhysRevD.64.012004, PhysRevLett.86.3712} and CDF experiments \cite{PhysRevD.78.012002, Aaltonen:2008vt} at the Tevatron, and by the H1 experiment \cite{PhysLettB.602.14, Aaron:2008aa} at HERA. Such searches have also been performed at the LHC by the ATLAS Collaboration \cite{Aaboud:2018ufy}, and preliminary results have been reported by the CMS Collaboration based on the \music approach using the $\Pp\Pp$ collision data set collected during the year 2010 at $\sqrt{s} = 7\TeV$ \cite{CMS-PAS-EXO-10-021}, and during 2012 at 8\TeV \cite{CMS-PAS-EXO-14-016}.

This paper describes the \music analysis that is performed with the full CMS data set of $\Pp\Pp$ collisions  at $\sqrt{s} = 13\TeV$ collected during 2016, corresponding to an integrated luminosity of $35.9\fbinv$. The increased centre-of-mass energy and much larger amount of data analysed compared to the previously reported results significantly extend the regions of BSM phase space than can be probed. 

We begin with the description of the CMS detector and object reconstruction in Section~\ref{sec:cms_objectreco}, followed by a summary of the data set and simulated samples along with the object and event selection in Sections~\ref{sec:data_mc} and~\ref{sec:selection}. The \music search strategy is presented in Section~\ref{sec:searchalgorithm}, and systematic uncertainties are discussed in Section~\ref{sec:systematic_uncertainties}. After selected sensitivity studies are presented in Section~\ref{sec:sensitivity}, the results are shown in Section~\ref{sec:results_2016}, before the paper is summarised in Section~\ref{sec:conclusion}. Additional figures complementing the results are shown in Appendix~\ref{sec:app_process_group_plots}.

\section{The CMS detector and object reconstruction}
\label{sec:cms_objectreco}

The central feature of the CMS apparatus is a superconducting solenoid of 6\unit{m} internal diameter, providing a magnetic field of 3.8\unit{T}. Within the solenoid volume are a silicon pixel and strip tracker, a lead tungstate crystal electromagnetic calorimeter (ECAL), and a brass and scintillator hadron calorimeter (HCAL), each composed of a barrel and two endcap sections. Forward calorimeters extend the pseudorapidity ($\eta$) coverage provided by the barrel and endcap detectors.
 Muons are detected in gas-ionisation chambers embedded in the steel flux-return yoke outside the solenoid.
A more detailed description of the CMS detector, together with a definition of the coordinate system used and the relevant kinematic variables, can be found in Ref.~\cite{Chatrchyan:2008zzk}.

Events of interest are selected using a two-tiered trigger system~\cite{Khachatryan:2016bia}. The first level, composed of custom hardware processors, uses information from the calorimeters and muon detectors to select events at a rate of around 100\unit{kHz} within a time interval of less than 4\mus. The second level, known as the high-level trigger (HLT), consists of a farm of processors running a version of the full event reconstruction software optimised for fast processing, and reduces the event rate to around 1\unit{kHz} before data storage.

At CMS, the global event reconstruction (also called the particle-flow event reconstruction~\cite{CMS-PRF-14-001}) aims to reconstruct and identify each individual particle in an event, using an optimised combination of all subdetector information. In this process, the identification of the particle type (muon, electron, photon, charged or neutral hadron) plays an important role in the determination of the particle direction and energy. Photons are identified as ECAL energy clusters not linked to the extrapolation of any charged particle trajectory to the ECAL. Electrons are identified as a charged-particle track and potentially many ECAL energy clusters corresponding to this track extrapolation to the ECAL and to possible bremsstrahlung photons emitted along the way through the tracker material. Muons are identified as tracks in the central tracker consistent with either a track or several hits in the muon system, and associated with calorimeter deposits compatible with the muon hypothesis. Charged hadrons are identified as charged-particle tracks neither identified as electrons, nor as muons. Finally, neutral hadrons are identified as HCAL energy clusters not linked to any charged-hadron trajectory, or as a combined ECAL and HCAL energy excess with respect to the expected charged-hadron energy deposit.

The energy of photons is obtained from the ECAL measurement. The energy of electrons is determined from a combination of the track momentum at the main interaction vertex, the corresponding ECAL cluster energy, and the energy sum of all bremsstrahlung photons associated with the track. The candidate vertex with the largest value of summed physics-object $\pt^2$ is taken to be the primary $\Pp\Pp$ interaction vertex, where \pt denotes the transverse momentum. The physics objects are the jets, clustered using the jet finding algorithm~\cite{Cacciari:2008gp,Cacciari:2011ma} with the tracks assigned to candidate vertices as inputs, and the associated missing transverse momentum, taken as the negative vector sum of the \pt of those jets. The energy of muons is obtained from the corresponding track momentum. The energy of charged hadrons is determined from a combination of the track momentum and the corresponding ECAL and HCAL energies, corrected for zero-suppression effects and for the response function of the calorimeters to hadronic showers. Finally, the energy of neutral hadrons is obtained from the corresponding corrected ECAL and HCAL energies.

In the barrel section of the ECAL, an energy resolution of about 1\% is achieved for unconverted or late converting photons in the tens of GeV energy range. Other photons detected in the barrel have a resolution of about 1.3\% up to $\abs{\eta} = 1.0$, rising to about 2.5\% at $\abs{\eta} = 1.4$. In the endcaps, the resolution of unconverted or late-converting photons is about 2.5\%, whereas other photons detected in the endcap have a resolution between 3 and 4\%~\cite{CMS:EGM-14-001}. 
 The momentum resolution for electrons with $\pt \approx 45\GeV$ from $\PZ \to \Pe^+ \Pe^-$ decays ranges from 1.7 to 4.5\%. It is generally better in the barrel region than in the endcaps, and also depends on the bremsstrahlung energy emitted by the electron as it traverses the material in front of the ECAL~\cite{Khachatryan:2015hwa}. 

Muons are measured in the range $\abs{\eta} < 2.4$, with detection planes made using three technologies: drift tubes, cathode strip chambers, and resistive-plate chambers. The single muon trigger efficiency exceeds 90\% over the full $\eta$ range, and the efficiency to reconstruct and identify muons is greater than 96\%. Matching muons to tracks measured in the silicon tracker results in a relative transverse momentum resolution for muons with \pt up to 100\GeV of 1\% in the barrel and 3\% in the endcaps. The \pt resolution in the barrel is better than 7\% for muons with \pt up to 1\TeV~\cite{Sirunyan:2018}.

For each event, hadronic jets are clustered from these reconstructed particles using the anti-\kt algorithm~\cite{Cacciari:2008gp, Cacciari:2011ma} with a distance parameter of 0.4. Jet momentum is determined as the vectorial sum of all particle momenta in the jet, and is found from simulation to be, on average, within 5 to 10\% of the true momentum over the whole \pt spectrum and detector acceptance. Additional $\Pp\Pp$ interactions within the same or nearby bunch crossings (pileup) can contribute additional tracks and calorimetric energy depositions to the jet momentum. To mitigate this effect, charged particles identified to be originating from pileup vertices are discarded and an offset correction is applied to correct for remaining contributions~\cite{CMS-PRF-14-001}. Jet energy corrections are derived from simulation to bring, on average, the measured response of jets to that of particle level jets. In situ measurements of the momentum balance in dijet, $\text{photon}+\text{jet}$, $\PZ+\text{jet}$, and multijet events are used to account for any residual differences between the jet energy scale in data and in simulation~\cite{Khachatryan:2016kdb}. The jet energy resolution amounts typically to 15\% at 10\GeV, 8\% at 100\GeV, and 4\% at 1\TeV. Additional selection criteria are applied to each jet to remove jets potentially dominated by anomalous contributions from various subdetector components or reconstruction failures~\cite{CMS:2017wyc}.
Jets originating from \PQb quarks are identified as \PQb-tagged jets using the combined secondary vertex algorithm (v2) described in Ref.~\cite{BTV-16-002}.

The missing transverse momentum vector \ptvecmiss is computed as the negative vector sum of the transverse momenta of all the particle-flow candidates in an event, and its magnitude is denoted as \ptmiss~\cite{CMS-PAS-JME-17-001}. The \ptvecmiss is modified to account for corrections to the energy scale of the reconstructed jets in the event.

\section{Data set and simulated samples}
\label{sec:data_mc}

The analysis presented in this paper is performed on the data sample collected by the CMS experiment during 2016, based on $\Pp\Pp$ collisions at a $\sqrt{s} = 13\TeV$, corresponding to an integrated luminosity of $35.9\fbinv$.

{\tolerance=800 The MUSiC analysis aims to find deviations in the data when compared to the SM predictions, and hence an inclusive description of the SM with a full set of simulated samples covering the entire range of SM processes that are expected to be detected by the CMS experiment is required to have a good estimate of the SM expectation in each final state. MC simulated events from the generators \PYTHIA8.212 \cite{pythia8}, \MGvATNLO version 2.2.2 \cite{amcatnlo} with MLM \cite{madgraph_mlm} or FxFx \cite{madgraph_fxfx} matching schemes, \POWHEG v2 \cite{powheg, powheg2, powheg3, Alioli:2008gx, Re:2010bp, Alioli:2009je, Alioli:2008tz, Nason:2009ai, Melia:2011tj, Nason:2013ydw, Campbell:2014kua, Bagnaschi:2011tu}, and \SHERPA 2.1.1 \cite{sherpa, sherpa1} are combined to model each SM process of relevance in the studied energy regime, with the \textsc{NNPDF3.0} \cite{Ball:2014uwa} parton distribution functions (PDFs) being used for most of the simulated samples. Simulation of the parton shower and hadronisation process is done with \PYTHIA8.205 \cite{pythia8}, with the underlying event tune \textsc{CUETP8M1} \cite{Khachatryan:2015pea}. The detector response is simulated using the \GEANTfour package \cite{Agostinelli:2002hh}. The presence of pileup in data is incorporated in simulated events by using additional inelastic events generated with \PYTHIA with the same underlying event tune as the main interaction that are superimposed on the hard-scattering events.\par}

{\tolerance=800 When available, higher order cross section estimates are used to normalise the MC simulated samples. The cross sections of the inclusive $\PW (\to \ell \PGn)+\text{jets}$ and $\PZ (\to \ell^+ \ell^-)+\text{jets}$ processes were obtained at next-to-next-to-LO (NNLO) in QCD using \textsc{FEWZ} 3.1.b2 \cite{Li:2012wna} and at next-to-LO (NLO) electroweak (EW) precision using \textsc{MCSANC} 1.01 \cite{Bondarenko:2013nu}, while that for the $\PZ\to \PGn \PGn$ process was calculated at NLO in QCD using \MCFM 6.6 \cite{mcfm}. Cross sections for the $\PW\PW\to \ell\PGn\Pq\Pq$ and $\PW\PW\to 2\ell 2\PGn$ processes were also obtained at NNLO in QCD using Ref.~\cite{Gehrmann:2014fva}. The $\PQt\PAQt$ cross section was calculated at NNLO in QCD including resummation of next-to-next-to-leading logarithmic soft-gluon terms with  \textsc{Top++2.0} \cite{Czakon:2011xx}, and the single top quark cross section was obtained at NLO in QCD with  \textsc{Hathor v2.1} \cite{Aliev:2010zk,Kant:2014oha}. The cross sections for the SM Higgs boson (\PH) processes were obtained at NLO, NNLO, or next-to-NNLO (N3LO) from Ref.~\cite{deFlorian:2016spz}, depending on the specific process. \par}

{\tolerance=800
A summary of the SM simulation samples can be found in Table~\ref{tab:mc_sets}. Considering their perturbative accuracy, for the processes that are expected to be dominant in several final states, such as the \PW, \PZ, diboson, and top quark processes, it is preferred to use samples generated at NLO or better. However, simulated samples at LO are also used to improve the statistical precision in the tails of the kinematic phase space. 
The choice of MC generators in some cases also reflects the limited availability of simulated samples for this analysis. 
Given the analysis requirement of the presence of at least a single isolated lepton, the QCD multijet, $\PGg$+jets, and diphoton processes are not expected to be the dominant processes although they have large cross sections. For such processes, simulated samples generated at LO are used.  
Both the $\PW$+jets and the $\PZ$+jets processes in leptonic final states are simulated with \MGvATNLO at NLO precision with up to two additional partons, with the FxFx scheme used for merging, and additional samples generated with \PYTHIA at LO precision and with \POWHEG at NLO precision are used  to improve the statistical precision at high masses for the $\PW$+jets process and the $\PZ$+jets process, respectively. For the $\PZ$+jets process with neutrinos in the final state, \MGvATNLO samples produced at LO are used. Simulated samples for the $\PGg$+jets process are generated at LO using \MGvATNLO. The $\PQt\PAQt$ process is simulated using \POWHEG at NLO, and \MGvATNLO simulation at NLO is used for top pair production in association with vector bosons. The $\PQt\PAQt\PQt\PAQt$ process is simulated at NLO using  \MGvATNLO.  Single-top processes are simulated at NLO using \POWHEG and single-top production in association with gauge bosons is simulated at NLO using \MGvATNLO. 
At NLO in perturbative QCD, the $\PQt\PW$ single-top process interferes with the $\PQt\PAQt$ process, and this effect is accounted for using the "diagram removal'' approach to correct the simulated samples for the $\PQt\PW$ single-top production process \cite{Frixione:2008yi}. 
Diboson processes are simulated at NLO using a combination of  \MGvATNLO and \POWHEG, with the exception of the $\PW\PGg$ and $\PGg\PGg$ processes where simulated samples at LO generated with \MGvATNLO and \SHERPA are used in certain phase space regions. QCD multijet events are simulated by \MGvATNLO at LO precision. Triboson processes are simulated at NLO with \MGvATNLO. The different Higgs boson production processes are simulated at NLO using a combination of  \MGvATNLO and \POWHEG. 
\par}

For the samples listed in Table~\ref{tab:mc_sets}, kinematic overlaps, resulting from additional samples used to increase the statistical precision, are removed. For most of the SM processes, the statistical precision of the number of simulated events corresponds to an integrated luminosity much larger than the analysed data set. This is not the case, in general, for the QCD multijet and the $\PGg$+jets MC samples; however, this analysis considers only final states that contain at least one isolated electron or muon, and the contribution of these SM processes is predicted to be small in such final states. For SM processes that are expected to have significant contributions in several different final states, such as the \PW, \PZ, diboson, and top quark processes, the number of simulated events correspond to a range of 3 to 10,000 times the number of expected events based on the integrated luminosity of the data set analysed.

\begin{table*}
\centering
\topcaption{Summary of standard model simulated samples. The generator described in the table corresponds to the matrix element generator.}
\cmsTable{
\begin{tabular}{l  l  l l l}
\hline
Process & Details & Generator    & Generator    & Cross section    \\
 & &  & order & order \\
\hline
$\PZ(\to \ell^+ \ell^-)$ + jets~ & $M_{\ell^+\ell^-} > 10 \GeV$ &  \MADGRAPH~~~ & NLO & NNLO  \\
  & $\pt(\PZ) > 50 \GeV$ &  \MADGRAPH & NLO & NNLO  \\
  & $M_{\ell^+\ell^-} > 120 \GeV$ &  \POWHEG & NLO &  NNLO  \\[\cmsTabSkip]
$\PZ(\to \PGn \PGn)$ + jets &  &  \MADGRAPH & LO & NLO  \\[\cmsTabSkip]
$\PW(\to \ell \PGn)$ + jets & Inclusive &  \MADGRAPH & NLO & NNLO  \\
  & $\pt(\PW) > 100 \GeV$ &  \MADGRAPH & NLO & NNLO  \\
  & $M_{\ell \PGn} > 200 \GeV$ & \PYTHIA8 & LO & NNLO  \\[\cmsTabSkip]
$\PGg$ + jets &  &   \MADGRAPH & LO & LO  \\[\cmsTabSkip]
$\PQt\PAQt$ &  Inclusive &   \POWHEG & NLO & NNLO  \\
  &  $M_{\ttbar} > 700 \GeV$ &  \POWHEG & NLO & NNLO  \\
  &  $\PQt\PAQt\PGg$ &  \MADGRAPH & NLO & NLO  \\
  &  $\PQt\PAQt\PW$ &  \MADGRAPH & NLO & NLO \\
  &  $\PQt\PAQt\PZ$ &  \MADGRAPH & NLO & NLO  \\
  &  $\PQt\PAQt\PGg\PGg$ &  \MADGRAPH & NLO & NLO  \\[\cmsTabSkip]
$\PQt\PAQt\PQt\PAQt$ &  &   \MADGRAPH & NLO & NLO  \\[\cmsTabSkip]
Top & $\PQt$ ($\PQt\PW$-channel)  &   \POWHEG & NLO &   NLO  \\
  &  $\PQt$ ($t$-channel)  &  \POWHEG & NLO &   NLO  \\
  &  $\PQt$ ($s$-channel)  &  \MADGRAPH & NLO & NLO  \\
  &  $\PQt\PGg$ &  \MADGRAPH & NLO &  NLO  \\
  &  $\PQt\PZ\PQq$ &  \MADGRAPH & NLO & NLO  \\[\cmsTabSkip]
$\PZ(\to 2\ell)\PGg$ &  &  \MADGRAPH & NLO & NLO  \\[\cmsTabSkip]
$\PW(\to \ell \PGn)\PGg$ & $\pt(\PGg) > 40 \GeV$ &   \MADGRAPH & LO & LO  \\
  &  $\pt(\PGg) > 130 \GeV$ &  \MADGRAPH & NLO & NLO  \\[\cmsTabSkip]
$\PZ\PZ$ & $\PZ\PZ\to 4\ell$ &   \MADGRAPH & NLO & NLO  \\
  &  $\PZ\PZ\to 2\ell 2\PQq$ &  \MADGRAPH & NLO & NLO  \\
  &  $\PZ\PZ\to 2\ell 2\PGn$ &   \POWHEG & NLO & NLO  \\[\cmsTabSkip]
$\PW\PW$ & $\PW\PW\to \ell \PGn \PQq\PQq$ &   \POWHEG & NLO &  NNLO  \\
  &  $\PW\PW\to 4\PQq$ &  \MADGRAPH & NLO & NLO \\
  &  $\PW\PW\to 2\ell 2\PGn$ &   \POWHEG & NLO &  NNLO \\[\cmsTabSkip]
$\PW\PZ$ & $\PW\PZ\to \ell \PGn 2\PQq$ &   \POWHEG & NLO &  NLO \\
  &  $\PW\PZ\to 3\ell \PGn$ &  \MADGRAPH & NLO & NLO  \\
  &  $\PW\PZ\to 2\ell 2\PQq$ &    \MADGRAPH & NLO & NLO  \\
  &  $\PW\PZ\to 1\ell 3\PGn$ &    \MADGRAPH & NLO & NLO \\
  &  $\PW\PZ\to 1\ell 1\PGn 2\PQq$ &    \MADGRAPH & NLO & NLO \\[\cmsTabSkip]
$\PGg\PGg$ &  $M_{\PGg\PGg} > 40 \GeV$ &   \SHERPA & LO &  LO \\
  &  $M_{\PGg\PGg} > 80 \GeV$ &  \MADGRAPH & NLO & NLO  \\
  & $M_{\PGg\PGg} > 200 \GeV$, $p_{\text{T}}^{\PGg} > 70 \GeV$ &   \SHERPA & LO  & LO  \\[\cmsTabSkip]
QCD multijet &  &   \MADGRAPH & LO & LO  \\[\cmsTabSkip]
Triboson  &  $\PZ\PZ\PZ$ &  \MADGRAPH & NLO & NLO  \\
  &  $\PW\PGg\PGg$ &  \MADGRAPH & NLO & NLO  \\
  &  $\PW\PZ\PZ$ &  \MADGRAPH & NLO & NLO  \\
  &  $\PW\PZ\PGg$ &  \MADGRAPH & NLO & NLO  \\
  &  $\PW\PW\PW$ &  \MADGRAPH & NLO & NLO  \\
  &  $\PW\PW\PZ$ &  \MADGRAPH & NLO & NLO  \\
  &  $\PW\PW\PGg$ &  \MADGRAPH & NLO & NLO  \\[\cmsTabSkip]
Higgs boson  &  $\Pg\Pg\PH \to \PQb\PAQb$, $\PGg\PGg$ &  \MADGRAPH & NLO & N3LO  \\
  &  $\Pg\Pg\PH \to \PGt\PAGt$, $\PZ\PZ(4\ell)$, $\PW\PW$($2\ell2\PGn$), $\PZ\PGg$ &  \POWHEG & NLO & N3LO  \\
  &  VBF ($\PH\to \PQb\PAQb$, $\PGt\PAGt$, $\PW\PW$, $\PZ\PZ$, $\PZ\PGg$) & \POWHEG & NLO & NNLO \\
  &  VBF ($\PH\to\PGg\PGg$) &  \MADGRAPH & NLO & NNLO \\
  &  VH (not $\PH\to \PQb\PAQb$) &   \MADGRAPH & NLO & NNLO  \\
  &  VH ($\PH\to \PQb\PAQb$) &  \POWHEG & NLO & NNLO  \\
  &  $\PQt\PAQt\PH$ &  \POWHEG & NLO & NLO  \\
\hline
\end{tabular}
}
\label{tab:mc_sets}
\end{table*}

\section{Object and event selection}
\label{sec:selection}

It  is necessary to determine the physics object content of each event unambiguously, which includes the identification of each reconstructed object and the removal of overlap between the individual objects. 
Since the analysis relies on simulation for the SM background prediction, tight selection criteria for the different objects are used to minimise the effect of misidentification while still retaining a reasonably high efficiency for selecting the objects. 
A summary of the object selection criteria, discussed in this Section, is shown in Table~\ref{tab:ids}.

\begin{table*}
  \centering
    \topcaption{Summary of object selection criteria discussed in Section~\ref{sec:selection}.}
    \begin{tabular}{ l l l}
      \hline
      Object                            & \pt [\GeVns{}]   &  Pseudorapidity    \\
      \hline
      Muon                              & $>$25         &  $\abs{\eta} < 2.4$     \\
      Electron                          & $>$25         &  $0 < \abs{\eta} < 1.44$ or $1.57 < \abs{\eta} < 2.50$ \\
      Photon                            & $>$25         &  $\abs{\eta} < 1.44$  \\
      Jet                               & $>$50         &  $\abs{\eta} < 2.4$     \\
      $\PQb$-tagged jet                 & $>$50         &  $\abs{\eta} < 2.4$     \\
      Missing transverse momentum       & $>$100        &  \NA                  \\
      \hline
   \end{tabular}
   \label{tab:ids}
\end{table*}

Events are required to be triggered by one of several single-lepton, dilepton, or single-photon triggers. 
The single-photon trigger improves the electron trigger efficiency at high electron momenta when used in combination with the single-electron trigger. 
Selected events are required to contain the reconstructed objects that correspond to the associated trigger for the event, and have a value of \pt that is above the trigger requirement. Overlap between triggers is removed, such that events triggered by two or more triggers are not counted multiple times. Details of the trigger selection are given in Table~\ref{tab:triggers}.

\begin{table*}
  \centering
    \topcaption{Summary of online and offline criteria.}
    \begin{tabular}{l  l  l}
      \hline
      Trigger                           & Trigger level requirement       &  Analysis requirement    \\
      \hline
      Single-muon trigger               & $1 \PGm$ with $\pt > 50 \GeV$     &  ${\geq}1 \PGm$ with $\pt > 53 \GeV$      \\
      Single-electron trigger           & $1 \Pe$ with $\pt > 115 \GeV$     &  ${\geq}1 \Pe$ with $\pt > 120 \GeV$       \\
      Dimuon trigger                    & $1 \PGm$ with $\pt > 17 \GeV$,    &  ${\geq}2 \PGm$, each with $\pt > 20 \GeV$  \\
                                        & second $\PGm$ with $\pt > 8 \GeV$ &                                        \\
      Dielectron trigger                & $2 \Pe$, each with $\pt > 33 \GeV$ &  ${\geq}2 \Pe$, each with $\pt > 40 \GeV$   \\
      Single-photon trigger             & $1 \PGg$ with $\pt > 175 \GeV$  &  ${\geq}1 \PGg$ with $\pt > 200 \GeV$  \\
      \hline
   \end{tabular}
   \label{tab:triggers}
\end{table*}

Muons are required to have $\pt > 25\GeV$ and $\abs{\eta} < 2.4$. Two dedicated selection criteria are used to select well-reconstructed, isolated muons: ``Tight muon ID'' for muons with \pt up to 200\GeV, and ``high-momentum muon ID'' for muons with $\pt > 200\GeV$, as described in Refs.~\cite{Sirunyan:2018, CMS-MUO-17-001}.  The efficiency for the selection of muons with such criteria has been measured to be between 96 and 98\%, whereas the  probability of pions (kaons) to be misidentified as muons is about 0.1 (0.3)\% \cite{Sirunyan:2018}.

Selected electrons need to fulfill $\pt > 25 \GeV$ and $\abs{\eta} < 2.5$, excluding electrons in the barrel endcap transition region of the CMS ECAL ($1.44 < \abs{\eta} < 1.57$). Two dedicated selection criteria are applied: the ``tight'' selection criteria are used for electrons with \pt up to 200\GeV, and the ``HEEP'' electron selection is used for electrons with $\pt > 200\GeV$ \cite{Khachatryan:2015hwa,Sirunyan:2018exx}. Detailed studies of efficiency and misidentification probabilities for the electron reconstruction are presented in Ref.~\cite{Khachatryan:2015hwa}.

Photons with $\pt > 25\GeV$ and $\abs{\eta} < 1.44$ in the barrel region of the CMS ECAL, where the misidentification rate is low, are selected if they pass the dedicated ``tight'' photon identification requirements that have been introduced in Ref.~\cite{CMS:EGM-14-001} and adapted for the present data set.

Jets must have $\pt > 50\GeV$ and $\abs{\eta} < 2.4$. These criteria select well-reconstructed jets within the coverage of the CMS tracking system in the high-pileup environment of the 2016 data taking, with an average of 22 $\Pp\Pp$ interactions per bunch crossing.
For $\PQb$-tagged jets, the chosen ``tight'' working point corresponds to about 41\% efficiency in identifying $\PQb$ jets and about 0.1\% misidentification rate for light-flavour and gluon jets \cite{BTV-16-002}.
The missing transverse momentum \ptmiss in the event is included as an object in the event classification if $\ptmiss > 100\GeV$. The distribution of \ptmiss at small values is strongly affected by resolution effects, and for most cases the value of \ptmiss associated with BSM phenomena is large. Thus, if $\ptmiss < 100\GeV$, this variable is not used in the selection process.

A reconstructed object may be identified as more than one particle. It is also possible for some detector signals to be used for different reconstructed objects, e.g.~an electron and a photon overlapping in the calorimeters. Possible ambiguities are resolved as follows. First, the list of particles is prioritised in the order of muons, electrons, photons, and finally jets, assumed to correspond to the order of purity. Then, in the case of an ambiguity such as the reconstruction of multiple electrons or photons based on the same calorimeter energy deposit, or in the case of an object overlapping with a jet, the particle with the highest priority is selected. Other particles close in $\Delta R = \sqrt{\smash[b]{(\Delta\eta)^2+(\Delta\phi)^2}}$ are removed from the event (where $\phi$ is the azimuthal angle in radians), using the threshold of $\Delta R = 0.5$ for jets and $\Delta R = 0.4$ for all other particles.

Events passing the above criteria are then categorised into event classes based on the event content. Event classes containing at least one lepton (electron or muon) are considered in the analysis.

\section{The \music search algorithm}
\label{sec:searchalgorithm}
The \music analysis is designed to be robust, unbiased by specific BSM physics models, and as inclusive as possible.
Every region is treated as a potential signal region.
The modelling of the known SM background processes is based solely on MC simulation.
No techniques based on control samples in data are employed to estimate the background expectation, since this would result in losing some kinematic regions in the data. 

The main steps of the \music search algorithm are described below, starting with the classification of events, then the introduction of the kinematic distributions of interest, followed by a description of the scanning procedure and the strategy to account for the look-elsewhere effect (LEE) using pseudo-data generation, and finally with the concept of the global overview of the scan results.

\subsection{Classification of events}
\label{sec:classification}

Events in data and simulated samples are assigned to different classes (final states) based on the physics object content of each event.
To determine the object content of an event unambiguously, all selection criteria described in detail in Section~\ref{sec:selection} are applied both to the observed data and the simulation, resulting in a defined number of well-reconstructed objects in the final state of the event. 

Each event is sorted into three different types of event classes:
\begin{enumerate}
  \item \textit{Exclusive} event classes for events containing only those selected objects that are specified for the event class and no additional final state objects. Thus each event is assigned to just one exclusive class.
  \item \textit{Inclusive} event classes contain events that include a nominal set of selected objects, but may contain additional objects. An event is assigned to all inclusive event classes that can be constructed from the selected objects. For example, events containing two muons and any number of additional objects would be classified into the $2 \Pgm +X$ inclusive event class.
  \item \textit{Jet-inclusive} event classes are defined as inclusive classes but restrict additional allowed objects to jets. High jet multiplicities are not expected to be accurately described in the simulation, and thus all exclusive classes with five or more jets are instead assigned to the $X + 5\text{jets} + \text{Njets}$ class, which includes events with at least five jets and is inclusive in terms of the number of additional jets that might be present. The threshold of five jets was chosen based on studies described in Section~\ref{sec:commisssioningstudies}.
\end{enumerate}

\begin{figure}[h!]
  \centering
    \includegraphics[width=0.49\textwidth]{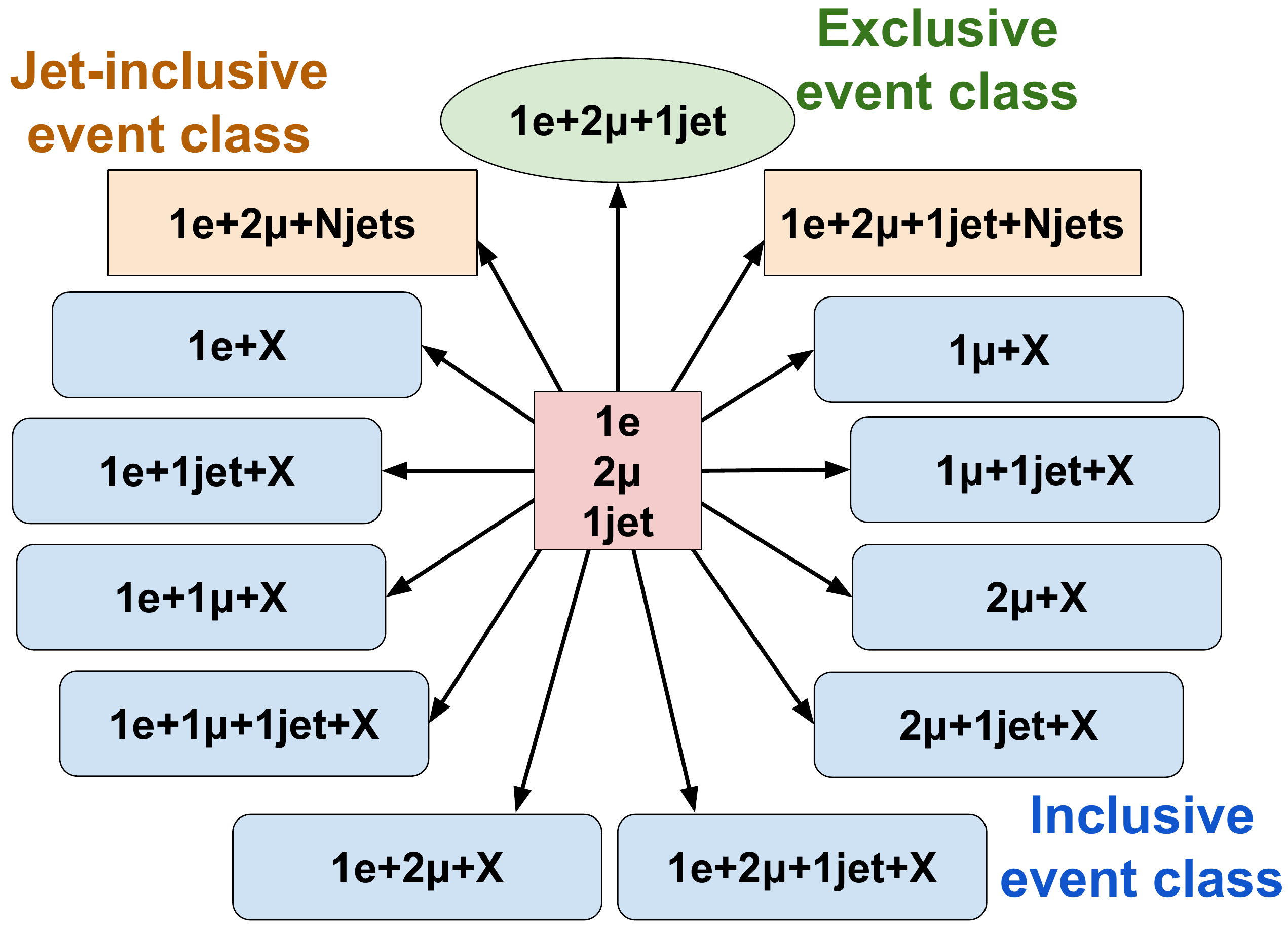}
    \caption{Illustrative example of classification of a single event (red square) containing one electron, two muons, and one jet. This event will contribute to precisely one exclusive (green), and several inclusive (blue) and jet-inclusive (orange) event classes.}
    \label{fig:event_class}
\end{figure}

There is no explicit limit placed on the number of objects, and, consequently, on the number of event classes, except for the case of jets, where it is set to
five. Events with greater than five jets can still enter the inclusive and jet-inclusive event classes. The construction of event classes from the physics object content of the final state, using the example of an event containing $1\Pe+2\Pgm+1\text{jet}$, is illustrated in Fig.~\ref{fig:event_class}.

All exclusive event classes are statistically independent of each other and can be regarded as uncorrelated (counting) experiments. This is not the case for the inclusive event classes, where a single event will generally end up in more than one event class. The resulting direct correlations are included while performing the statistical analysis, with the exception of correlations in the statistical uncertainties in the simulated events, which are assumed to be negligible. 
In the presence of a possible signal, it is a priori unknown how
the same events populate different inclusive and jet-inclusive event
classes, and therefore further interpretation of the results of the
statistical analysis would need to include the possible consequences
of such an effect.

\subsection{Kinematic distributions of interest}
\label{sec:kinDist}
Although signs of new physics can become visible in the distributions of many different kinematic variables, three are chosen for this analysis that seem especially promising in terms of sensitivity to phenomena at high \pt predicted by a large number of BSM scenarios.
This choice also prevents the analysis from being overly complex, as might result from the addition of more kinematic distributions. The three chosen kinematic distributions are:
\begin{enumerate}
  \item \ST: The \pt sum of all the physics objects that are considered for that event class, defined as
\begin{linenomath}
\begin{equation}
\ST = \sum_{i}\abs{\vec{p}_{T,i}}
\label{eq:st}
\end{equation}
\end{linenomath}
where the sum is over the particles that make up the event class. It is the most general variable of the three. It is calculated for every event passing the analysis requirements, and includes \ptmiss when applicable. The BSM physics is often expected to involve new heavy particles, the effects of which would show up predominantly in the tails of the \ST distributions.
  \item \minv or \MT: The combined mass \minv is the invariant mass calculated from all physics objects considered for the event class. For classes with \ptmiss, the transverse mass \MT is used instead of \minv, because the longitudinal component of the missing momentum is unknown.
Here, \MT is defined as
\begin{linenomath}
\begin{equation}
\MT = \sqrt{\Bigl(\sum_{i}E_{i}\Bigr)^2-\Bigl(\sum_{i}p_{x,i}\Bigr)-\Bigl(\sum_{i}p_{y,i}\Bigr)}
\label{eq:mt}
\end{equation}
\end{linenomath}
where the sum is over the particles that make up the event class, $E_{i}$ is the energy, and $p_{x,i}$ and $p_{y,i}$ are the $x$ and $y$ components of the momenta of the particle with index $i$.
This distribution is important for cases where a new massive particle is produced as a resonance and the mass distribution of its decay products is a prominent place to look for a deviation. All events in the event classes containing at least two objects are used to evaluate the combined mass.
  \item \ptmiss: For classes with significant \ptmiss of at least 100\GeV, it is an indicator of the energy of particles escaping detection. Only events with a substantial amount of \ptmiss are considered here, since the low-\ptmiss region is dominated by detector resolution effects and SM processes containing neutrinos. High values for \ptmiss can be associated with new particles with large \pt that do not interact with the detector.
\end{enumerate}

For a given exclusive event class, object types and multiplicities are identical in all events. Variables for the distributions of interest are calculated from the kinematic properties of all final-state objects. Since inclusive and jet-inclusive event classes also include events with more objects than those associated with the corresponding exclusive event class, an ambiguity must be resolved to ensure that the same event property is evaluated for all events in the distribution. Hence, only the objects stated explicitly in the name of the event class are used to derive the kinematic properties. For example, in the case of the $1\Pe+2\Pgm+\ptmiss+X$ event class, only the four mentioned objects (one electron, two muons, and \ptmiss) contribute to the \ST distribution, in each case considering the ones with the highest \pt if more than the mentioned number of a particular object are present. 

{\tolerance=800 The bin widths for the kinematic distributions probed are chosen as a compromise between a relatively large bin width, which is favorable in terms of computation time but detrimental in terms of sensitivity to potential narrow signals, and a small bin width, where random fluctuations will gain in importance and possibly mask the actual deviations of interest. An optimal choice is made in an automated way based on the typical total detector resolution of all objects in each specific kinematic region, leading to a larger value for the bin width at higher energies. All bin widths are integer multiples of 10\GeV. \par}

\subsection{Scan for deviations: region of interest scan}
\label{sec:music_roi_scan}
A statistical analysis is performed to identify deviations between data and the SM prediction by initially comparing the event yields in the event classes, followed by a complete scan of the kinematic distributions in the different event classes, referred to as the region of interest (RoI) scan. The procedure is described in two parts below, beginning with a discussion of the \pvalue definition that is used to quantify any observed deviation, followed by the description of the construction of the regions within which the algorithm searches for deviations. 

The measure for deviations is a \pvalue that describes the agreement between simulation and data using a hybrid Bayesian-frequentist approach, where the statistical fluctuations are assumed to follow a Poisson distribution, and nuisance parameters are modelled using a Gaussian prior function. Both excesses and deficits are taken into account. The \pvalue $p_{\text{data}}$ is defined as:
\begin{linenomath}
\ifthenelse{\boolean{cms@external}}
{
\begin{equation}
   p_{\text{data}} =\\
      \begin{cases}\displaystyle
         \sum^{\infty}_{i=N_{\text{data}}}
         C \int^{\infty}_{0} \rd \lambda\,\exp{\left(-\frac{(\lambda - N_{\text{SM}})^2}{2\,\sigma^2_{\text{SM}}}\right)}
         \frac{\re^{-\lambda}\,\lambda^i}{i!},\\ 
         \hfil\text{if $N_{\text{data}} \geq N_{\text{SM}}$},\\[6pt]
         \displaystyle
         \sum^{N_{\text{data}}}_{i=0}
          C \int^{\infty}_{0} \rd \lambda\,
         \exp{\left(-\frac{(\lambda - N_{\text{SM}})^2}{2\,\sigma^2_{\text{SM}}}\right)} 
         \frac{\re^{-\lambda}\,\lambda^i}{i!},\\
         \hfil\text{if $ N_{\text{data}} < N_{\text{SM}}$},
      \end{cases}
\label{eq:pvalue}
\end{equation}
}
{
\begin{equation}
   p_{\text{data}} =
      \begin{cases}
        \displaystyle
         \sum\limits^{\infty}_{i=N_{\text{data}}}
         C \int\limits^{\infty}_{0} \rd \lambda\,\exp{\left(-\frac{(\lambda - N_{\text{SM}})^2}{2\,\sigma^2_{\text{SM}}}\right)}
         \frac{\re^{-\lambda}\,\lambda^i}{i!},
         &\qquad \text{if $N_{\text{data}} \geq N_{\text{SM}}$},\\
         \displaystyle
         \sum\limits^{\hphantom{i}N_{\text{data}}\hphantom{=}}_{i=0}
          C \int\limits^{\infty}_{0} \rd \lambda\,
         \smash{\vphantom{\int\limits^{\infty}_{0}}\exp{\left(-\frac{(\lambda - N_{\text{SM}})^2}{2\,\sigma^2_{\text{SM}}}\right)}} 
         \smash{\vphantom{\int\limits^{\infty}_{0}}\frac{\re^{-\lambda}\,\lambda^i}{i!}},
         &\qquad \text{if $ N_{\text{data}} < N_{\text{SM}}$},
      \end{cases}
\label{eq:pvalue}
\end{equation}
}
\end{linenomath}
where $N_{\text{data}}$ is the number of observed events, $N_{\text{SM}}$ is the number of expected events from SM simulation, and $\sigma_{\text{SM}}$ denotes the uncertainty in $N_{\text{SM}}$, combining the statistical uncertainty arising from the number of generated MC events and systematic uncertainties. The probability distribution is summed up from $i = N_{\text{data}}$ to infinity for the case of an excess in observed data compared with the expectation, and from $i = 0$ to $N_{\text{data}}$ for the case of a deficit in observed data compared with the expectation. The Gaussian distribution is truncated at zero and normalised to unity with a factor $C$.

\begin{figure}[htp]
   \centering
   \includegraphics[width=\cmsFigWidth]{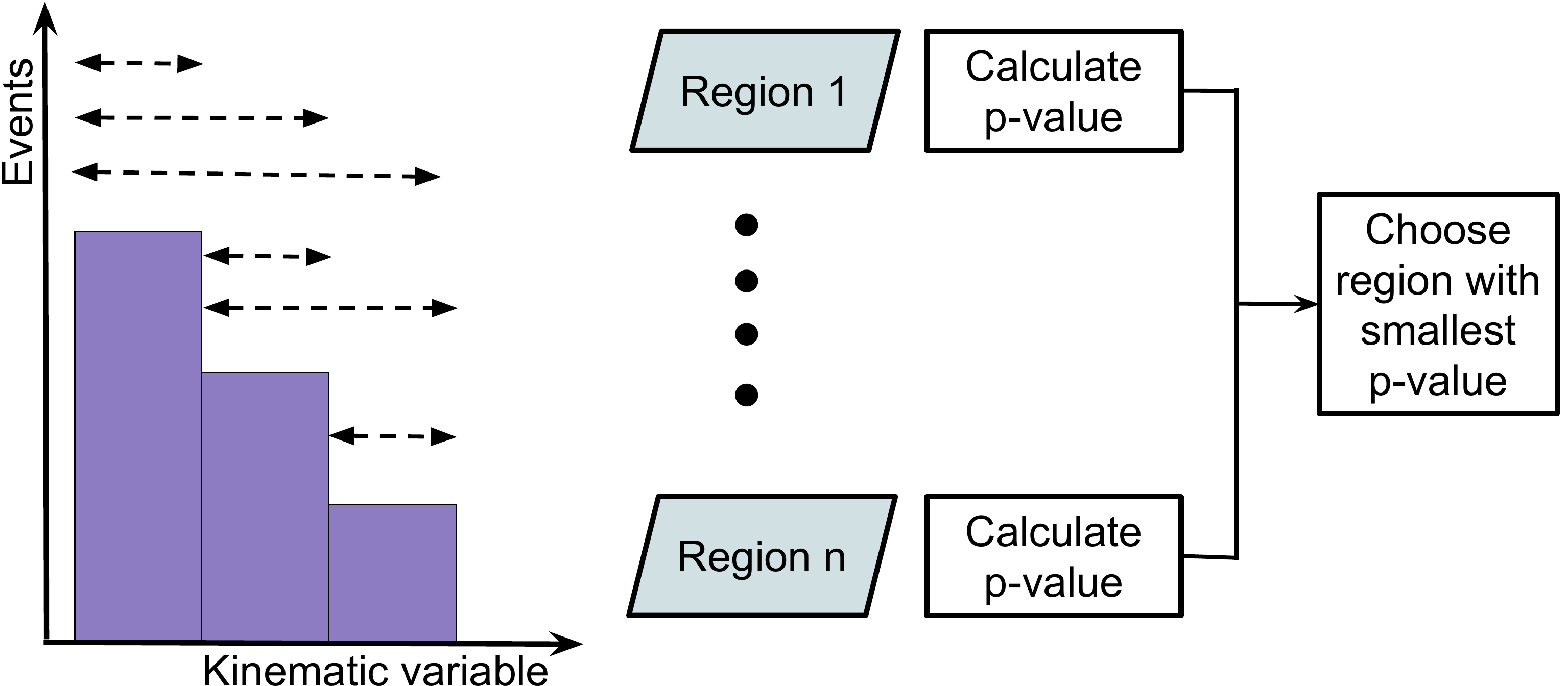}
\caption{Illustration for the calculation of \pvalues in different regions and the selection of the RoI as the region with the smallest \pvalue. The dashed lines with arrowheads represent the different possible continuous combinations of bins that are referred to as regions.}
\label{fig:regions}
\end{figure}

A region is defined as any contiguous combination of bins. Since several regions can contain the same bins, they are not disjoint, and a distribution with $N_{\mathrm{bins}}$ bins will result in $N_{\text{bins}} (N_{\text{bins}} + 1) / 2$ connected regions. 
All bins in question are then successively combined into regions by adding up their individual contributions, and a \pvalue is calculated. The smallest \pvalue ($p^{\text{data}}_{\text{min}}$) defines the RoI. This process is illustrated in Fig.~\ref{fig:regions}. This procedure, referred to as the RoI algorithm, is performed for all distributions in all classes.

The minimum number of bins within a region is required to be three for the \ST and \ptmiss distributions, since in these cases narrow deviations of less than three bins would be indicative of statistical fluctuations. Regions with a single bin are allowed for the mass distributions. Regions where statistical accuracy is poor due to the limited number of simulated events leading to an unreliable background estimate are removed, effectively considering only regions composed of a larger number of bins in such cases. Separate monitoring is in place to ensure that no potentially interesting signal is missed because an event class that contains data events fails to reach the required number of simulated events, particularly checking for event classes with more than one event in the recorded data where the expectation from simulation is below a threshold of 0.1 events.

\subsection{Post-trial probability (\ptilde) and look-elsewhere effect}
\label{sec:lee}

The chosen definition of the \pvalue serves as a measure of the probability of observing a deviation in a single region, while the algorithm is intended to give a measure of finding a deviation of equal or lesser compatibility anywhere in the distribution. We define the probability \ptilde to observe such a deviation in any of the considered regions throughout the distribution. The transition from a per-region to a per-distribution \pvalue, sometimes referred to as a post-trial probability, is a requirement to allow the comparison of observed deviations between many different distributions. A given \pvalue can be translated into a \ptildevalue by the LEE effect correction, which describes the increased probability to observe a significant deviation if a large number of regions is considered.

{\tolerance=800 An analytical calculation of the required correction is difficult because of correlations between bins and the irregular shape of systematic uncertainties, but the LEE correction can be determined using pseudo-experiments. Pseudo-experiments are generated in a randomized manner according to the background-only hypothesis, varying the prediction of the simulation according to the associated uncertainties. \par}
The RoI is not necessarily at the same position as the one found in the data.
The number of trials resulting in a local \pvalue ($p_{\text{min}}$) smaller or equal to the one found in the data to simulation comparison ($p^{\text{data}}_{\text{min}}$) is determined and divided by the full number of trials to get the fraction \ptilde:
\begin{linenomath}
\begin{equation}
   \label{eq:p-tilde}
   \ptilde = \frac{N_{\text{pseudo}}(p_{\text{min}} < p^{\text{data}}_{\text{min}})}{N_{\text{pseudo}}}.
\end{equation}
\end{linenomath}
This fraction is the post-trial \pvalue (\ptilde), representing a statistical estimate of how probable it is to see a deviation at least as strong as the observed one in any region of the distribution. While optimising for the computation time and also ensuring a robust measurement, and at the same time taking into account correlations across the different event classes, a total of 10,000 trials are conducted for each event class. 

Pseudo-experiments for a single distribution in a single class require generating randomised values for each bin $n$ to closely resemble the ensemble of expected values given the background-only (null) hypothesis. The systematic uncertainties in the null hypothesis are represented by a set of nuisance parameters $\nu_j$, which are expected to be fully correlated across all bins. This assumption requires that systematic uncertainties have been separated to a level at which the underlying processes responsible for the uncertainty remain similar for the complete range considered in a distribution. The effect of each nuisance parameter is modelled with a Gaussian centred on the mean expectation value for each bin $n$. To include these correlation effects, a random number $\kappa_{j}$, following a standard normal distribution, is generated for each nuisance parameter, excluding the statistical uncertainty of the MC samples.
The mean expectation value $\langle N_{n} \rangle$ in each bin is then shifted according to
\begin{linenomath}
\begin{equation}
  \label{eq:systshift}
  \langle N_{n,{\text{shifted}}}\rangle = \langle N_{n}\rangle + \sum_{j} \kappa_{j} \,\delta_{\nu_{j,n}},
\end{equation}
\end{linenomath}
where $\langle N_{n,{\text{shifted}}}\rangle$ is the shifted mean in each bin, and $\delta_{\nu_{j,n}}$ denotes the symmetrised 68\% confidence interval for $\nu_j$ in bin $n$. 
The value of $\langle N_{n,{\text{shifted}}}\rangle$ is further spread using a Poisson distribution to model the expected statistical variations. 

The procedure employed here has been developed considering the limitations arising from the absence of control regions and the requirement to keep the algorithm consistent over the vast range of final states probed. The procedure has been validated using toy simulations for the various statistical configurations that are expected in the analysis of the data set.

\subsection{Global overview of the RoI scan}
\label{sec:global_overview}

{\tolerance=800 Considering the large number of different event classes and kinematic distributions scanned, a convenient global overview of the scan is required.
The RoI scanning algorithm gives a RoI and its associated \ptildevalue separately for each event class and each kinematic distribution. 
To create the global overview, the results of the RoI scanning algorithm for the different event classes are grouped and presented together, separately for each of the three kinematic distributions scanned (\ST, \minv, and \ptmiss) for a particular type of event class (exclusive, inclusive, or jet-inclusive). 
For each such grouping, the observed distribution of deviations in the different event classes is compared with the expectation for the same distribution from the SM-only hypothesis, obtained from the pseudo-experiments. Any unexpected deviation from the scans would become apparent in such a comparison.
This would also probe certain BSM signals that show smaller deviations spread out over several different final states, in addition to such scenarios where the signature is a large deviation in specific final states. \par}

\begin{figure*}[h!tp]
   \centering
   \includegraphics[width=\textwidth]{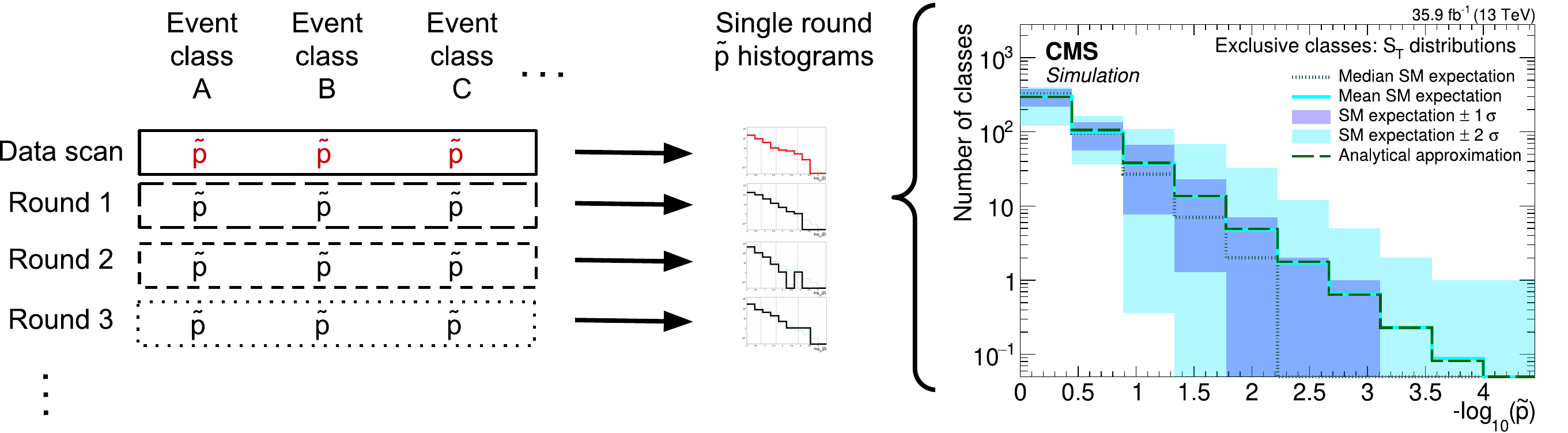}
\caption{Illustrative example of a \ptildevalue distribution for different event classes (final states) based on a RoI scan of an \ST distribution. Histograms of the number of event classes corresponding to a bin in $-\log_{10}(\ptilde)$ for the different pseudo-experiment iterations (shown on the left) are used to create the global overview plot for a scan of each particular kinematic distribution for each event class type (shown on the right for an \ST distribution scan in exclusive event classes, without showing the observed deviations from data here). The mean and the median distributions of \ptildevalues obtained from the different pseudo-experiments are shown as solid cyan and dotted grey lines. The distribution estimated from the analytic calculation is shown as a green dashed line. The 68\% ($\pm 1 \sigma$) and 95\% ($\pm 2 \sigma$) uncertainty bands are displayed as dark and light blue areas, respectively.}
\label{fig:global_overview}
\end{figure*}

{\tolerance=800 To produce a global overview of all event classes, the \ptildevalues calculated for each kinematic distribution are summarised in a single histogram. A \ptildevalue can be calculated for any particular pseudo-experiment in the same way as is done for collision data, by dividing the number of such experiments with \pvalues that are smaller than the \pvalue for the particular pseudo-experiment under consideration by their total number. This represents the \ptildevalue for one pseudo-experiment, and similarly \ptildevalues can be calculated for all the different pseudo-experiments. The resulting histogram of \ptildevalues from the different pseudo-experiments shows the expected distribution of \ptildevalues for the simulation-only hypothesis. The \ptildevalue distribution obtained from the observed distribution (from collision data) is then compared with the one obtained from pseudo-experiments, taking the median of the distributions from the different pseudo-experiments as the central value for the SM-only hypothesis. Furthermore, $\pm 1 \sigma$ ($\pm 2 \sigma$) uncertainty bands around the median expectation are obtained corresponding to the bands within which the distributions from 68 (95)\% of the pseudo-experiments are contained. This is summarised with an illustrative example in Fig.~\ref{fig:global_overview}. \par}

In addition to this numerical calculation, the \ptildevalue distribution can also be determined analytically. Since \ptildevalues are distributed uniformly, the content of each bin ($N_{\text{bin}}$) can be evaluated from the edges of the bin and the number of event class distributions ($N_{\mathrm{dist}}$) contributing to the bin. For the double-logarithmic scale used to plot the \ptildevalue distribution (as in the nominal distribution shown in Fig.~\ref{fig:global_overview}), the content of each bin is given by
\begin{linenomath}
\begin{equation}
   N_{\text{bin}} = \left(10^{-B_{\text{low}}} - 10^{-B_{\text{up}}}\right) N_{\text{dist}},
\label{eq:global_overview}
\end{equation}
\end{linenomath}
where $B_{\text{low}}$ and $B_{\text{up}}$ are the lower and upper bin edges, respectively. This analytic description is depicted by a green dashed line in the \ptildevalue distribution shown in Fig.~\ref{fig:global_overview}. The analytical and numerical distributions are found to agree.
The approach to estimate systematic uncertainties, as implemented in this analysis and described later, can affect the \ptildevalue distribution with more event classes showing smaller deviations and appearing in the bin with the smallest deviations. 
The size of the uncertainty bands of the expected \ptildevalue distributions has been increased, motivated by studies with pseudo-data that include the potential effect of overestimating the systematic uncertainties by up to 50\%.  Since this effect is limited to the first few bins in the \ptildevalue distributions, which correspond to small deviations, the modification does not have a significant impact on the part of the distribution where possible statistically significant deviations are expected to appear. 

\section{Systematic uncertainties}
\label{sec:systematic_uncertainties}

Estimates of all major known sources of systematic uncertainties are incorporated. In particular, uncertainties in the following quantities are included: integrated luminosity, contributions of pileup interactions, total cross sections of SM processes, PDFs, energy and momentum scale of all objects, reconstruction efficiencies, resolutions, misidentification probabilities, and the number of simulated events. The uncertainties arising from the finite size of the simulated samples are uncorrelated between bins. The effect of each of the other sources of uncertainty is fully correlated across all bins and event classes. 
Systematic uncertainties that influence kinematic properties are evaluated by variations of such variables, which might shift some particles in and out of the acceptance for the selection. This effect is included by allowing uncertainty contributions to cause migrations between different event classes. A summary of the systematic uncertainties is presented in Table~\ref{tab:systematics}.

\begin{table*}
  \centering
   \topcaption{Summary of systematic uncertainties in the analysis.}
    \begin{tabular}{l  l}
      \hline
      Source of uncertainty & Typical values   \\
      \hline
      Integrated luminosity & 2.5\% \\
      Pileup & $<$5\% \\
      Cross sections of SM processes & For processes calculated at LO: 50\%\\
                                    & For higher-order calculations: varies\\
      Parton distribution functions   & Varies, following PDF4LHC \cite{pdf4lhc2015} recommendations\\
      Value of $\alpS$ & Varies, variations of $\pm0.0015$ around central value ($0.118$) \\
      Electron, muon, and photon energy scales  & 0.15--7.00\% \\
      Jet energy scale and resolution & 3--5\% \\
      Unclustered energy & Varies, typically 0--15 \GeV \\
      Reconstruction and identification efficiency  & Varies, $<$10\% \\
      Misidentification uncertainties & 50\% \\
      MC statistical uncertainty  & Varies, up to 30\% \\
      \hline
   \end{tabular}
   \label{tab:systematics}
\end{table*}

The uncertainty in the value of the integrated luminosity is 2.5\% \cite{LUM-17-001}, and this uncertainty is propagated into the analysis as a normalisation uncertainty in all simulated events in each region. Since the pileup conditions assumed in the sample generation are not identical to the data, simulated samples are corrected to reproduce the pileup distribution in data, which has an average number of $\Pp\Pp$ interactions per bunch crossing of approximately 22 for the 2016 data sample. The associated uncertainties because of the estimation of the pileup distribution in data are propagated through the individual event weights for the MC samples in the analysis.

The uncertainties in the total cross sections for individual SM physics processes are included, although not all cross sections are known to the same order of perturbation theory. Uncertainties in individual simulated samples of a single physics process, \eg samples of different phase space regions or QCD multijet samples enriched in heavy flavours, are assumed to be fully correlated. The total cross section uncertainty is evaluated from all contributing physics processes assuming their separate uncertainties are uncorrelated.
For processes generated at LO we apply an uncertainty of 50\% in the value of the cross section.
For higher-order calculations the effect of missing higher-order corrections is estimated using coherent variations
in the factorisation and renormalisation scales in the MC simulation by factors of $2.0$ and $0.5$ up and down,
respectively.

To estimate the uncertainties corresponding to the PDFs, the procedure outlined in the PDF4LHC recommendations for the LHC Run 2 \cite{pdf4lhc2015} is used. These uncertainties are treated as arising from a single source, and hence are fully correlated  over all bins and event classes. 
For the assumed central value of the strong coupling $\alpS = 0.118$, variations of $\pm0.0015$ are used. 
This simplified approach is used here, considering the complexities associated with the variety of event classes and the number of bins corresponding to the different kinematic observables that are used for the scan in each event class probed. While the correlations are treated in a simplistic manner in this approach, the limitations of this approach are not expected to have a significant impact on the results considering the impact of other systematic and statistical uncertainties that are taken into account in this analysis.

Uncertainties in the energy or momentum measurement of the different physics objects are estimated by varying the measured kinematic observables such as \pt and $\eta$ within their uncertainties. For all variations, the full analysis is performed and the uncertainty in the event yield is derived from the resulting difference in each kinematic distribution. The effect of these variations in the measured \ptmiss is also included.
The uncertainty in the muon momentum scale has a dependence on \pt and $\eta$ that is taken into account. For $1\TeV$ muons in the central region of the detector, the uncertainty is 7\%~\cite{Sirunyan:2018}.
Uncertainties in the energy scale for electrons and photons have been estimated separately for the barrel and endcap regions, and are  
0.2\% (barrel) and 0.3\% (endcap) for low-energy electrons \cite{Khachatryan:2015hwa}, 0.15\% (barrel) and 0.30\% (endcap) for low-energy photons \cite{CMS:EGM-14-001}, and 2\% for high-energy electrons and photons \cite{Khachatryan:2015hwa}.
Corrections are applied to the energy scale of reconstructed jets to account for effects from pileup, simulated true jet response, and residual data and simulation scale factors, as summarised in Ref.~\cite{Khachatryan:2016kdb}. The associated uncertainties range from 3--5\%, depending on the jet \pt and $\eta$.
Although the jet energy corrections are not constant throughout the entire detector, they will be similar for jets close to each other. For this analysis it is assumed that jet energy scale uncertainties are fully correlated.
All energy deposits measured in the CMS detector and not assigned to a reconstructed physics object are summed up and referred to as unclustered energy, with its uncertainty propagated to the \ptmiss uncertainty \cite{CMS-PAS-JME-17-001}. 

Scale factors, in general close to one and depending on \pt and $\eta$, correct for differences in the efficiencies for reconstruction and identification of the objects in data and simulation. The uncertainties arising from the employed methods or limited size of the analysed data sets used to calculate the efficiencies are included. 
Uncertainties are assumed to be fully correlated for objects of the same type and  uncorrelated for objects of different type. 
For $\PQb$ tagging uncertainties we use \pt and jet hadron flavour-dependent scale factors along with their uncertainties, which are derived from data \cite{BTV-16-002}.

Although the selection criteria have been chosen to minimise misidentification, residual amounts of misidentified objects remain. The fraction of misidentified objects is determined in the simulation with generator-level information by matching generated particles to the reconstructed objects. 
To cover the uncertainty in misidentified objects (i.e.~those not matched to a generated particle of the same type), we apply a 50\% uncertainty in the simulation. Misidentification is mainly relevant when jets are wrongly identified as charged leptons or photons, whereas the uncertainty in the inverse process, i.e.~leptons misidentified as jets, is usually negligible compared to the reconstruction efficiency and scale factor uncertainties. The uncertainty assigned has been validated in the studies described later in Section~\ref{sec:commisssioningstudies}.
Misidentification uncertainties are assumed to be fully correlated for objects of the same type, and uncorrelated for objects of different types.

The statistical uncertainty in the number of generated events and the total event weight, arising from the limited number of events produced for each simulated process, is included.
Contributions from different simulated data sets are uncorrelated when constructing regions. 

\section{Sensitivity studies}
\label{sec:sensitivity}

To illustrate the capability of the \music analysis to identify deviations in the comparison of measured data and the SM simulation, two separate approaches are followed. In the first approach, a simulated BSM signal in addition to the SM simulation is injected into the analysis and compared to the SM simulation alone. In the second approach, the measured data are used, but a particular process is removed from the SM simulation. In both cases the \music algorithm is shown to find event classes in significant tension with the SM expectation. Examples for both approaches are discussed here.

A dedicated BSM signal is used to test the ability of the \music algorithm to identify possible new physics signals. A simulation of the BSM signal is added to the SM simulation, and pseudo-data are generated that can be scanned with the \music algorithm against the SM-only background.
The signal described here is expected to introduce a localised excess of events in individual final states: a new heavy vector boson $\PWpr$ is produced and promptly decays into a charged lepton and a neutrino, as predicted by the sequential standard model (SSM) \cite{Altarelli:1989ff}. In the CMS detector, such a signature is reconstructed as an event containing a single isolated, energetic charged lepton, substantial \ptmiss, and any number of jets originating from initial- or final-state radiation. Simulated samples for the $\PWpr$ process are produced at LO using \PYTHIA8.212, and the cross sections are obtained at NNLO QCD using \textsc{FEWZ} for different values of the $\PWpr$ boson mass.
Distributions of pseudo-data with an additional $\PWpr$ boson signal are generated 200 times per event class and kinematic variable. This ensures a statistically stable outcome. The \pvalue between signal-induced pseudo-data and SM expectation is calculated for each of the pseudo-experiments, and the pseudo-data generation corresponding to the median \pvalue is chosen as the representative event class distribution.
Up to 10,000 pseudo-experiments are generated under the SM-only hypothesis to account for the LEE in each distribution and event class.

\begin{figure}[htb!]
  \centering
    \includegraphics[width=\cmsFigWidth]{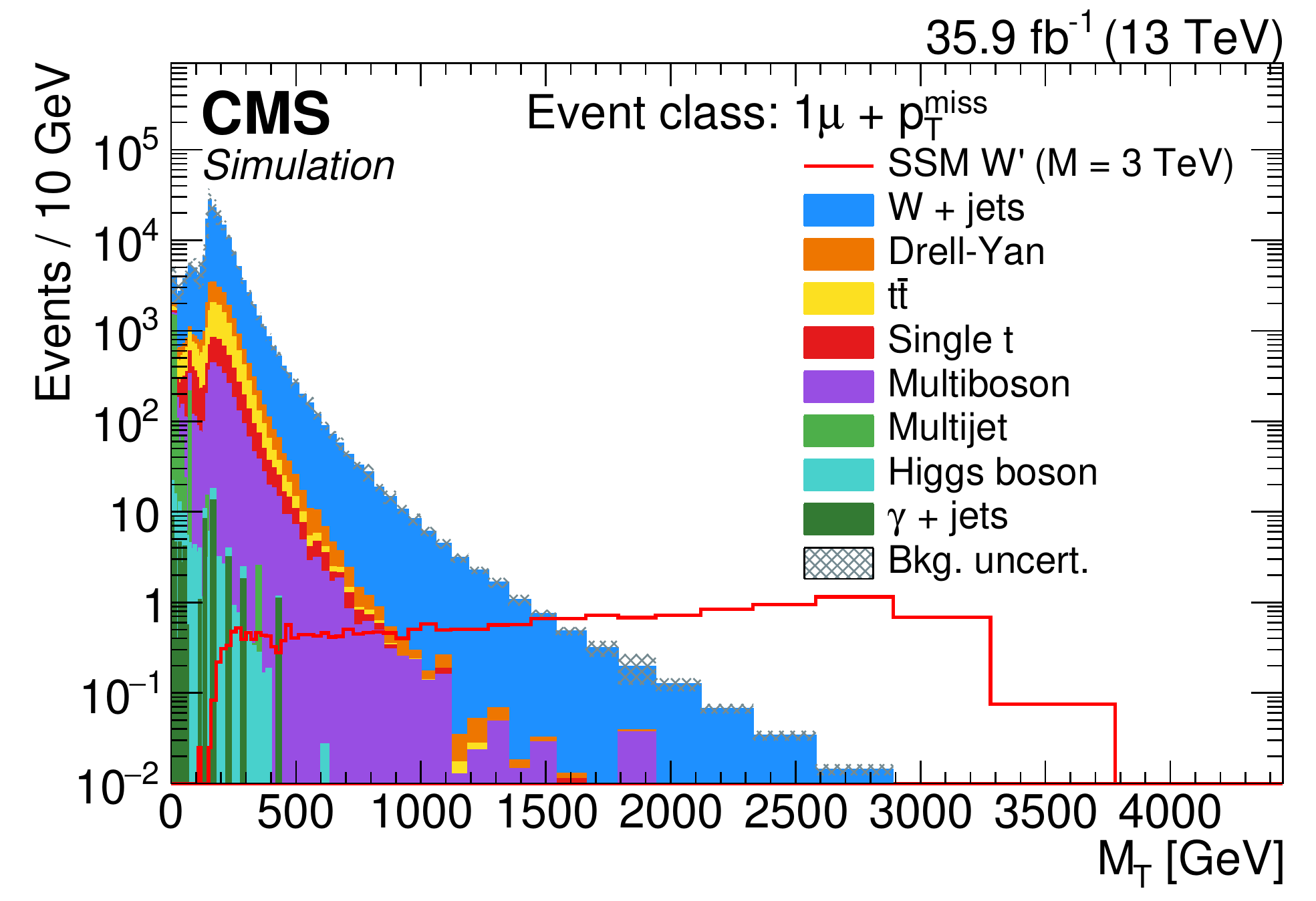}
    \caption{Distribution of the transverse mass for the $1\Pgm+\ptmiss$ exclusive class with a hypothetical SSM $\PWpr$ boson (with mass of 3 \TeV) along with the SM simulation.}
    \label{fig:sensitivity_Wprime_signal}
\end{figure}

\begin{figure*}[htbp!]
   \centering
     \includegraphics[width=0.495\textwidth]{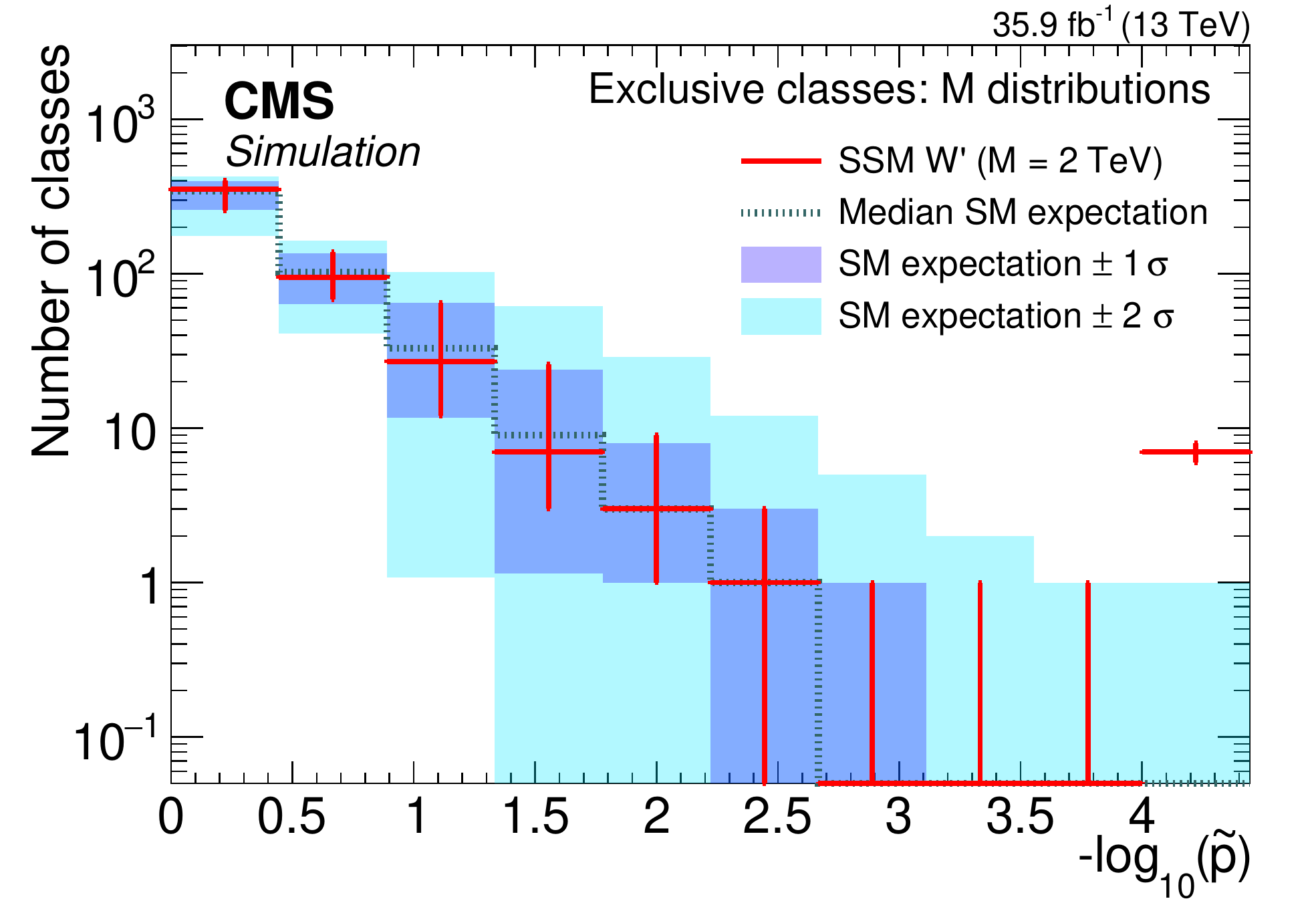}
     \includegraphics[width=0.495\textwidth]{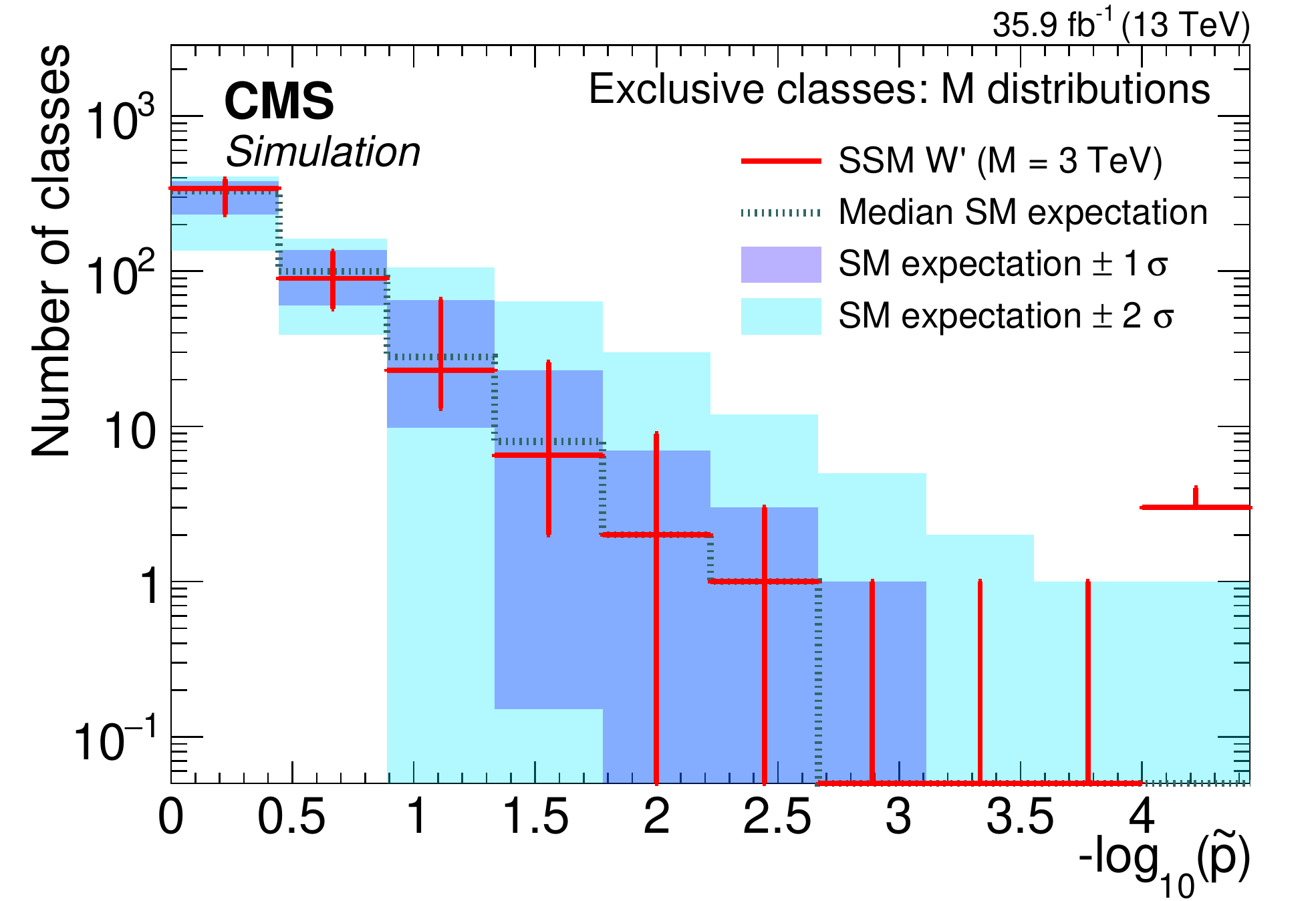}\\
     \includegraphics[width=0.495\textwidth]{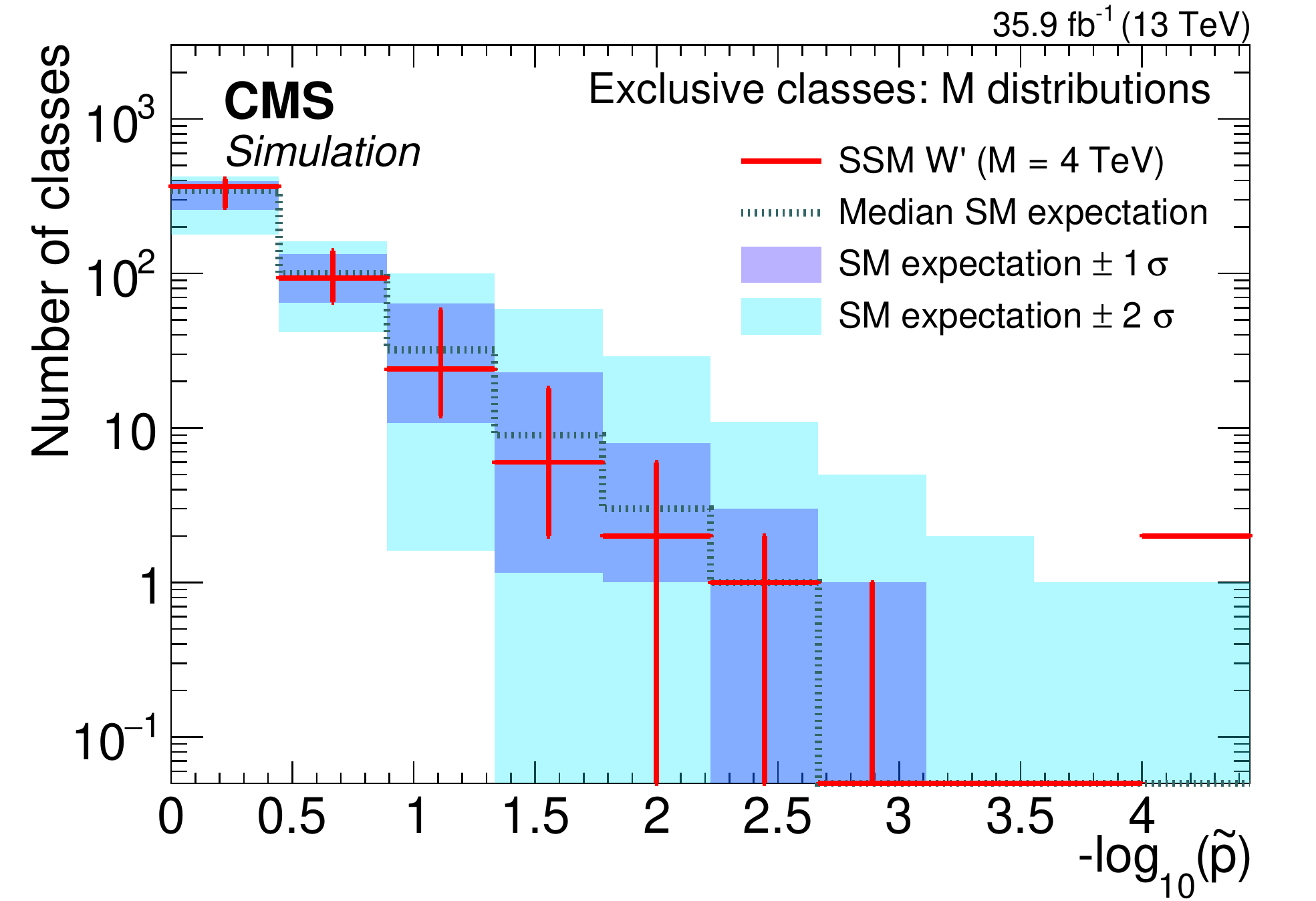}
     \includegraphics[width=0.495\textwidth]{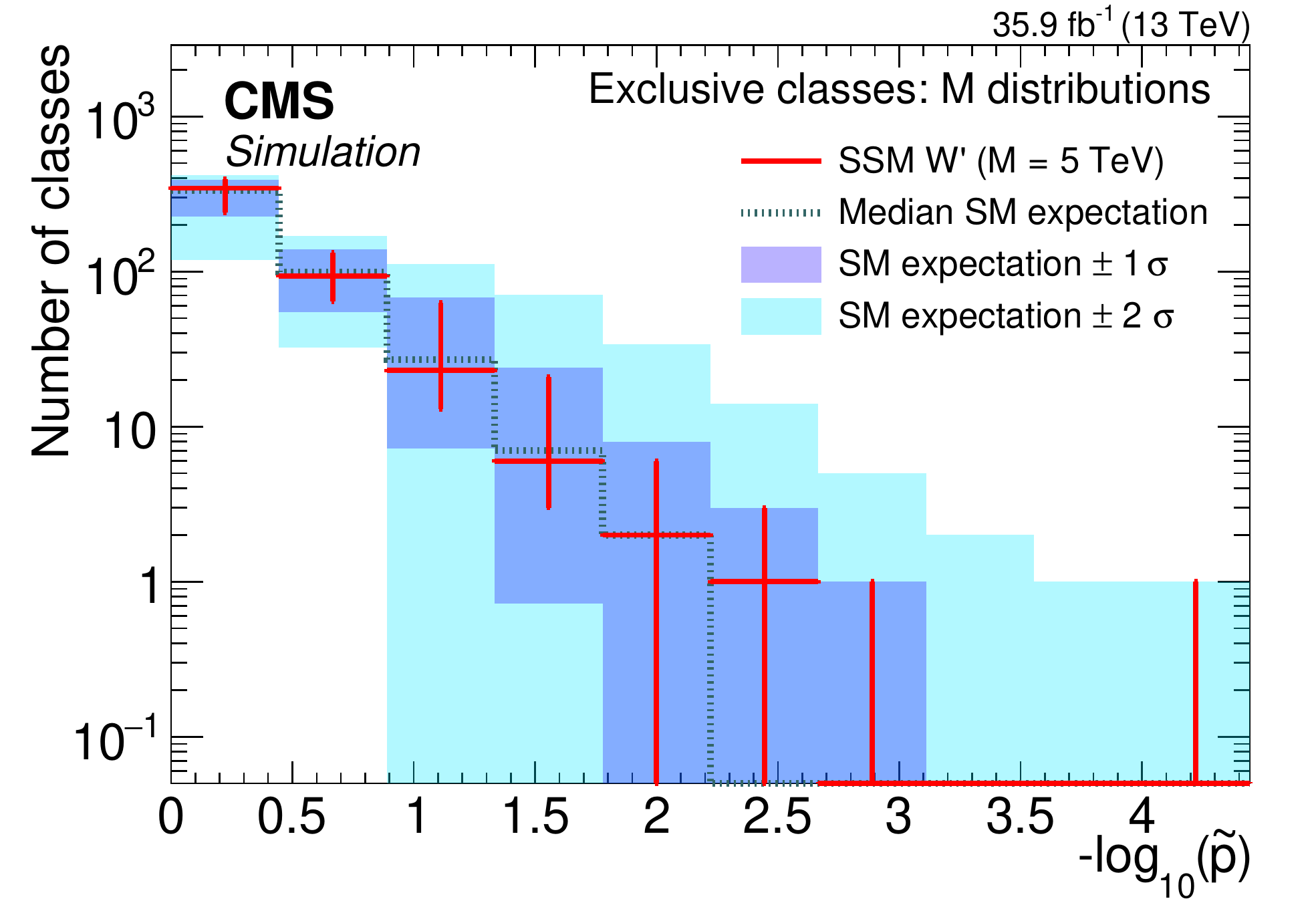}
     \caption{Distribution of \ptildevalues for the RoI scan in exclusive classes for the invariant mass (transverse mass for classes with \ptmiss) with assumed values for the mass of the SSM $\PWpr$ boson of 2 (upper left), 3 (upper right), 4 (lower left), and 5\TeV (lower right). The uncertainty in the distribution of \ptildevalues for the signal is obtained from the variations in the pseudo-data performed with the $\PWpr$ signal simulation.}
    \label{fig:sensitivity_Wprime_ptilde600}
\end{figure*}

As a representative example, a scan of the (transverse) invariant mass distribution in exclusive event classes is presented here. The distribution of a signal and the SM background for the $1\Pgm+\ptmiss$ event class is shown in Fig.~\ref{fig:sensitivity_Wprime_signal} for a $\PWpr$ boson with mass of 3 \TeV. For $\PWpr$ masses of 2, 3, 4, and 5\TeV, the final $\tilde{p}$ distributions for the scan of the invariant (transverse) mass in exclusive event classes are shown in Fig.~\ref{fig:sensitivity_Wprime_ptilde600}.
Two or more final states with significant deviations from the SM simulation beyond the expectation are found, as seen in the entries in the final bin of each distribution that lie outside of the uncertainty bands of the SM-only expectation, thus illustrating the ability of \music to identify deviations arising from a signal. The observation of such deviations would prompt dedicated studies in the event classes of interest probing the possibility of BSM physics phenomena. The signal corresponding to 4\TeV leads to significant deviations only in the
$1\Pe+\ptmiss$ and $1\Pgm+\ptmiss$ event classes. As shown in Fig.~\ref{fig:sensitivity_Wprime_ptilde600}, the scan performed for the $\PWpr$ boson with a mass of 5\TeV did not show any significant deviations.
These results are consistent with the dedicated analysis of the same data set \cite{Sirunyan:2018mpc}, where stronger exclusion limits for the mass of the $\PWpr$ boson of 4.9\TeV were placed at 95\% confidence level based on the individual analyses in the $1\Pe+\ptmiss$ and $1\Pgm+\ptmiss$ channels, respectively, and 5.2\TeV for the combination of both channels.

\begin{figure}[htb!]
   \centering
     \includegraphics[width=0.495\textwidth]{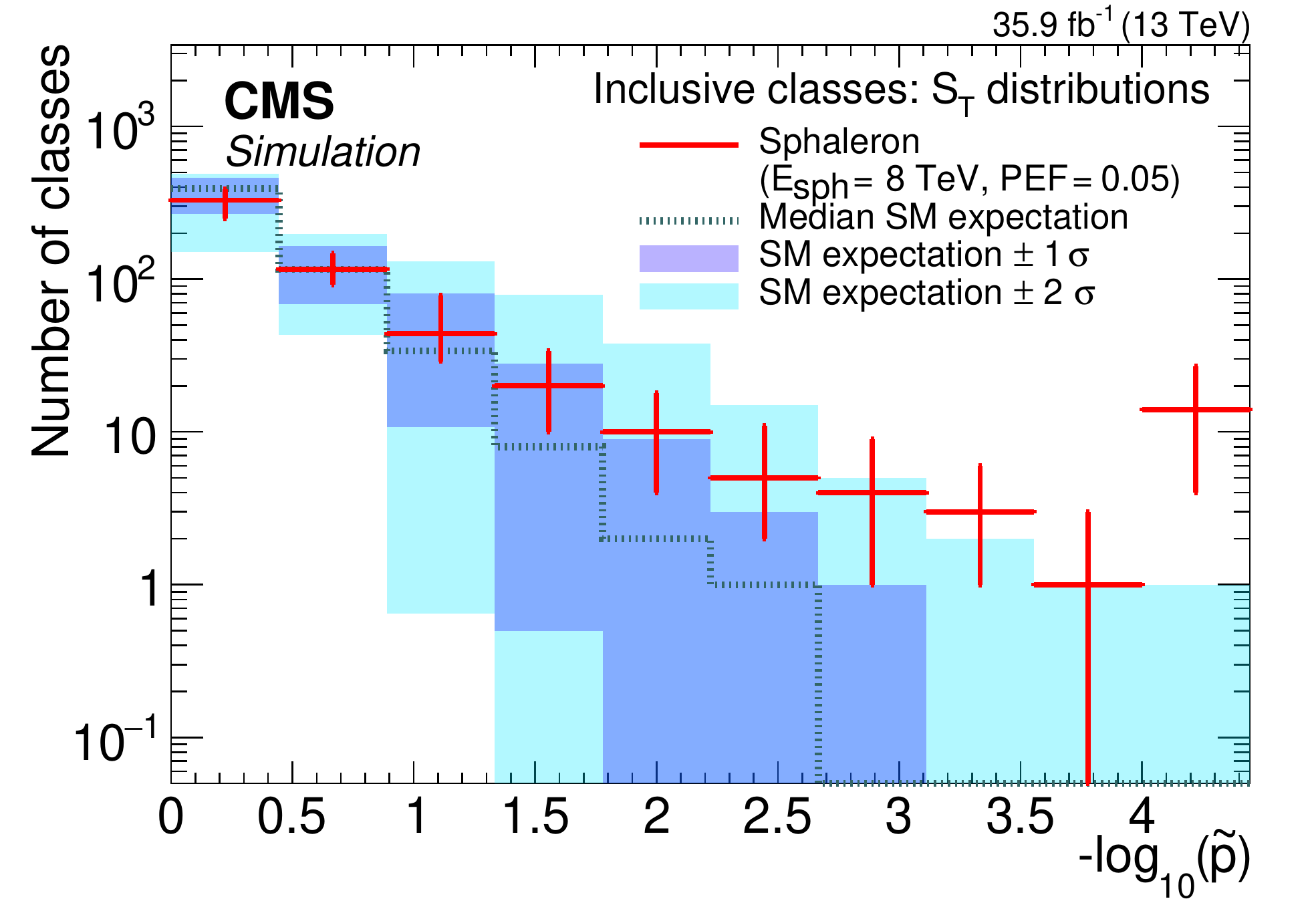}
     \caption{Distribution of \ptildevalues for the RoI scan in inclusive classes for the \ST distributions for a sphaleron signal with $E_{\text{sph}} = 8\TeV$ and PEF = 0.05. The uncertainty in the distribution of \ptildevalues for the signal is obtained from the variations in the pseudo-data performed with the sphaleron signal simulation.}
    \label{fig:sensitivity_Sphaleron_ptilde}
\end{figure}

Another hypothetical BSM signal that has been used to test the capabilities of \music is the EW production of sphalerons \cite{PhysRevLett.37.8, PhysRevD.30.2212, Ringwald:1989ee, Ellis:2016ast}. This model is based on a possible nonperturbative solution to the SM Lagrangian of the EW sector,  which
includes a vacuum transition referred to as a ``sphaleron''. It plays an important role in the EW baryogenesis theory \cite{RevModPhys.71.1463}, which can explain the matter-antimatter asymmetry of the universe.
The CMS experiment has published the results of a search for sphalerons in inclusive final states that are dominated by jets associated with the QCD multijet process \cite{Sirunyan:2018xwt}. No analysis targeting leptonic final states has been performed to date. 
The sphaleron signal sample used for the sensitivity study is generated at LO with the \textsc{BaryoGEN} v1.0 generator \cite{Bravo:2018dkg} with the CT10 LO PDF set \cite{Lai:2010vv} using a threshold energy $E_{\text{sph}}$ = 8\TeV for the sphaleron transition. The cross section for sphaleron production is given by $\sigma$ = PEF $\sigma_{0}$ \cite{Ellis:2016ast}, where $\sigma_{0} = 121$~fb for $E_{\text{sph}} = 8\TeV$, and PEF denotes the pre-exponential factor, defined as the fraction of all quark-quark interactions above the sphaleron energy threshold $E_{\text{sph}}$  that undergo the sphaleron transition.
The result of the \music RoI scan for \ST distributions in inclusive event classes is shown in Fig.~\ref{fig:sensitivity_Sphaleron_ptilde}, where the simulation of the sphaleron production with PEF = 0.05 is used as the signal. Several event classes with large deviations beyond the expectation from the SM-only hypothesis are identified in the final bins of the \ptildevalues distribution. Among the most deviating event classes are the $1 \Pgm + 5 \text{jets} + \ptmiss\ +X$, $1 \Pe + 5 \text{jets} + \ptmiss\ +X$, $1 \Pgm + 1 \PQb + 2 \text{jets} + \ptmiss\ +X$, and $1 \Pe + 1 \Pgm + 3 \text{jets} + \ptmiss\ +X$ event classes. In the inclusive CMS analysis \cite{Sirunyan:2018xwt} based on the same data set, an upper limit of PEF = 0.002 was set at the 95\% confidence level.
This result demonstrates the sensitivity of \music to an example of BSM physics in final states where no previous search has been conducted by the CMS experiment.

\begin{figure}[h!]
  \centering
    \includegraphics[width=0.495\textwidth]{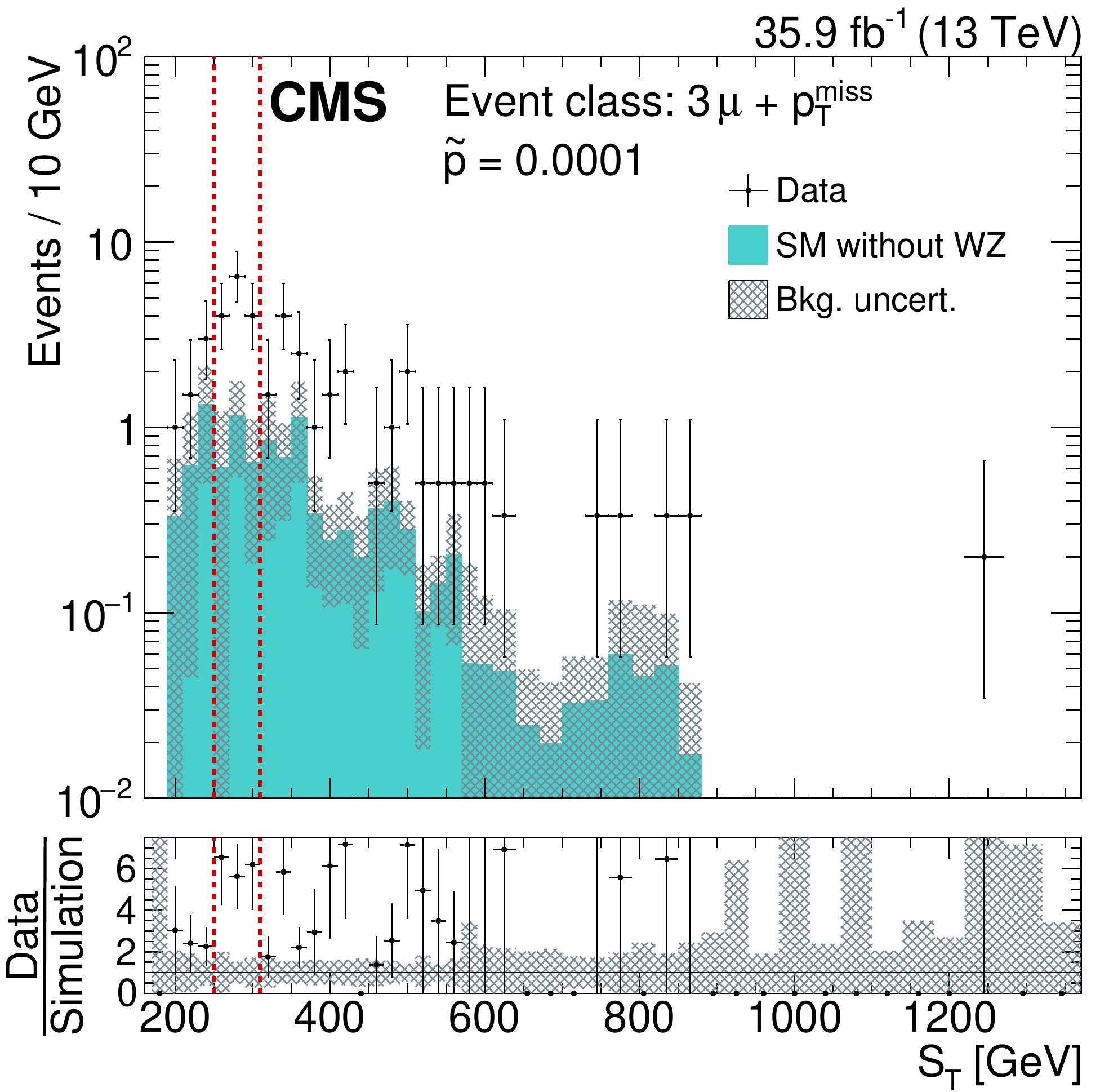}
    \includegraphics[width=0.495\textwidth]{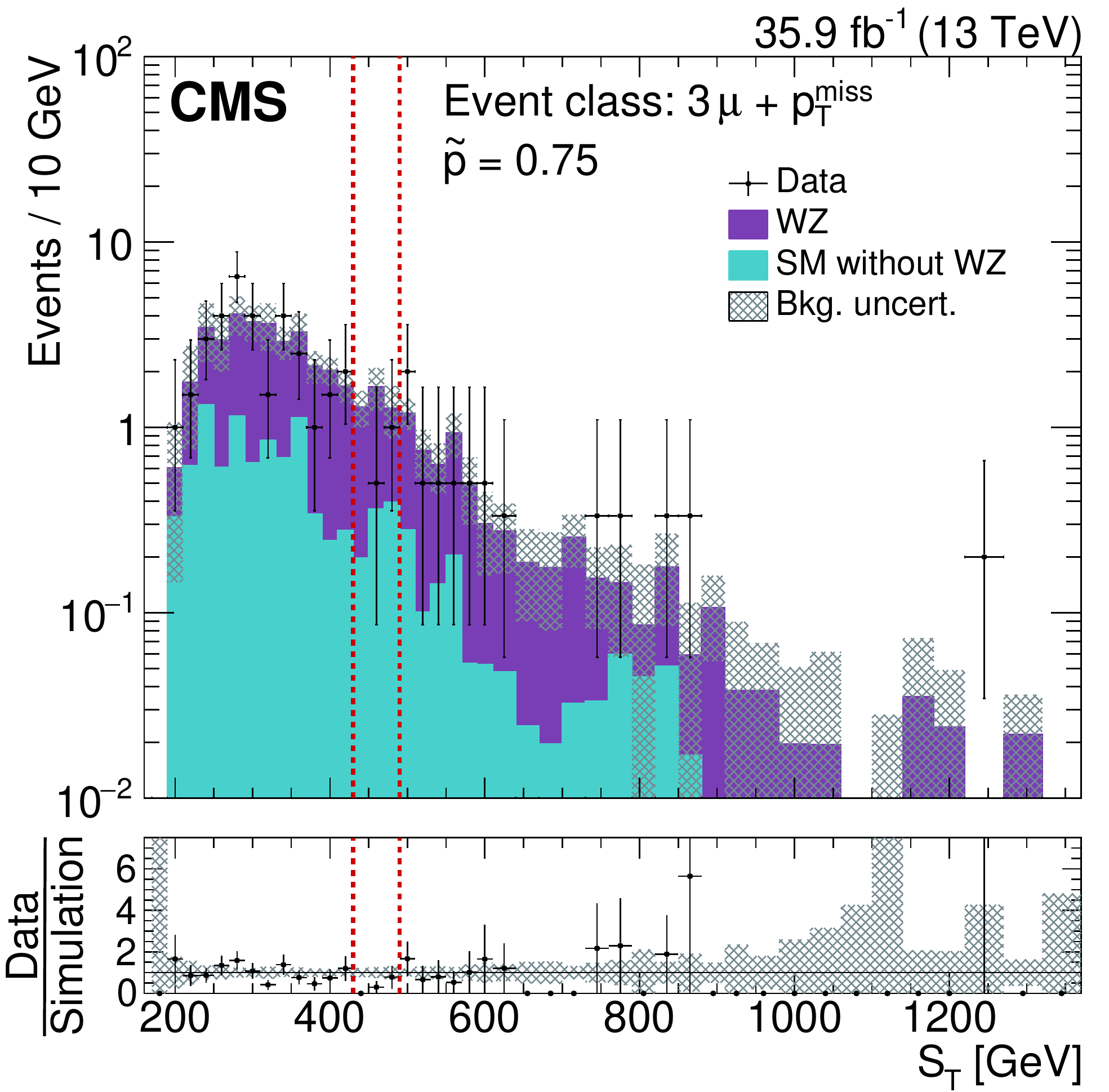}
    \caption{Distributions of \ST for the $3\Pgm+\ptmiss$ exclusive class without (\cmsLeft) and with (\cmsRight) $\PW\PZ$ production as part of the SM simulation. The data events are shown in black and the simulations of the SM processes are shown as coloured histograms. The region enclosed within the red dashed lines is the region of interest.}
    \label{fig:sensitivity_wz_MET}
\end{figure}

In a second approach to evaluate the sensitivity of the \music analysis, a single SM process is removed from the SM simulation, and the scanning algorithm is applied against the recorded CMS data using the modified SM simulation.
In the example shown here, the process of $\PW\PZ$ diboson production is removed. Several final states show large and significant deviations with $\tilde{p} < 0.0002$, compared to the prediction of having no final states showing such a deviation based on the simulation.
The most significant final states concern classes with three leptons, as well as three leptons and \ptmiss, corresponding to event classes where the $\PW\PZ$ process is expected to contribute, confirming the ability of the \music analysis to detect deviations corresponding to the missing $\PW\PZ$ process.
Figure~\ref{fig:sensitivity_wz_MET} shows the event class $3\Pgm+\ptmiss$ with and without the $\PW\PZ$ process as part of the SM simulation. 
The sensitivity has also been verified by removing other SM processes with smaller cross sections, such as $\PZ\PZ$ and $\PQt\PAQt\PZ$ production, from the SM simulation, leading to similar conclusions. 
For the case where the $\PQt\PAQt\PZ$ process is removed from the SM background, the $3 \Pe + 1 \PQb + 2 \text{jets} +\text{Njets}$ jet-inclusive event class shows the most significant deviation with $\tilde{p} < 0.0002$ for the RoI scan of the \ST distributions in jet-inclusive event classes.

These sensitivity studies emphasize the ability of the \music algorithm to identify deviations of the data from the simulated background.

\section{Results}
\label{sec:results_2016}

Using the \music classification procedure we observe 498 exclusive event classes and
571 (530) inclusive (jet-inclusive) event classes with at least one data event.
For the number of classes in the simulation, we
use a lower threshold of 0.1 on the expected total yield to make the number of classes
stable against small changes in the total number of simulated events, and to ensure well-defined
statistical properties for the comparison of deviations.
We did not find any event class that contained data but no simulated events at all. No event class with a total expected yield below 0.1 events from the simulation was found that contained more than one data event, which would have required further investigation.

Before the results of the scan algorithm are presented in detail, the overall performance to reconstruct
and identify objects and their multiplicities is discussed, based on a set of final states
where a single SM process is expected to dominate, and where contributions from a potential signal
are unlikely, based on previously published search and precision measurement results.

\subsection{Commissioning studies and vetoed event classes}
\label{sec:commisssioningstudies}

The final state $\PZ\to \ell\ell+X$ is defined by the presence of at least two same-flavour leptons (\Pe or \Pgm) and any additional number of particles.
For the total inclusive selection, the invariant mass of the lepton pair in the event is studied along with the distribution of the number of jets and the  $\Delta R$ distribution between the leading lepton and a jet, to verify the event cleaning introduced in Section \ref{sec:selection}. The distribution of the number of $\PQb$ jets is checked for $\PZ\to \ell\ell+X$ with at least two jets.
We choose events with electron or muon pairs within a $20\GeV$ window around the mass of the \PZ boson to further validate our ability to reliably reconstruct the lepton kinematic properties, using the $\pt$ distribution and angular distributions for $\eta$ and $\phi$ of the two leading leptons. The distributions are in agreement with the SM simulation within the uncertainties.
In addition, the global event properties \ptmiss, \ST, and  \HT (defined as the sum of $\pt$ of all jets and $\PQb$ jets in an event) are checked for events in the \PZ mass window, and are in agreement with the SM simulation. Events in the \PZ boson mass window along with one additional lepton of a different flavour and without substantial \ptmiss ($< 100 \GeV$) are selected, which form a region dominated by SM processes with relatively small cross sections and sensitive to possible misidentification of charged leptons. The \ST distribution for that selection is in agreement with the prediction within the uncertainties.

The final state $\PQt\PAQt \to \ell+2 \text{jets} + 2 \PQb + X$ is defined by the presence of at least one lepton (\Pe or \Pgm), two jets, two $\PQb$-tagged jets, and any additional number of final-state objects (i.e.~additional leptons, jets, or $\PQb$ jets). We use this final state to validate our ability to describe the kinematic properties in events with a complicated event topology and a larger contribution from misidentified objects. An overall good agreement is observed between the data and simulation. In addition to the inclusive selection, we require the mass of the jet pair to be within a $30\GeV$ window centered on the \PW boson mass, and the \MT of the lepton plus \ptmiss system to be larger than $60\GeV$. Within this selected region we check the hadronic activity \HT and find no significant deviations of the data from the expectation.

Kinematic distributions of photons are studied in photon-triggered events, using $\PGg$+jets events with one photon, one jet, and no leptons nor substantial \ptmiss. The kinematic distributions of the photons, such as the $\pt$, $\eta$, and $\phi$, are in reasonable agreement with the SM simulation.

A few final states have been found to be unsuitable for the present analysis, since they require special treatment of simulated samples in these specific final states, which cannot be applied generally, and are therefore removed from the analysis.
This is the case for the event classes containing two same-flavour leptons and one photon, but no additional leptons or photons. These classes are affected by the overlap between simulated samples for the inclusive $\PZ(\to\ell^+\ell^-)+\text{jets}$ process and specific samples for the SM production of a \PZ boson in association with a photon, leading to an overestimation of the background by the simulation. Since no consistent overlap removal could be performed in these event classes, they are removed from further analysis. Dedicated analyses of the same data set target such final states \cite{Sirunyan:2017hsb}.

\subsection{Total yield scans and object group representation}

\begin{figure*}[h!]
  \centering
    \includegraphics[width=0.75\textwidth]{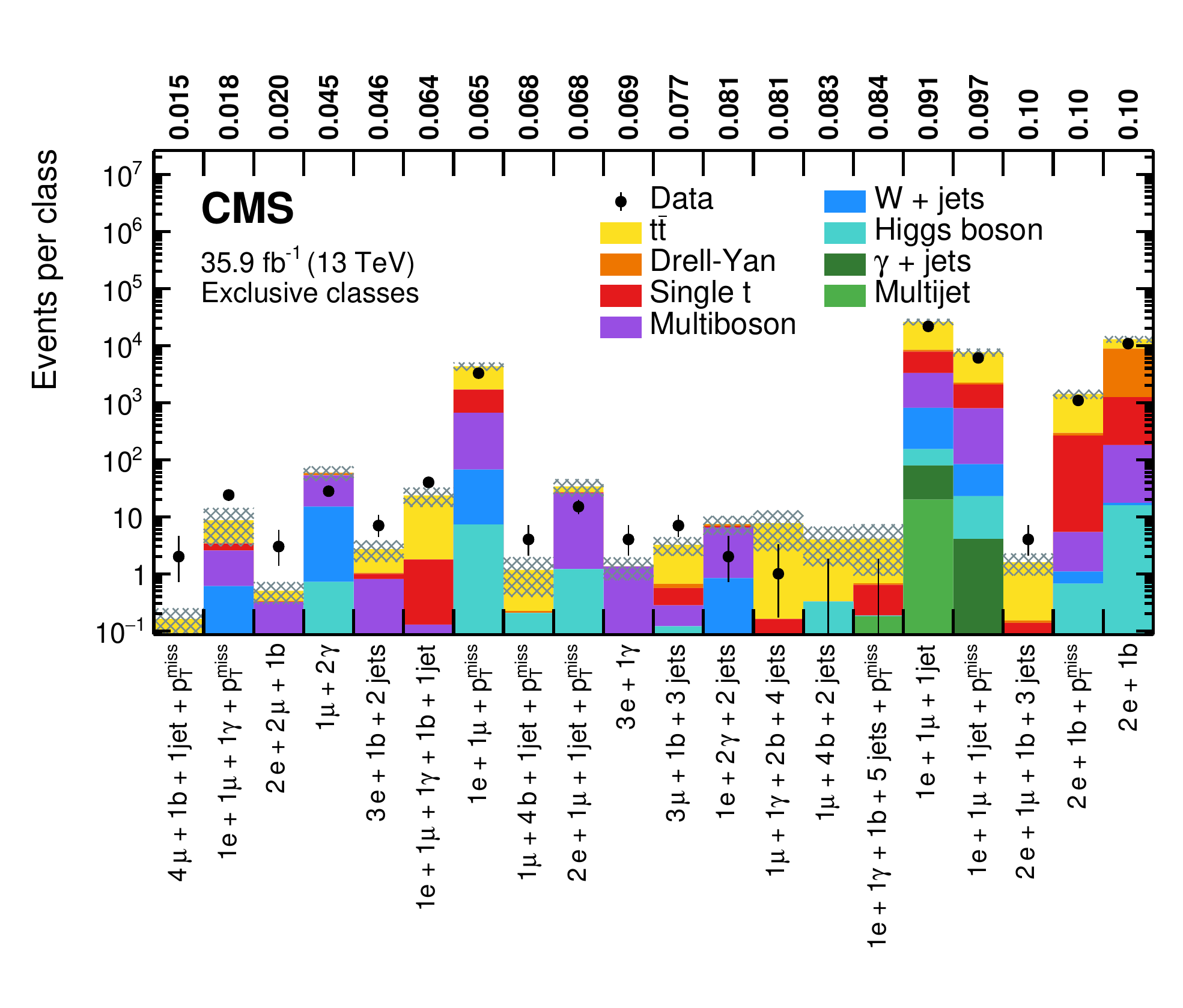}
    \caption{Data and SM predictions for the most significant exclusive event classes, where the significance of an event class is calculated in a single aggregated bin. Measured data are shown as black markers, contributions from SM processes are represented by coloured histograms, and the shaded region represents the uncertainty in the SM background. The values above the plot indicate the observed \pvalue for each event class.}
    \label{fig:simpsons-most-significant-exclusive}
\end{figure*}

Scans are performed  based on the total event yield in the different event classes between data and SM expectation, calculating the \pvalue for each event class based on the total yield. Broad agreement is observed between the data and simulation, with no particular event class being found to have a significant discrepancy between the data and the SM simulation. Selected results for the exclusive event classes are shown in Fig.~\ref{fig:simpsons-most-significant-exclusive},
 where the 20 most significant event classes are displayed, along with the \pvalue for each event class calculated based on the total event yield. The \pvalues for the most significant classes are within the expectations of the SM considering the number of classes.

\begin{figure*}[htb!p]
    \centering
        \includegraphics[width=0.75\textwidth]{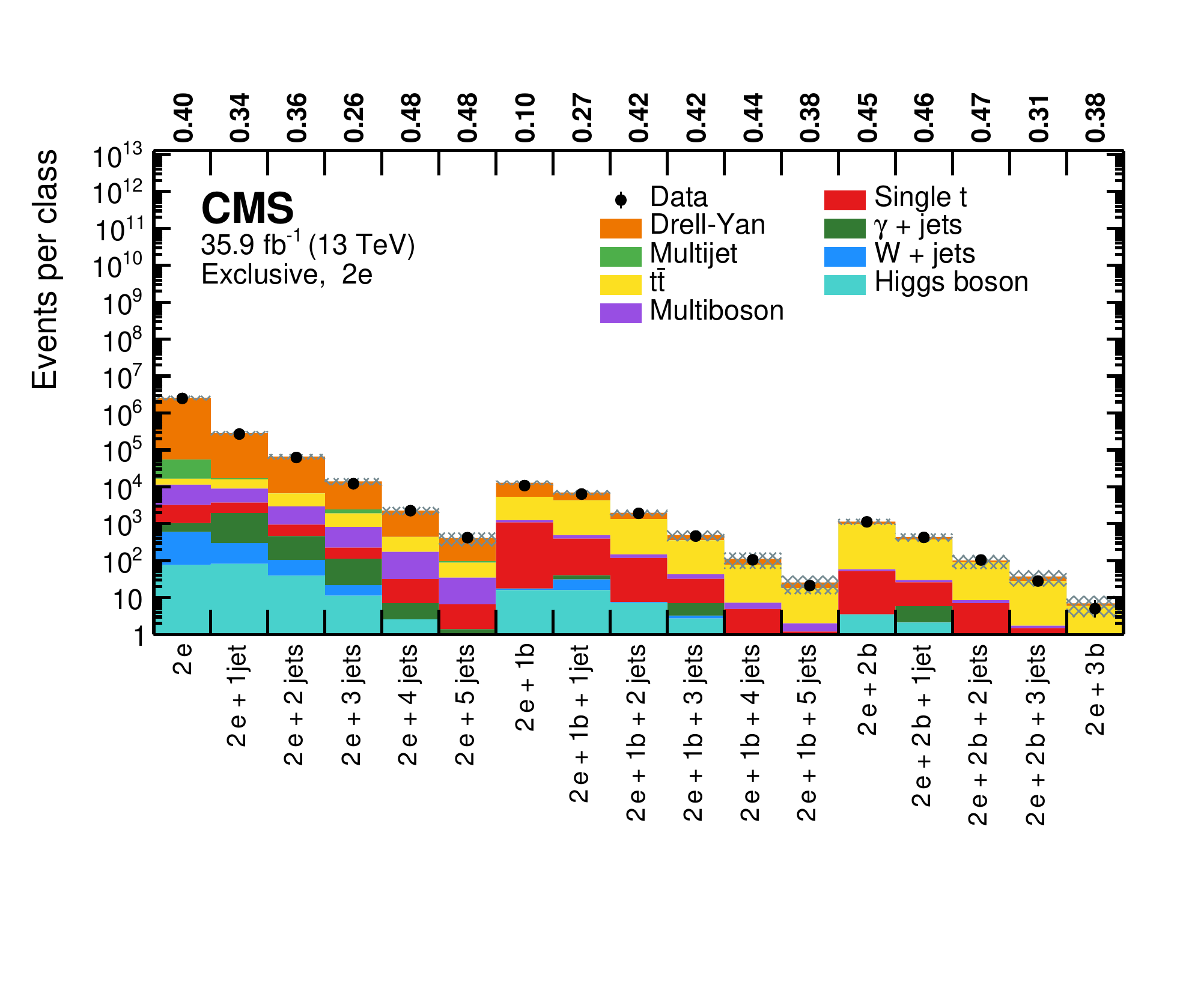}
        \includegraphics[width=0.75\textwidth]{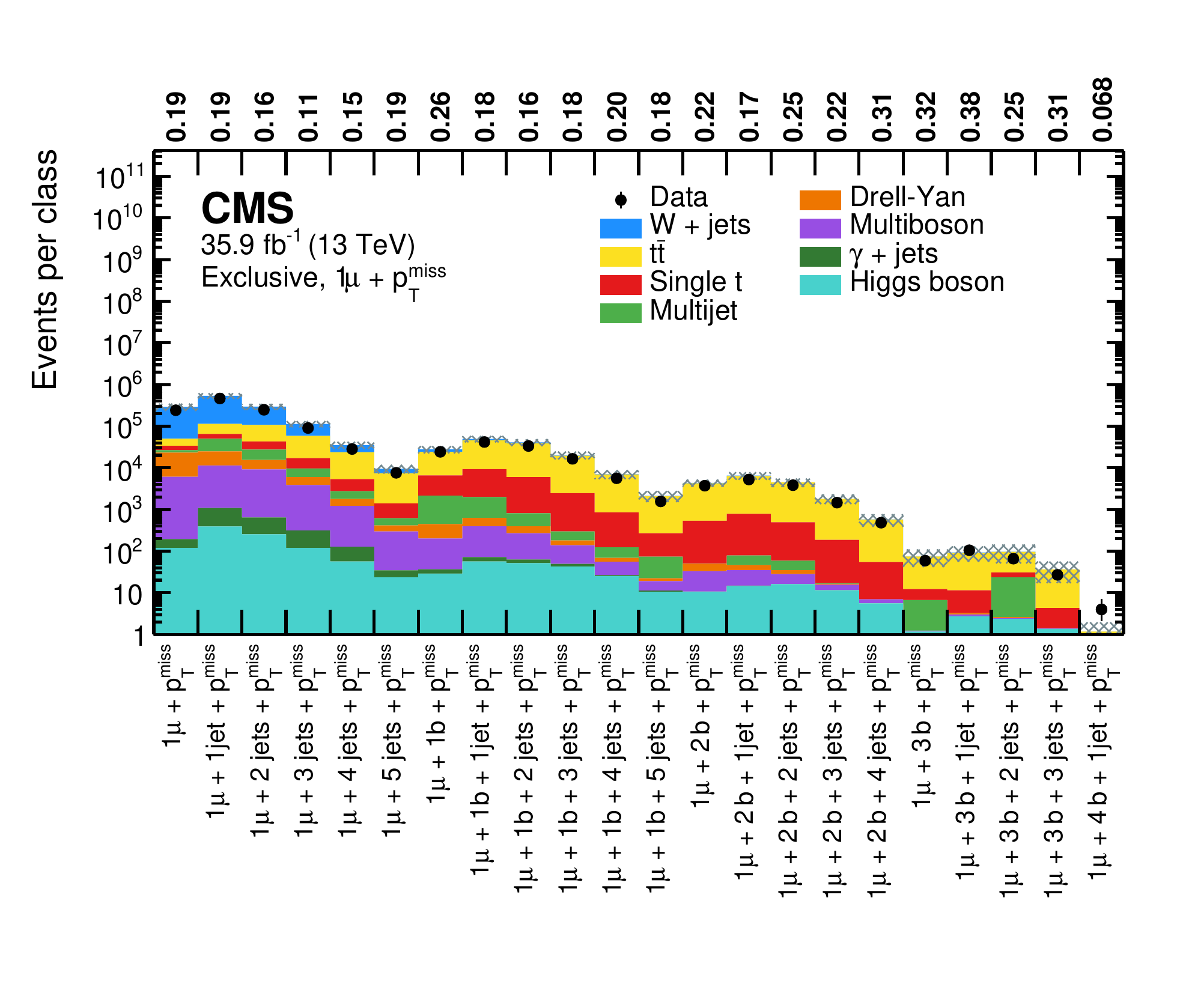}
        \caption{Overview of total event yields for event classes corresponding to the double-electron (upper) and for the single-muon + \ptmiss object groups (lower). Measured data are shown as black markers, contributions from SM processes are represented by coloured histograms, and the shaded region represents the uncertainty in the SM background. The numbers above each plot indicate the observed \pvalue for the agreement of data and simulation for the corresponding event class.}
        \label{fig:Simpsons-2Ele-example}
\end{figure*}

{\tolerance=800 Further results are presented of the comparison of the total event yield in event classes between data and the SM expectation, grouped by their object content.
The term ``object group'' is used to describe 
a set of classes based on the composition of its event content, e.g.~the double electron object group consists of all classes
with exactly two electrons and any number of jets (or $\PQb$ jets). Two examples are shown in Fig.~\ref{fig:Simpsons-2Ele-example}, for the double electron object group and
for the single-muon + \ptmiss group.  For a quantitative comparison of data and simulation,  the event classes are displayed along with the corresponding \pvalue for each event class.
Different jet multiplicities are overall well described, and the total event yields agree with the SM simulation within their uncertainties for different dominating processes, where \PZ and \PW boson decays dominate when additional light-flavour jets are present, whereas final states with additional $\PQb$ jets are dominated by $\PQt\PAQt$ production.
Figures for additional object groups can be found in Appendix~\ref{sec:app_process_group_plots}. \par}

\subsection{Results of the RoI scans}

\begin{figure*}[htbp!]
  \centering
    \includegraphics[width=0.4950\textwidth]{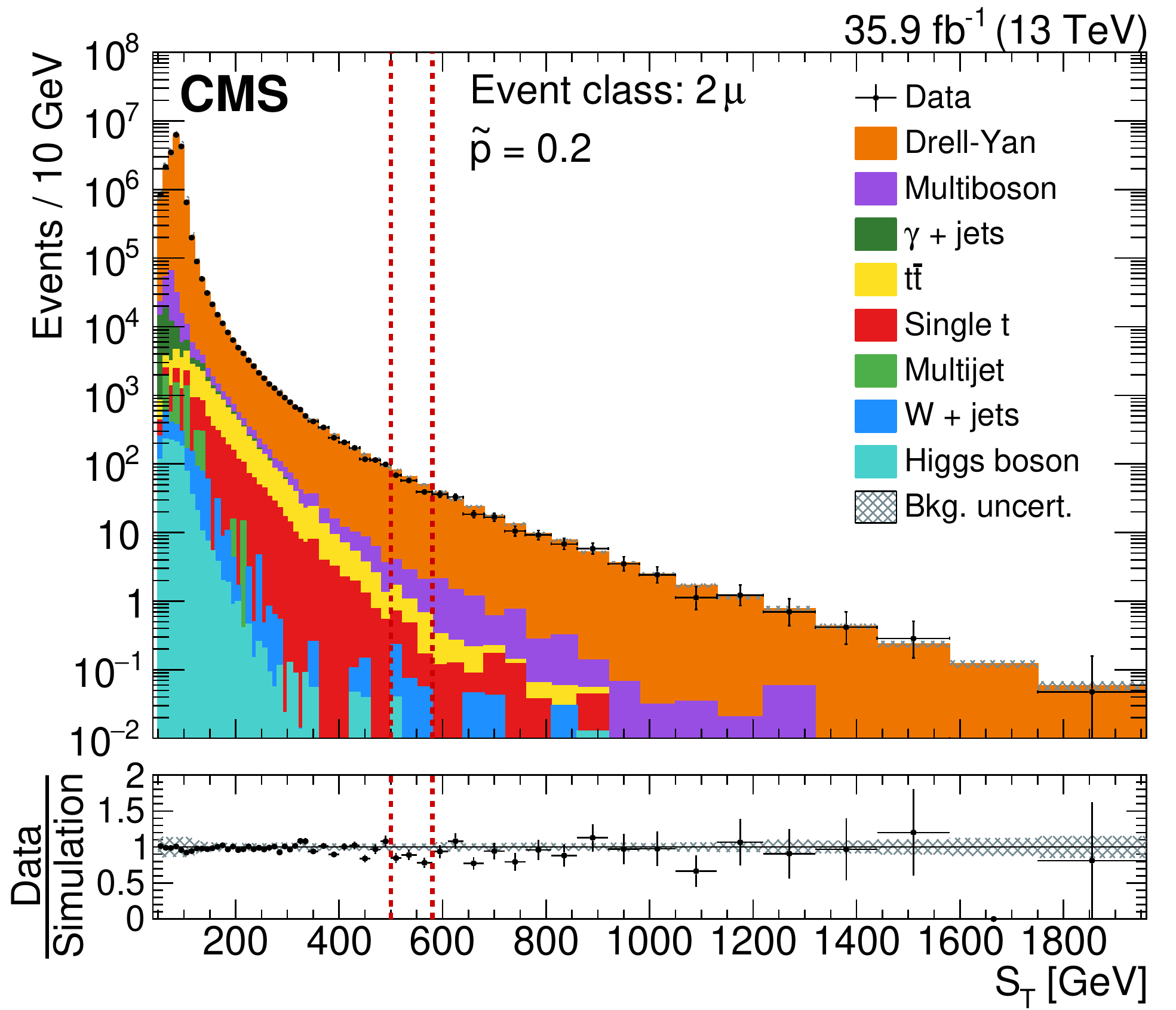}
    \includegraphics[width=0.4950\textwidth]{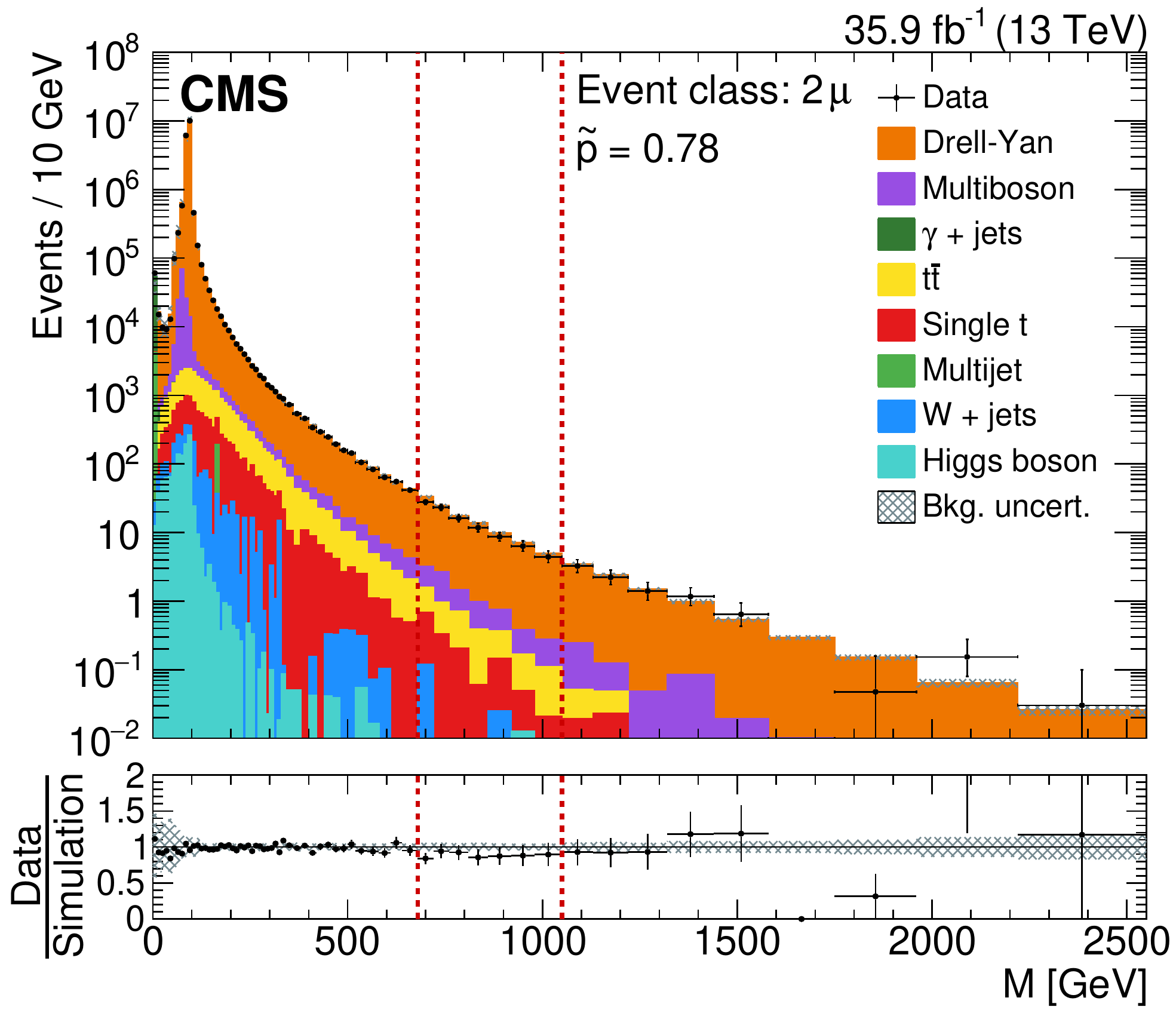}\\
    \includegraphics[width=0.4950\textwidth]{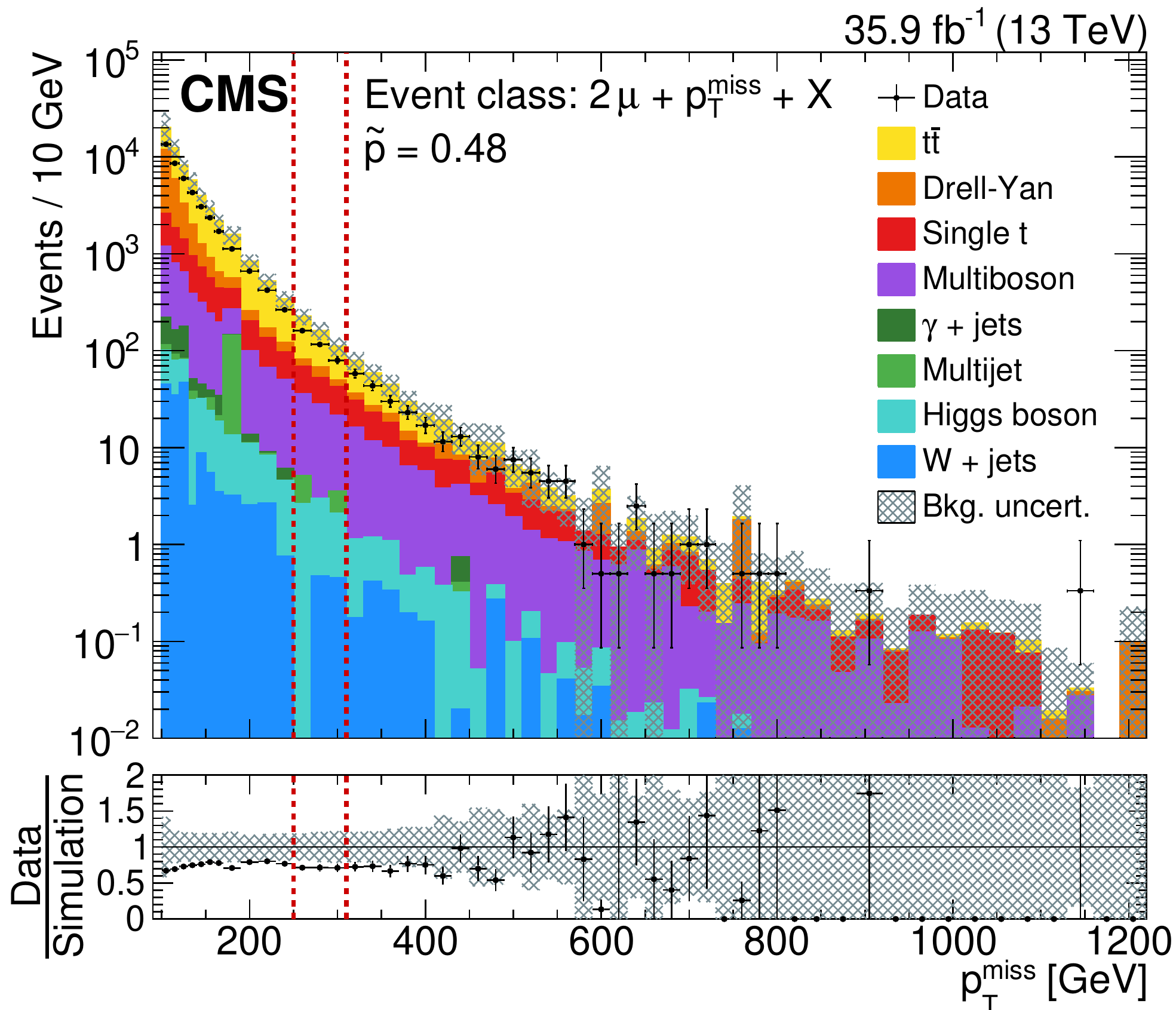}
    \caption{Example \ST (upper left) and \minv (upper right) distributions for the $2\Pgm$ exclusive event class, and the \ptmiss distribution for the $2\Pgm+\ptmiss+X$ inclusive event class (lower). Measured data are shown as black markers, contributions from SM processes are represented by coloured histograms, and the region enclosed by red dashed lines in each figure corresponds to the region of interest determined by the RoI algorithm described in Section~\ref{sec:searchalgorithm}.}
    \label{fig:evaluation_sumpt}
\end{figure*}

Some typical examples of kinematic distributions are shown. The distributions in Fig.~\ref{fig:evaluation_sumpt} for \ST and \minv belong to the $2\Pgm$ exclusive event class, and the \ptmiss distribution is from the $2\Pgm+\ptmiss+X$ inclusive event class. No significant deviations are found with respect to the SM expectations.
The aforementioned distributions illustrate the variable binning depending on the resolution, and the contributions of the different physics processes. They also show experimental features arising from a combination of the threshold effects, such as the trigger and the minimum \pt of the selected objects, along with effects related to the underlying physics, such as the peak associated with the \PZ boson.
In the \ptmiss distribution, a global offset between data and SM simulation is observed, covered by the uncertainties, which are mostly related to
\ptmiss and dominated by the uncertainties in the jet energy scale and resolution. In general, the observed differences between data and SM simulation are covered by the systematic uncertainties over the entire kinematic ranges, and the resulting \ptildevalues for the regions of interest indicate agreement between the two.

\begin{figure}[htb!]
  \centering
    \includegraphics[width=\cmsFigWidth]{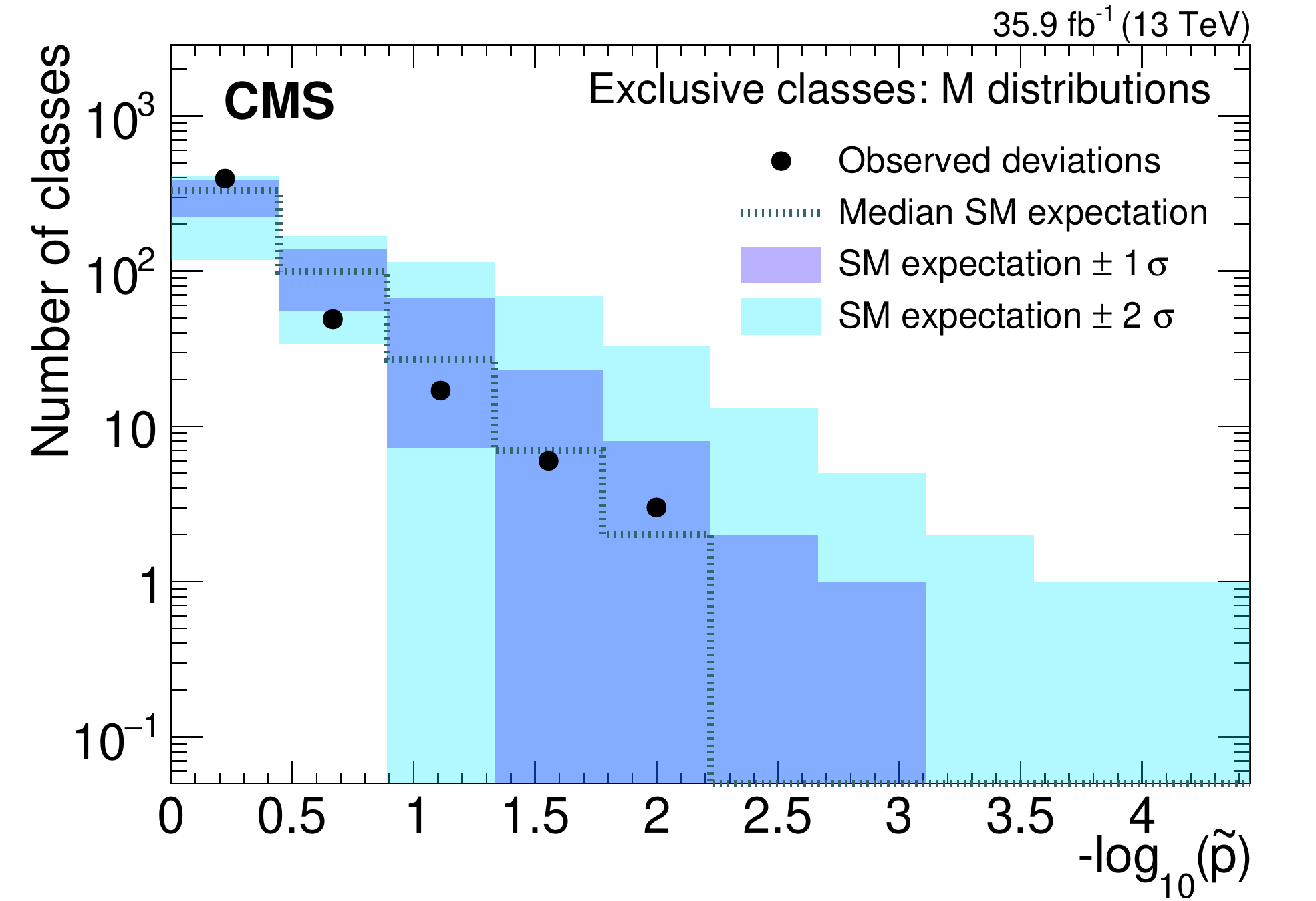}\\
    \includegraphics[width=\cmsFigWidth]{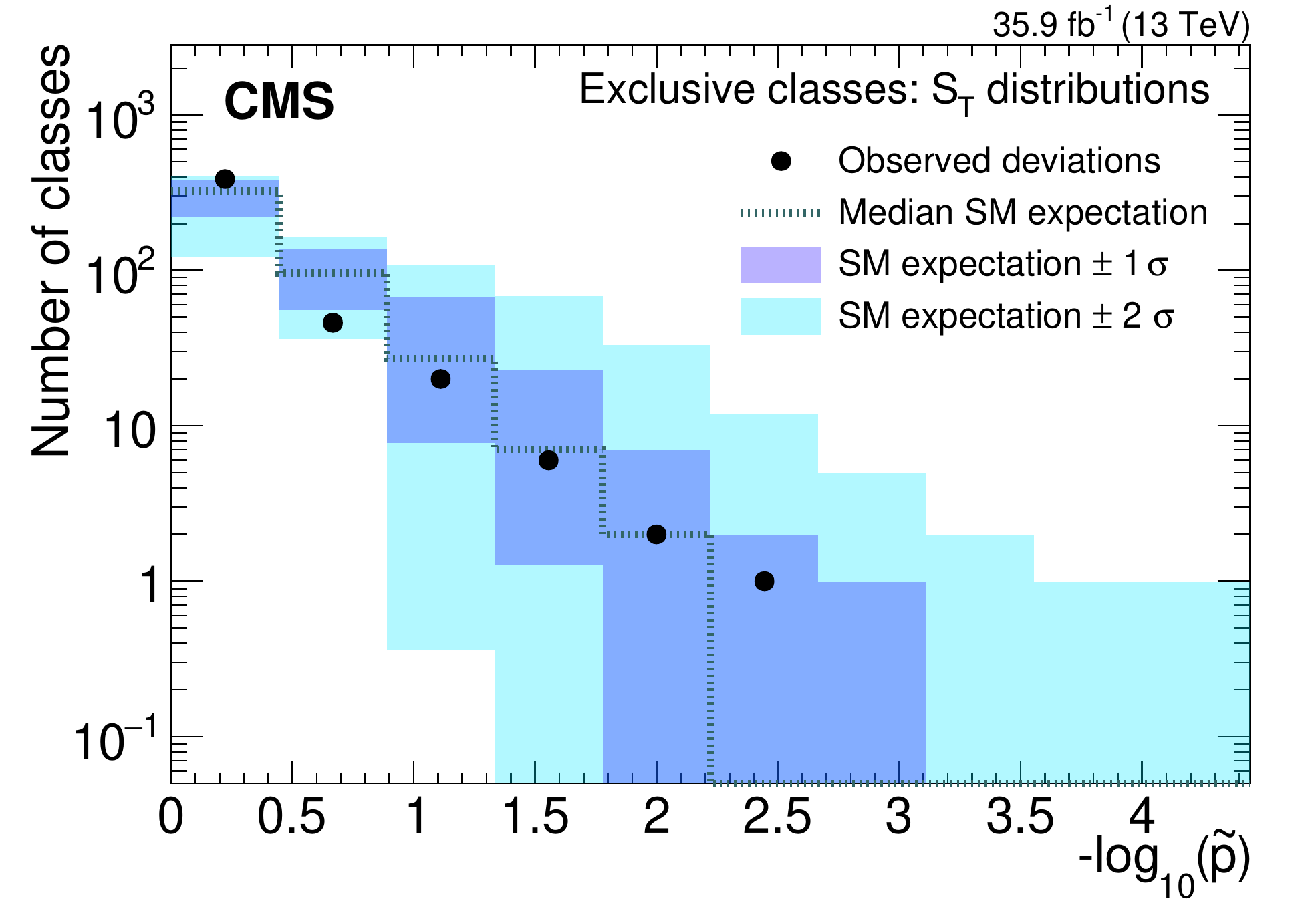}\\
    \includegraphics[width=\cmsFigWidth]{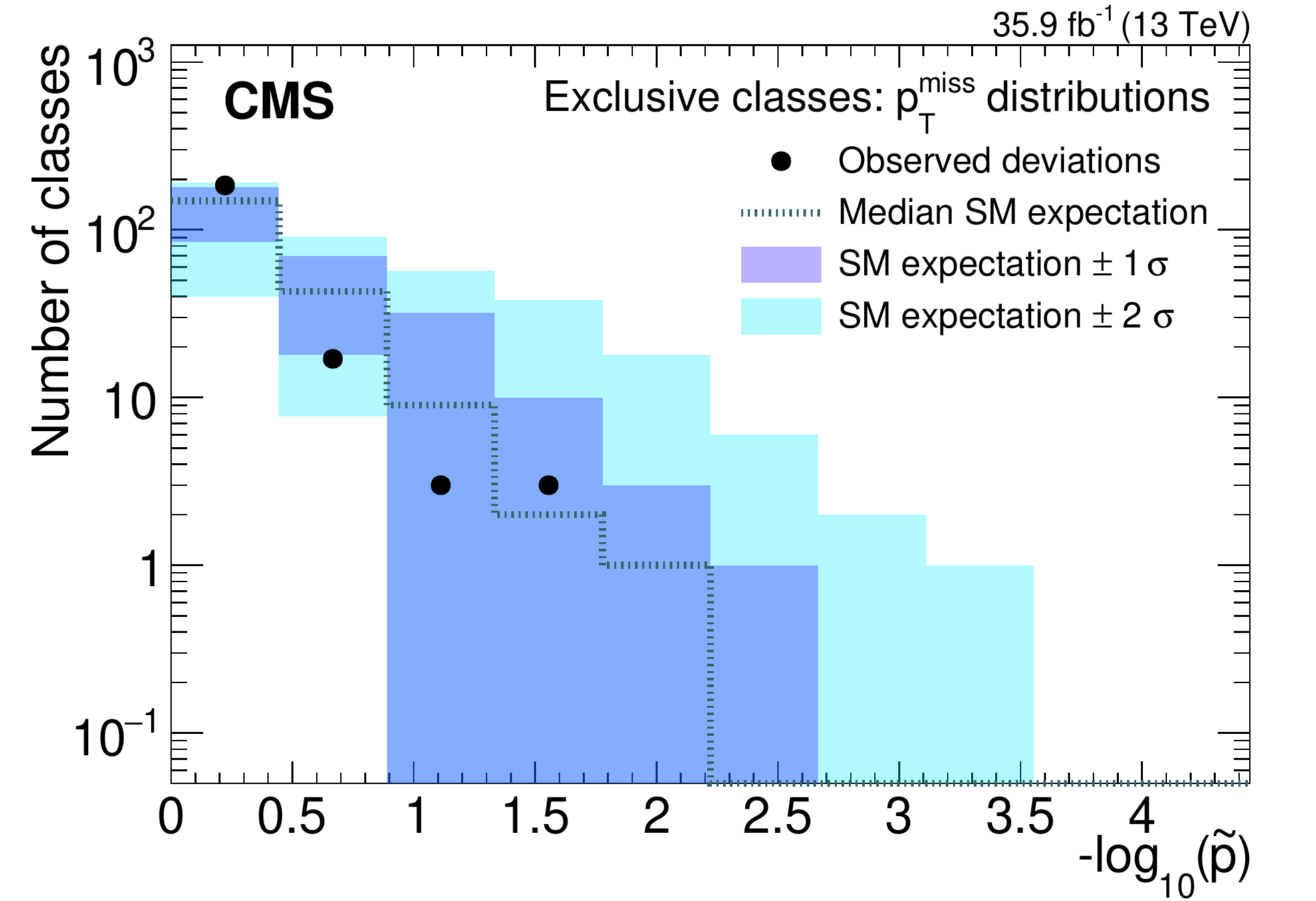}\\
    \caption{Distribution of \ptildevalues for the RoI scan in exclusive classes for the \minv (upper), \ST (middle), and \ptmiss (lower) distributions.}
    \label{fig:ptilde-exclusive}
\end{figure}

\begin{figure}[htb!]
  \centering
    \includegraphics[width=\cmsFigWidth]{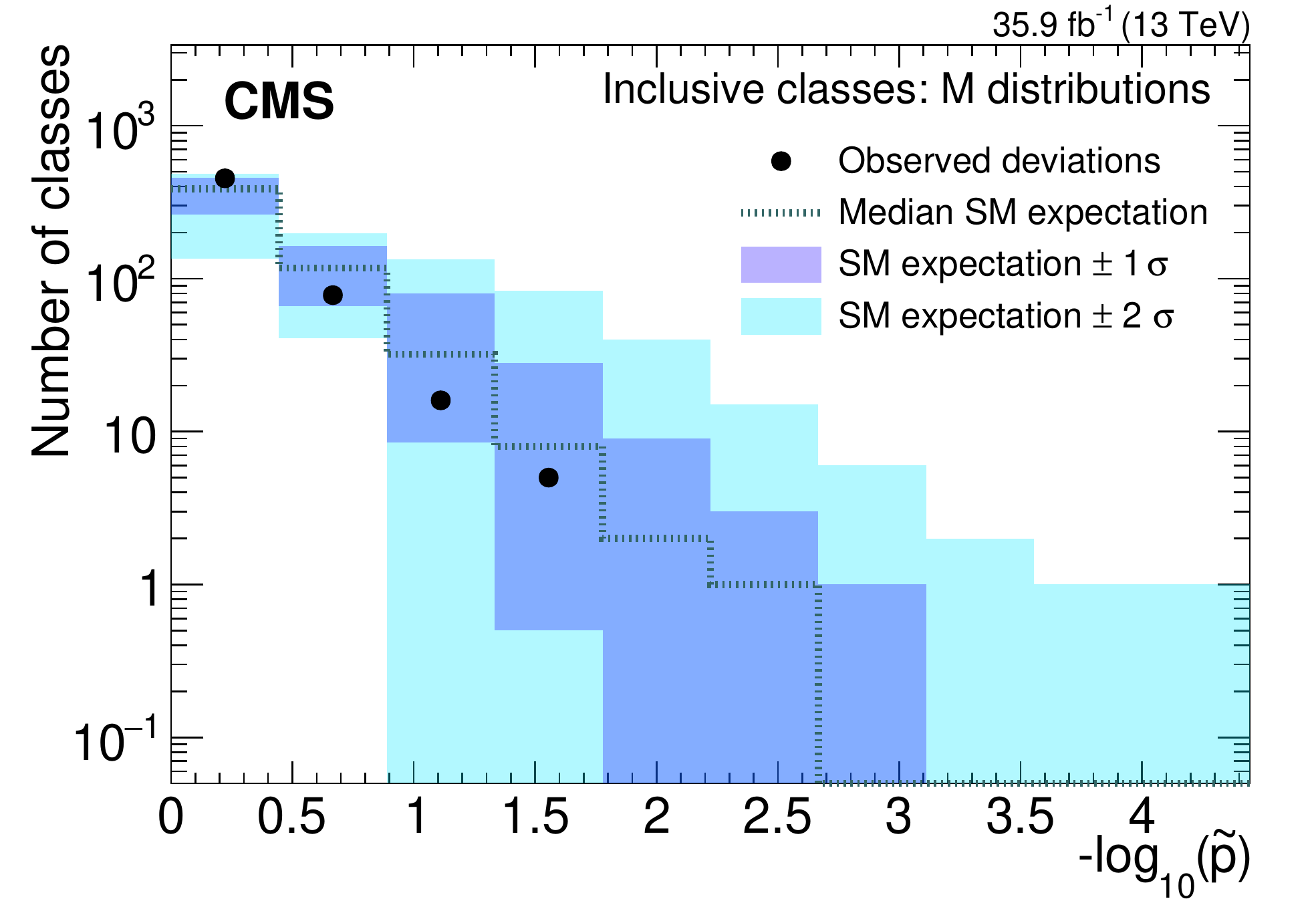}\\
    \includegraphics[width=\cmsFigWidth]{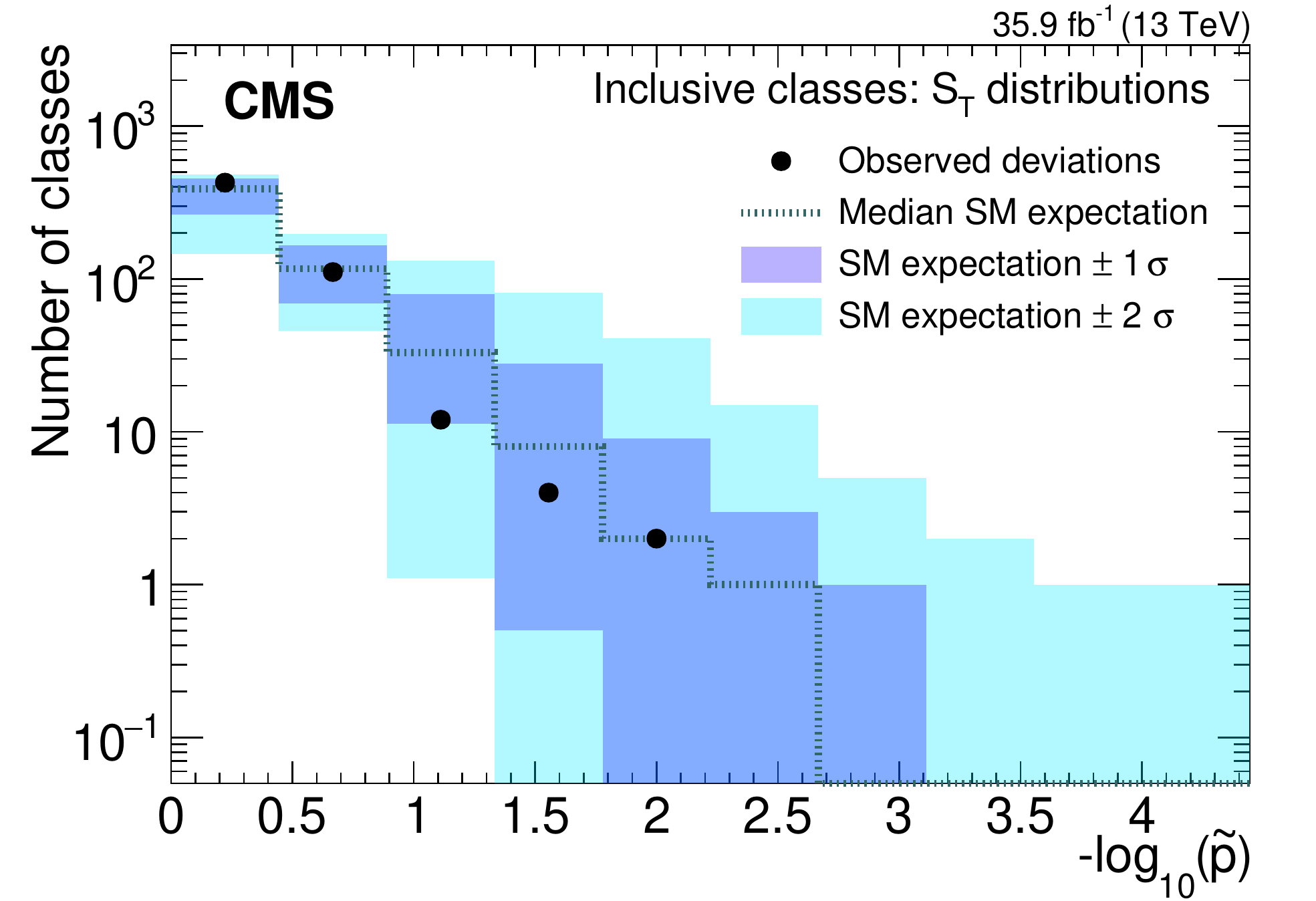}\\
    \includegraphics[width=\cmsFigWidth]{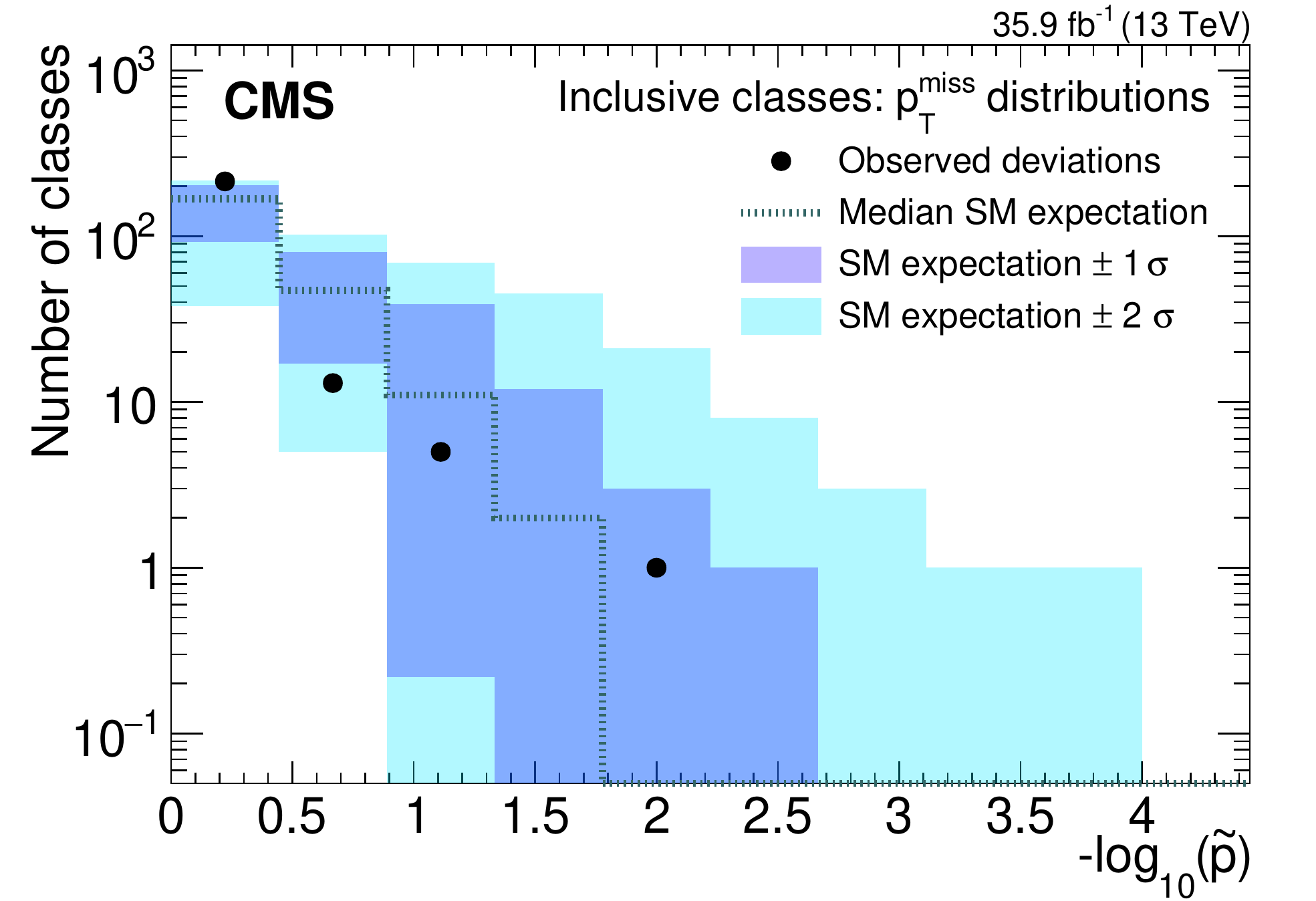}\\
    \caption{Distribution of \ptildevalues for the RoI scan in inclusive classes for the \minv (upper), \ST (middle), and \ptmiss (lower) distributions.}
    \label{fig:ptilde-inclusive}
\end{figure}

\begin{figure}[htb!]
  \centering
    \includegraphics[width=\cmsFigWidth]{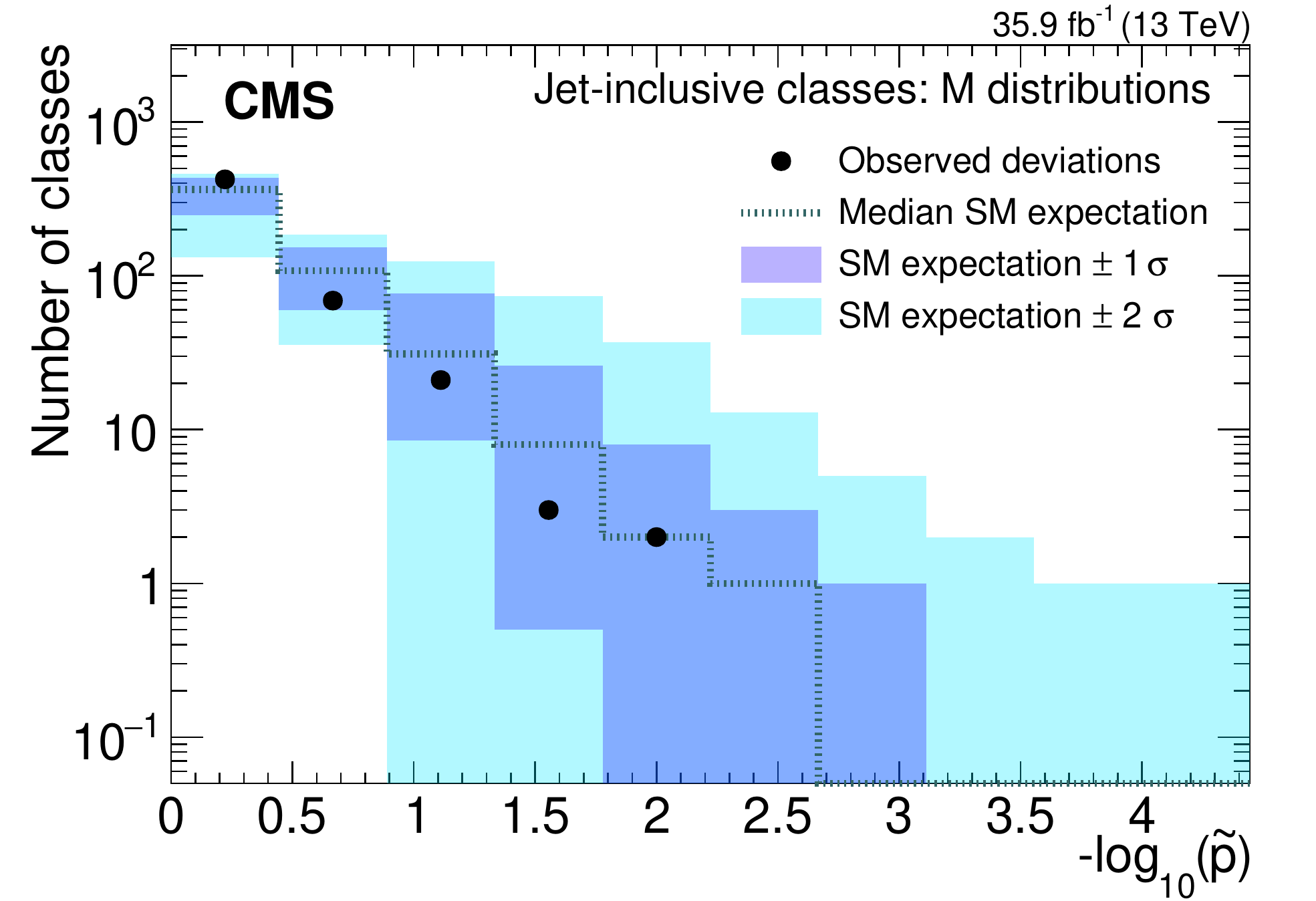}\\
    \includegraphics[width=\cmsFigWidth]{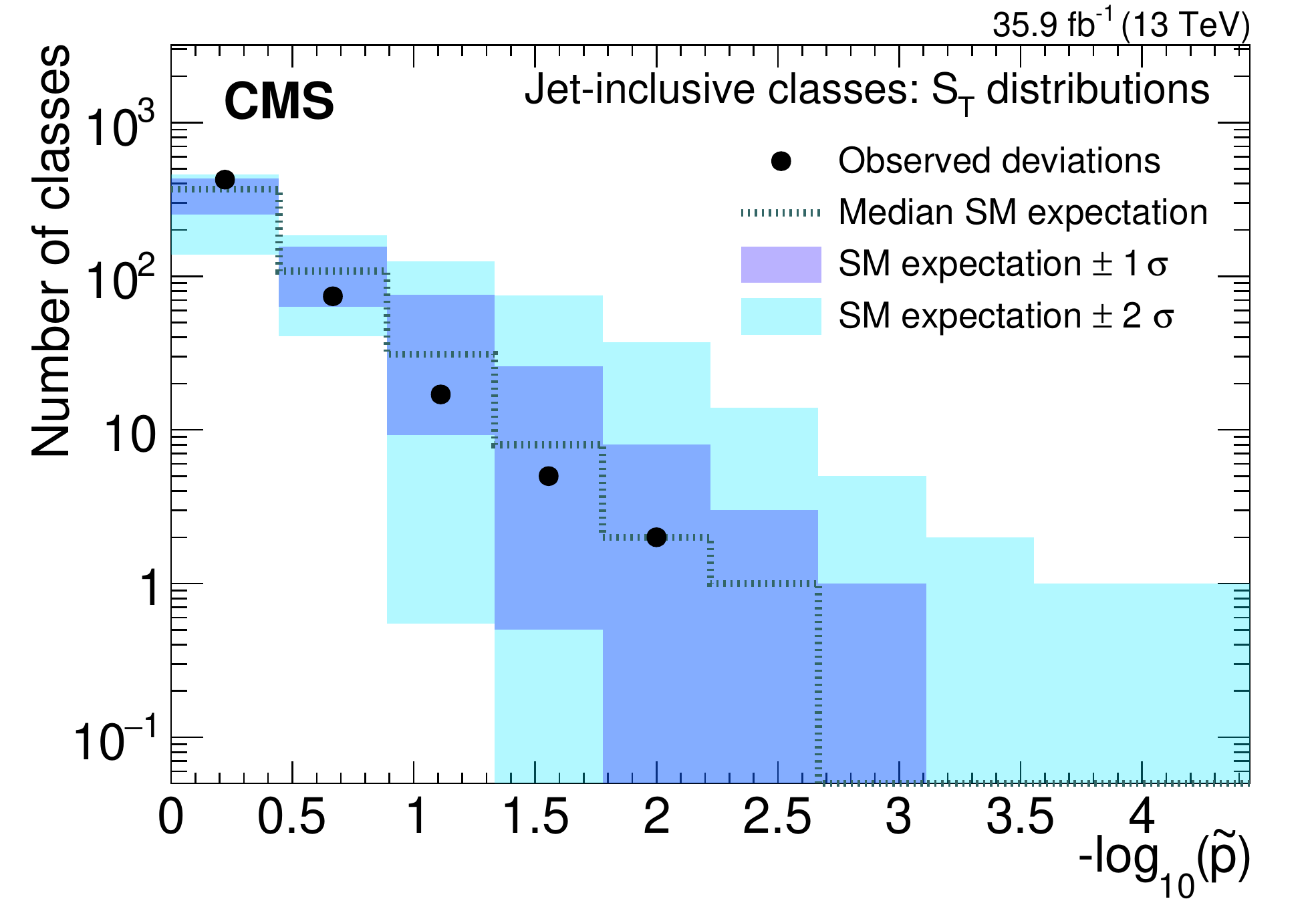}\\
    \includegraphics[width=\cmsFigWidth]{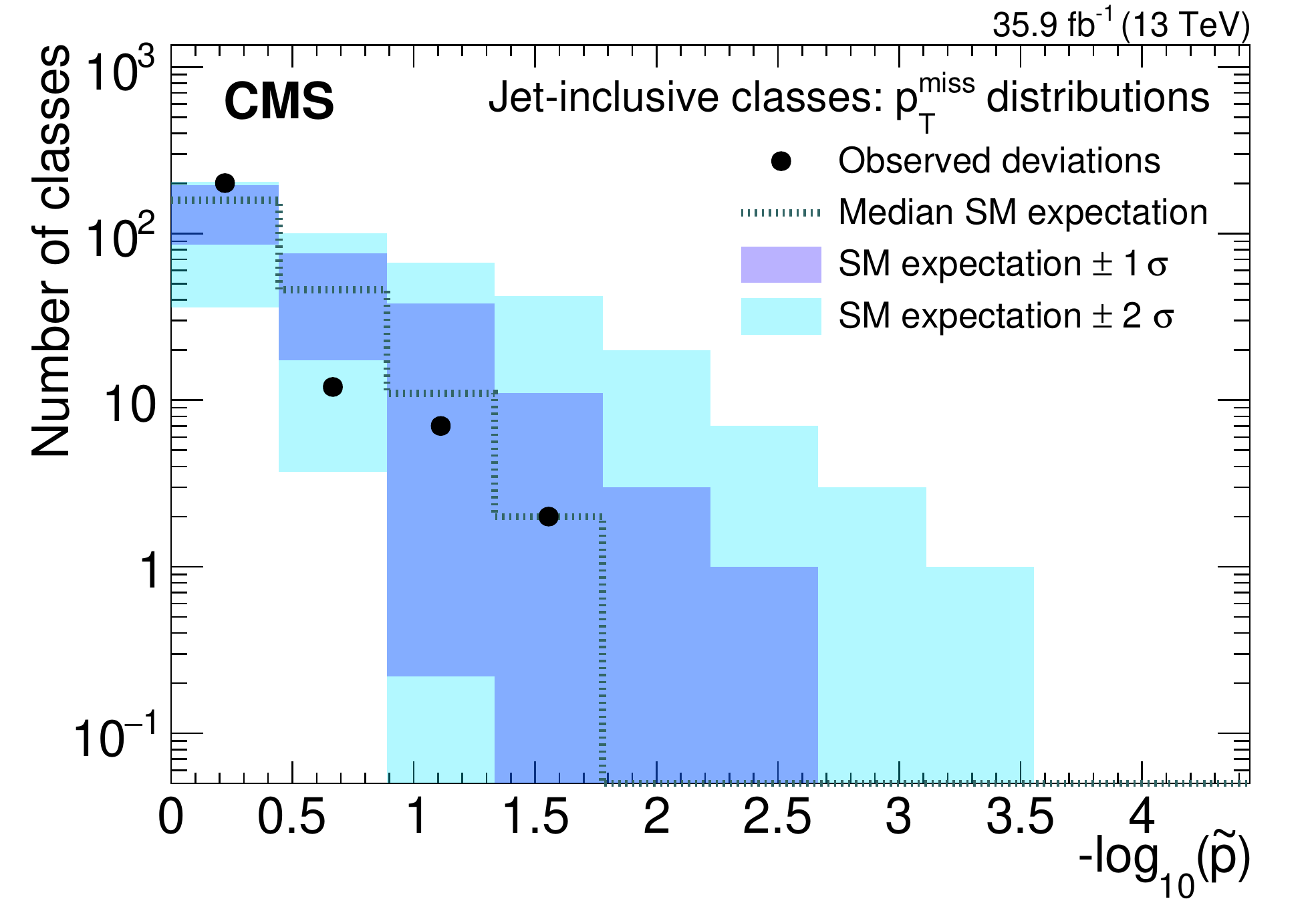}\\
    \caption{Distribution of \ptildevalues for the RoI scan in jet-inclusive classes for the \minv(upper), \ST (middle), and \ptmiss (lower) distributions.}
    \label{fig:ptilde-jet-inclusive}
\end{figure}

The global overview plots for the \minv, \ST, and \ptmiss RoI scans for the exclusive event classes are shown in Fig.~\ref{fig:ptilde-exclusive}. The corresponding plots for the inclusive and the jet-inclusive classes are shown in Figs.~\ref{fig:ptilde-inclusive} and \ref{fig:ptilde-jet-inclusive}, respectively. The distributions observed based on the scans of the data are consistent with the expectations based on simulation within the uncertainty bands.
In general, slightly fewer event classes are observed in data in the second bin of the distributions compared to the expectation, while there are more event classes in data in the first bin, where the observed deviation is smaller. This is a consequence of a possible overestimation of systematic uncertainties (see Section~\ref{sec:global_overview}).

No event classes with an outstanding deviation from the SM simulation beyond the expectation, which could be studied for signs of BSM physics, have been found in the analysed data set.
The largest deviations seen are consistent with the statistical analysis based on the SM-only hypothesis. 
The exact number of event classes and the corresponding values of \ptilde required to be considered significant deviations beyond the SM-only hypothesis depend on the kinematic distributions and the type of event class being probed, and can be inferred from the global overview plots (Figs.~\ref{fig:ptilde-exclusive},~\ref{fig:ptilde-inclusive} and~\ref{fig:ptilde-jet-inclusive}).
The two most significant
event classes from the RoI scan for each kinematic variable are described in Table~\ref{tab:most_significant}, separately for exclusive, inclusive, and jet-inclusive event classes, respectively.
The event classes showing the most significant deviations have been studied in detail, and no systematic trend in related or neighboring event classes has been found. Since the individual event classes do not show a deviation that is statistically significant compared to the expectation, a deeper inspection for possible signs  of BSM physics is not required as a part of this analysis.

\begin{table*}
  \centering
   \topcaption{Overview of the two most significant event classes in each RoI scan. Details of the RoI, the expectation from the SM simulation, and the number of data events within the RoI are shown along with the $p$- and \ptildevalues.}
    \begin{tabular}{l c c c c c}
      \hline
      Event class & RoI & $N_{\text{MC}}$    & $N_{\text{Data}}$       & $p$    & \ptilde   \\
      & [\GeVns{}] &  &  &  & \\
      \hline
      \multicolumn{6}{c}{Exclusive event classes: \minv} \\

      $1 \Pe + 1 \PGm + 1 \PGg + \ptmiss$ & 380--560 & 2.7 $\pm$ 2.5 & 14 & 0.0026 & 0.0061\\
      $4 \PGm + 1 \PQb + 1 \text{jet} + \ptmiss$ & 590--950 & 0.092  $\pm$ 0.044 & 2 & 0.0048 & 0.0072\\[\cmsTabSkip]

      \multicolumn{6}{c}{Exclusive event classes: \ST} \\

      $3 \Pe  + 1 \PQb + 2 \text{jets}$ & 340--540 & 0.84  $\pm$ 0.27 & 6 & 0.00053 & 0.0038\\
      $4 \PGm + 1 \PQb + 1 \text{jet} + \ptmiss$ & 590--950 & 0.092 $\pm$ 0.047 & 2 & 0.0052 & 0.0082\\[\cmsTabSkip]

      \multicolumn{6}{c}{Exclusive event classes: \ptmiss} \\

      $4 \PGm + 1 \PQb + 1 \text{jet} + \ptmiss$ & 100--390 & 0.16  $\pm$ 0.12 & 2 & 0.018 & 0.022\\
      $1 \PGm + 4 \PQb + 1 \text{jet} + \ptmiss$ & 140--330 & 0.57  $\pm$ 0.50 & 4 & 0.014 & 0.027\\[\cmsTabSkip]
      
      \multicolumn{6}{c}{Inclusive event classes: \minv} \\

      $4 \PGm + 1 \PQb + 1 \text{jet} + \ptmiss+X$ & 590--860 & 0.16  $\pm$ 0.10 & 2 & 0.016 & 0.022\\
      $4 \PGm + 1 \text{jet} + \ptmiss+X$ & 560--770 & 0.60  $\pm$ 0.24 & 4 & 0.0055 & 0.026\\[\cmsTabSkip]

      \multicolumn{6}{c}{Inclusive event classes: \ST}\\

      $2 \Pe + 1 \PGm + 1 \PQb + 5 \text{jets}+X$ & 740--890 & 0.062  $\pm$ 0.043 & 2 & 0.0028 & 0.0063\\
      $2 \PGm+X$ & 1050--6110 & 95.8  $\pm$ 6.8 & 58 & 0.00036 & 0.012\\[\cmsTabSkip]

      \multicolumn{6}{c}{Inclusive event classes: \ptmiss}\\

      $4 \PGm + 1 \text{jet} + \ptmiss+X$ & 130--160 & 0.46  $\pm$ 0.32 & 4 & 0.0045 & 0.012\\
      $3 \PGm + 4 \text{jets} + \ptmiss+X$ & 170--570 & 2.5  $\pm$ 1.3 & 8 & 0.021 & 0.048\\[\cmsTabSkip]
      
      \multicolumn{6}{c}{Jet-inclusive event classes: \minv}\\

      $2 \Pe + 1 \PGm + 5 \text{jets}+\text{Njets}$ & 1370--2030 & 0.37  $\pm$ 0.29 & 4 & 0.0028 & 0.0063\\
      $1 \Pe + 1 \PGm + 3 \PQb + 2 \text{jets} + \ptmiss+\text{Njets}$ & 1140--1700 & 0.79  $\pm$ 0.46 & 5 & 0.0050 & 0.0071\\[\cmsTabSkip]

      \multicolumn{6}{c}{Jet-inclusive event classes: \ST}\\

      $2 \Pe + 1 \PGm + 5 \text{jets}+\text{Njets}$ & 990--1780 & 0.39  $\pm$ 0.34 & 4 & 0.0039 & 0.0060\\
      $2 \Pe + 1 \PGm + 1 \PQb + 3 \text{jets}+\text{Njets}$ & 430--650 & 0.52  $\pm$ 0.26 & 5 & 0.00066 & 0.0070\\[\cmsTabSkip]

      \multicolumn{6}{c}{Jet-inclusive event classes: \ptmiss}\\

      $4 \PGm + 1 \PQb + 1 \text{jet} + \ptmiss+\text{Njets}$ & 100--150 & 0.19  $\pm$ 0.12 & 2 & 0.022 & 0.022\\
      $4 \PGm + 1 \text{jet} + \ptmiss+\text{Njets}$ & 130--160 & 0.36 $\pm$ 0.24  & 3 & 0.012 & 0.032\\
      \hline
   \end{tabular}
   \label{tab:most_significant}
\end{table*}

Some of these event classes have low numbers of events and high object multiplicity, such as the $4 \Pgm + 1 \PQb + 1 \text{jet} + \ptmiss$ event class with two data events compared to an overall expectation of $0.16 \pm 0.11$ events in the entire event class (the numbers displayed in Table~\ref{tab:most_significant} refer to the data events and simulated expectation within the RoI),  where the deviation can be attributed to a fluctuation. The events in this event class also contribute to the $4 \Pgm + 1 \PQb + 1 \text{jet} + \ptmiss\ +X$, $4 \Pgm + 1 \text{jet} + \ptmiss\ +X$, and  $4 \Pgm + 1 \PQb + 1 \text{jet} + \ptmiss\ +\textrm{Njets}$ event classes, which also appear among the event classes with the largest deviations for the inclusive and jet-inclusive categories.
There are high jet multiplicity event classes with relatively low numbers of events, particularly $2 \Pe + 1 \Pgm + 1 \PQb + 5 \text{jets} +X$, $2 \Pe + 1 \Pgm + 5 \text{jets} +\text{Njets}$, $1 \Pgm + 4 \PQb + 1 \text{jet} + \ptmiss$, $1 \Pe + 1 \Pgm + 3 \PQb + 2 \text{jets} + \ptmiss +\text{Njets}$, and $2 \Pe + 1 \Pgm + 1 \PQb + 3 \text{jets} +\text{Njets}$, most of which are also inclusive at least in terms of the number of jets. The \ptildevalues of these deviations are not very significant, and they can be ascribed either to fluctuations  or to inadequate modelling of the data by the simulation at high jet multiplicities.

The $3 \Pe + 1 \PQb + 2 \text{jets}$ event class is the event class with the smallest \ptildevalue, and it appears in the scan of the \ST distribution for exclusive event classes. The entire event class has seven data events compared to the expectation of $2.7 \pm 1.8$ from the simulation. The major contribution of SM processes in this event class is $\PQt\PAQt$ production in association with a vector boson. Related event classes were studied, including the corresponding inclusive and jet-inclusive event classes, the flavour counterpart $3 \Pgm + 1 \PQb + 2 \text{jets}$, and event classes with one object removed. None of those event classes show a significant deviation in the data from the simulated SM background predictions.
Similar studies were performed also for the $1 \Pe + 1 \Pgm + 1 \PGg + \ptmiss$ event class that shows the second-smallest \ptildevalue in exclusive event classes, as a result of the scan of the \minv distribution. Related event classes, such as the corresponding inclusive and jet-inclusive classes, and event classes where the number of physics objects has been reduced by one, were checked. Again, none of the related event classes show a large deviation from the simulated SM background predictions. The largest SM contribution in this event class corresponds to the $\PQt\PAQt$ process, and other event classes dominated by the same process are described well.
The low \ptildevalue in the $2 \Pgm +X$ event class identified by the scan of the \ST distribution for inclusive event classes corresponds to a deficit in the tail of the distribution. It is not found as a prominent deviation in the corresponding exclusive or jet-inclusive categories. The observed effect is not very significant, and was also seen during a dedicated analysis targeting this final state~\cite{Sirunyan:2018exx}.
The remaining event classes detailed in Table~\ref{tab:most_significant} show smaller deviations from the simulated SM background predictions.

In summary, the low \ptildevalues observed in the aforementioned individual event classes are not beyond the expectations from SM, and no systematic trends are observed.

\section{Summary}
\label{sec:conclusion}

The Model Unspecific Search in CMS (\music) analysis has been presented. The analysis is based on data recorded by the CMS detector at the LHC during proton-proton collisions at a centre-of-mass energy of $13\TeV$ in 2016 and corresponding to an integrated luminosity of $35.9\fbinv$.
The \music analysis searches for anomalies and possible hints of physics beyond the standard model in the data using a model-independent approach, relying solely on the assumptions of the well-tested standard model. 

Events from data and simulation containing at least one electron or muon have been sorted into event classes based on their final-state topology, defined by the number of electrons, muons, photons, jets and $\PQb$-tagged jets, and missing transverse momentum. The event yields were compared between the data and the expectation in a wide range of event classes. The kinematic distributions corresponding to the sum of transverse momenta, invariant (or transverse) mass, and missing transverse momentum in each of the event classes have been scanned using a region of interest algorithm. The algorithm identifies deviations of the data from the simulated standard model predictions, calculating a \pvalue of any observed deviation after correcting for the look-elsewhere effect. A global overview of the results from the different event classes and distributions has been presented.

The sensitivity and robustness of the analysis has been shown in a variety of different studies. No significant deviations from the standard model expectations were found in the data analysed by the \music algorithm. A wide range of final-state topologies has been studied, and there is agreement between data and the standard model simulation given the experimental and theoretical uncertainties. 
This analysis complements dedicated search analyses by significantly expanding the range of final states covered using a model independent approach with the largest data set to date to probe phase space regions beyond the reach of previous general searches. 

\clearpage
\begin{acknowledgments}
  We congratulate our colleagues in the CERN accelerator departments for the excellent performance of the LHC and thank the technical and administrative staffs at CERN and at other CMS institutes for their contributions to the success of the CMS effort. In addition, we gratefully acknowledge the computing centres and personnel of the Worldwide LHC Computing Grid for delivering so effectively the computing infrastructure essential to our analyses. Finally, we acknowledge the enduring support for the construction and operation of the LHC and the CMS detector provided by the following funding agencies: BMBWF and FWF (Austria); FNRS and FWO (Belgium); CNPq, CAPES, FAPERJ, FAPERGS, and FAPESP (Brazil); MES (Bulgaria); CERN; CAS, MoST, and NSFC (China); COLCIENCIAS (Colombia); MSES and CSF (Croatia); RIF (Cyprus); SENESCYT (Ecuador); MoER, ERC IUT, PUT and ERDF (Estonia); Academy of Finland, MEC, and HIP (Finland); CEA and CNRS/IN2P3 (France); BMBF, DFG, and HGF (Germany); GSRT (Greece); NKFIA (Hungary); DAE and DST (India); IPM (Iran); SFI (Ireland); INFN (Italy); MSIP and NRF (Republic of Korea); MES (Latvia); LAS (Lithuania); MOE and UM (Malaysia); BUAP, CINVESTAV, CONACYT, LNS, SEP, and UASLP-FAI (Mexico); MOS (Montenegro); MBIE (New Zealand); PAEC (Pakistan); MSHE and NSC (Poland); FCT (Portugal); JINR (Dubna); MON, RosAtom, RAS, RFBR, and NRC KI (Russia); MESTD (Serbia); SEIDI, CPAN, PCTI, and FEDER (Spain); MOSTR (Sri Lanka); Swiss Funding Agencies (Switzerland); MST (Taipei); ThEPCenter, IPST, STAR, and NSTDA (Thailand); TUBITAK and TAEK (Turkey); NASU (Ukraine); STFC (United Kingdom); DOE and NSF (USA).
   
  \hyphenation{Rachada-pisek} Individuals have received support from the Marie-Curie programme and the European Research Council and Horizon 2020 Grant, contract Nos.\ 675440, 752730, and 765710 (European Union); the Leventis Foundation; the A.P.\ Sloan Foundation; the Alexander von Humboldt Foundation; the Belgian Federal Science Policy Office; the Fonds pour la Formation \`a la Recherche dans l'Industrie et dans l'Agriculture (FRIA-Belgium); the Agentschap voor Innovatie door Wetenschap en Technologie (IWT-Belgium); the F.R.S.-FNRS and FWO (Belgium) under the ``Excellence of Science -- EOS" -- be.h project n.\ 30820817; the Beijing Municipal Science \& Technology Commission, No. Z191100007219010; the Ministry of Education, Youth and Sports (MEYS) of the Czech Republic; the Deutsche Forschungsgemeinschaft (DFG) under Germany's Excellence Strategy -- EXC 2121 ``Quantum Universe" -- 390833306; the Lend\"ulet (``Momentum") Programme and the J\'anos Bolyai Research Scholarship of the Hungarian Academy of Sciences, the New National Excellence Program \'UNKP, the NKFIA research grants 123842, 123959, 124845, 124850, 125105, 128713, 128786, and 129058 (Hungary); the Council of Science and Industrial Research, India; the HOMING PLUS programme of the Foundation for Polish Science, cofinanced from European Union, Regional Development Fund, the Mobility Plus programme of the Ministry of Science and Higher Education, the National Science Center (Poland), contracts Harmonia 2014/14/M/ST2/00428, Opus 2014/13/B/ST2/02543, 2014/15/B/ST2/03998, and 2015/19/B/ST2/02861, Sonata-bis 2012/07/E/ST2/01406; the National Priorities Research Program by Qatar National Research Fund; the Ministry of Science and Higher Education, project no. 02.a03.21.0005 (Russia); the Tomsk Polytechnic University Competitiveness Enhancement Program; the Programa Estatal de Fomento de la Investigaci{\'o}n Cient{\'i}fica y T{\'e}cnica de Excelencia Mar\'{\i}a de Maeztu, grant MDM-2015-0509 and the Programa Severo Ochoa del Principado de Asturias; the Thalis and Aristeia programmes cofinanced by EU-ESF and the Greek NSRF; the Rachadapisek Sompot Fund for Postdoctoral Fellowship, Chulalongkorn University and the Chulalongkorn Academic into Its 2nd Century Project Advancement Project (Thailand); the Kavli Foundation; the Nvidia Corporation; the SuperMicro Corporation; the Welch Foundation, contract C-1845; and the Weston Havens Foundation (USA).
\end{acknowledgments}

\bibliography{auto_generated} 

\appendix
\numberwithin{figure}{section}
\clearpage
\section{Total event yield distributions by object group for exclusive event classes}
\label{sec:app_process_group_plots}

\begin{figure*}[h]
    \centering
        \includegraphics[width=\cmsFigWidthApp]{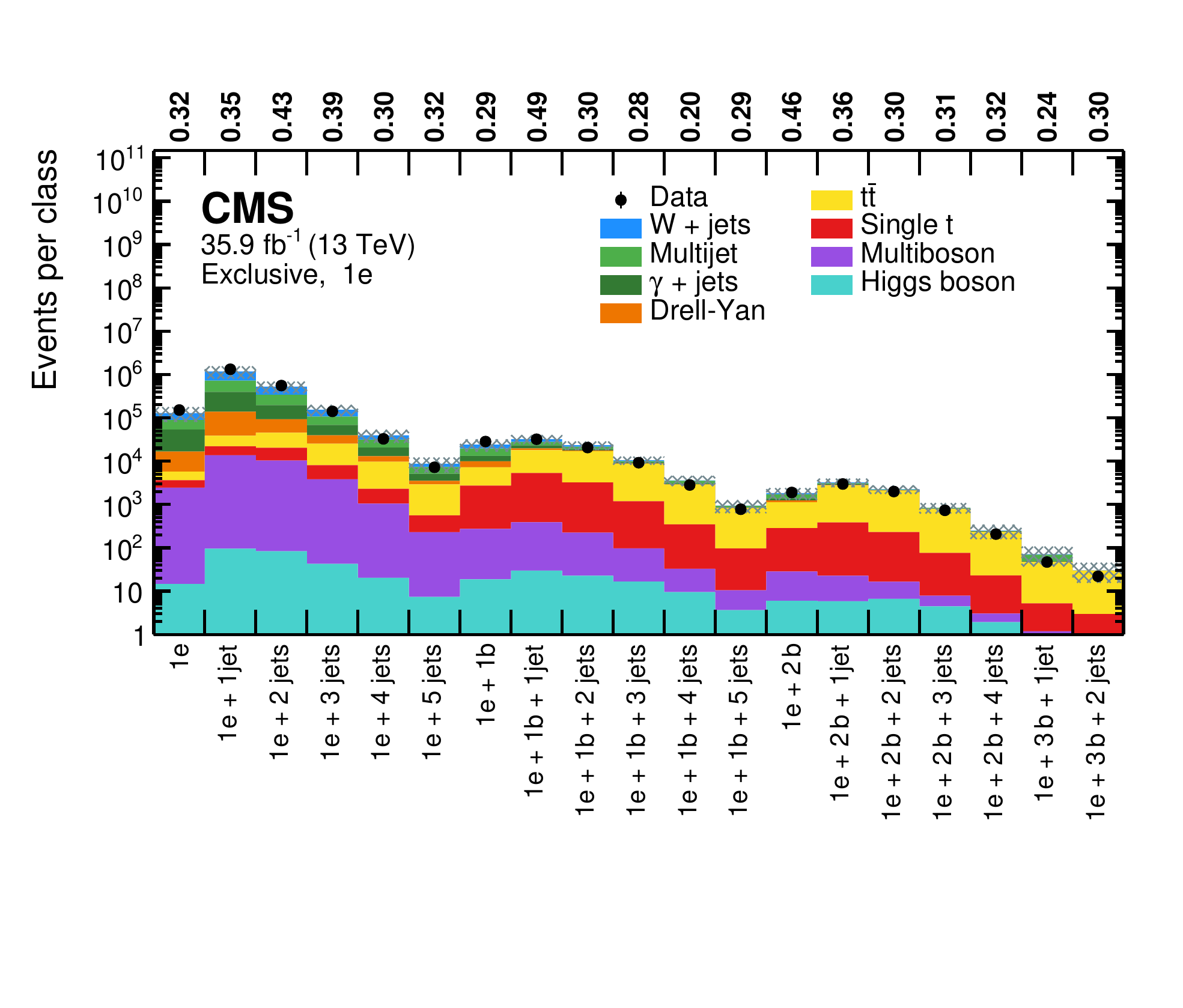}
        \includegraphics[width=\cmsFigWidthApp]{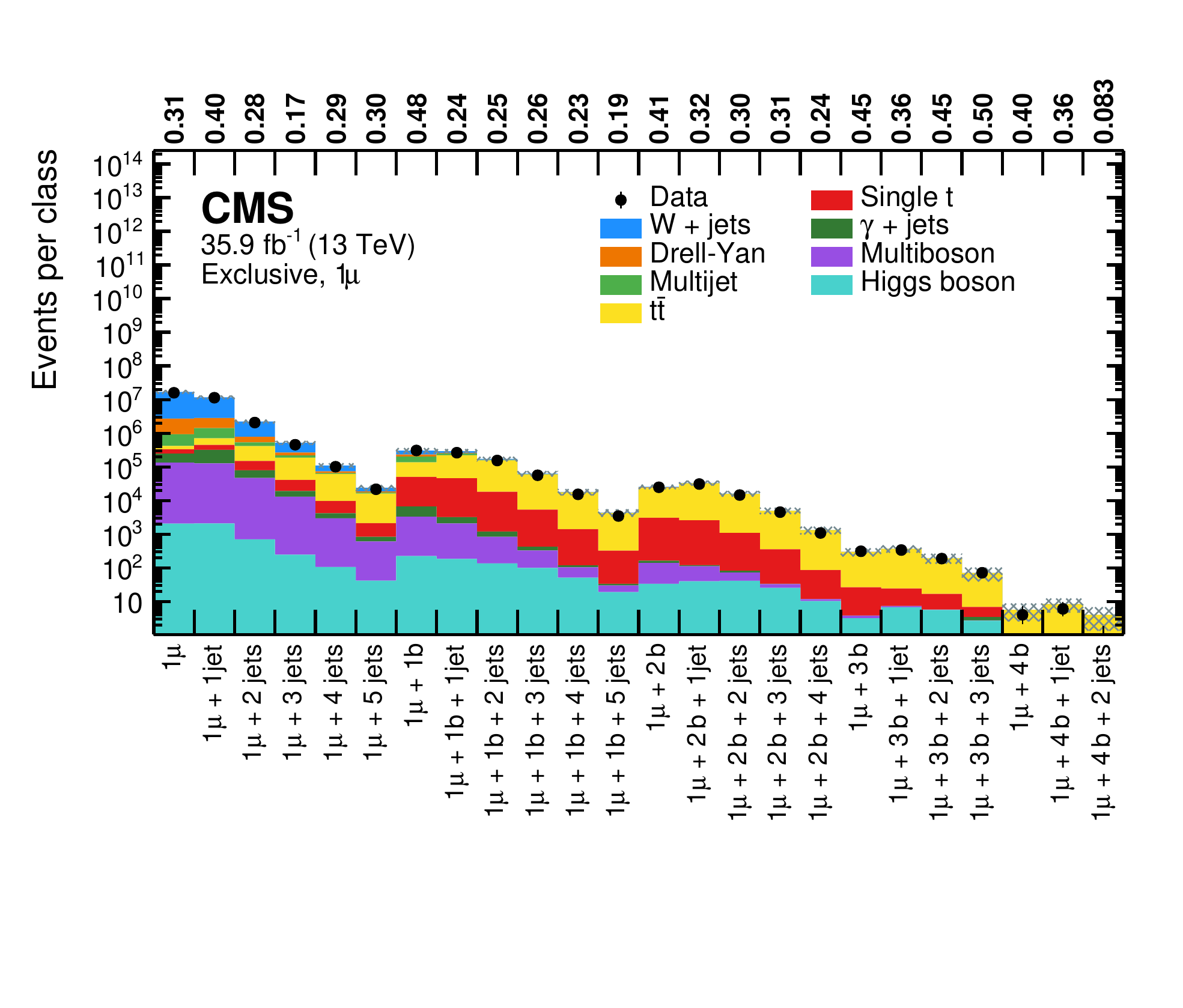}
        \caption{Overview of total event yields for the event classes of the single-electron (upper) and single-muon (lower) object groups. Measured data are shown as black markers, contributions from SM processes are represented by coloured histograms, and the shaded region represents the uncertainty in the SM background. The numbers above the plot indicate the observed \pvalue for the agreement of data and simulation.}
        \label{fig:Simpsons-1Ele-Simpsons-1Muon}
\end{figure*}

\begin{figure*}[h]
    \centering
        \includegraphics[width=\cmsFigWidthApp]{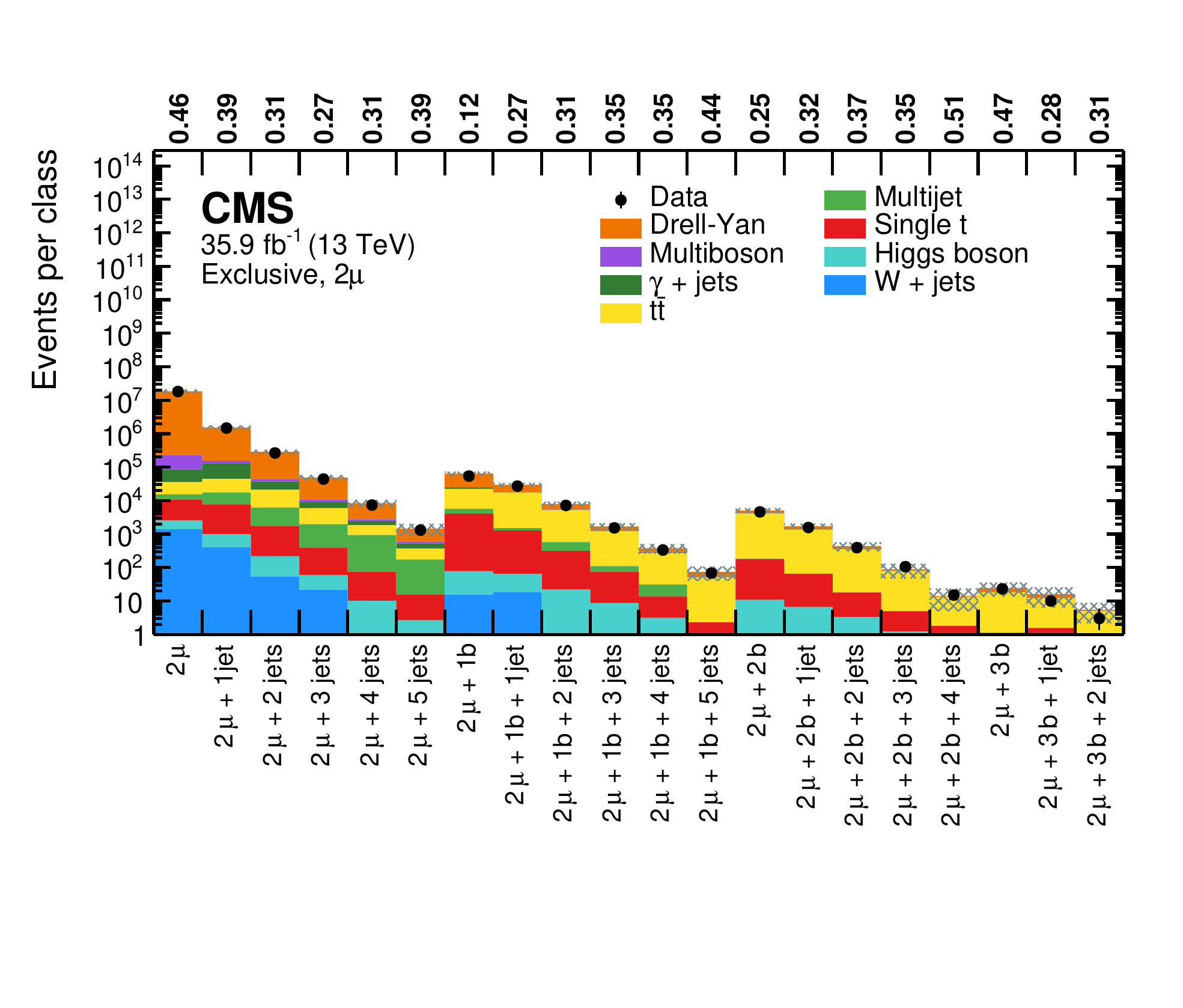}
        \includegraphics[width=\cmsFigWidthApp]{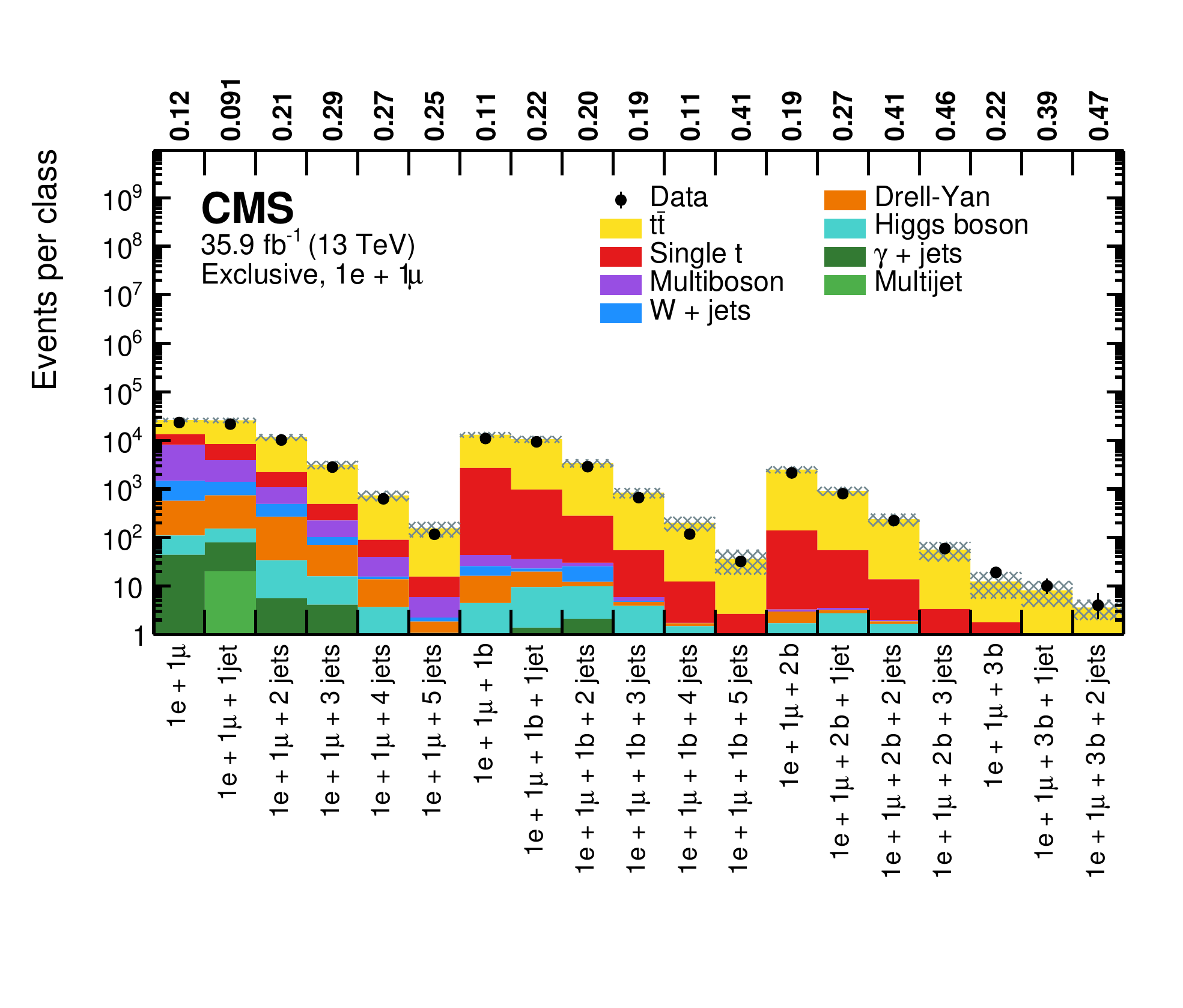}
        \caption{Overview of total event yields for the event classes of the double-muon (upper) and the electron $+$ muon (lower) object groups. Measured data are shown as black markers, contributions from SM processes are represented by coloured histograms, and the shaded region represents the uncertainty in the SM background. The numbers above the plot indicate the observed \pvalue for the agreement of data and simulation.}
        \label{fig:Simpsons-2Muon-Simpsons-1Ele_1Muon}
\end{figure*}

\begin{figure*}[h]
    \centering
        \includegraphics[width=\cmsFigWidthApp]{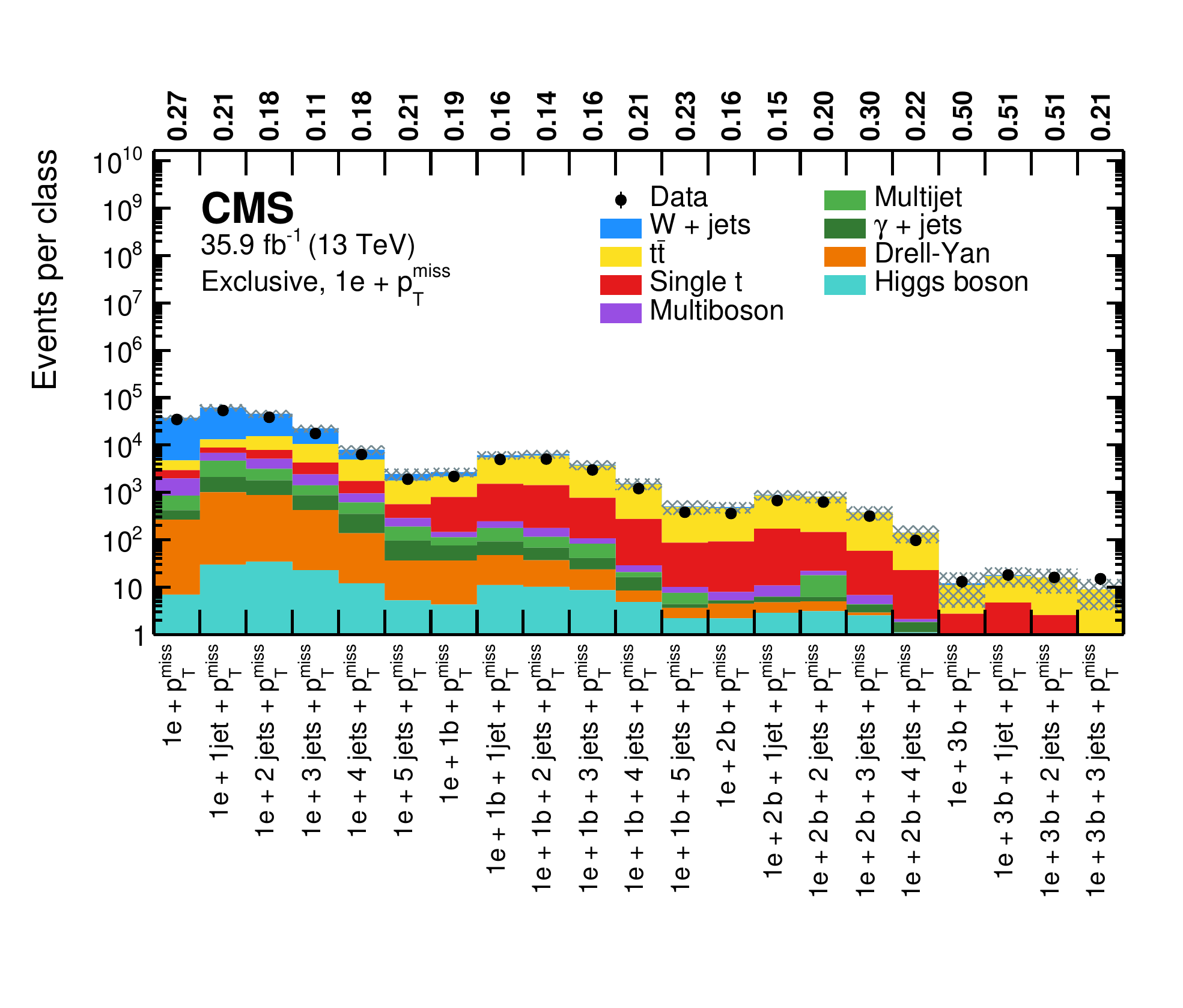}
        \includegraphics[width=\cmsFigWidthApp]{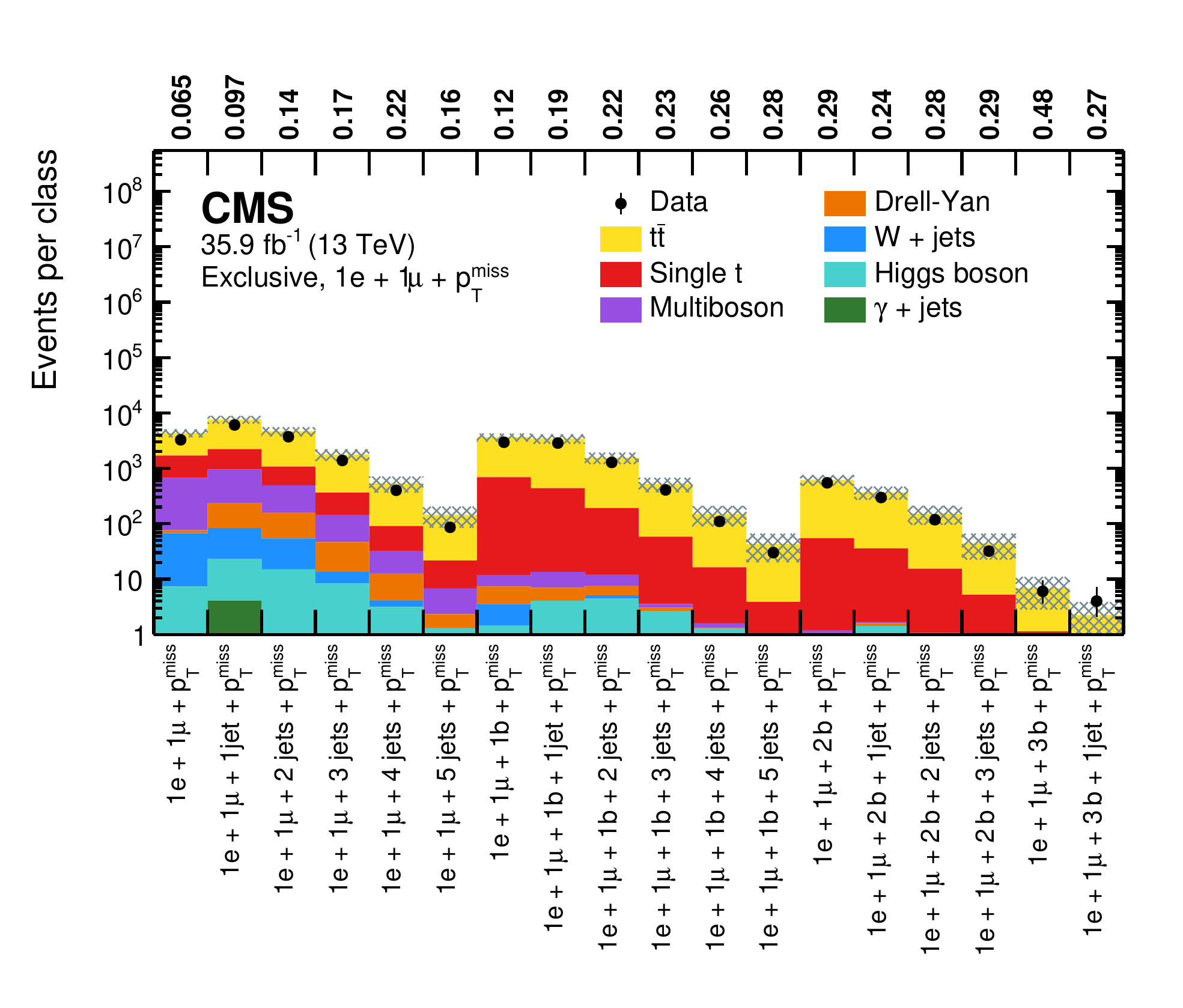}
        \caption{Overview of total event yields for the event classes of the single-electron $+$ \ptmiss (upper) and the single-electron $+$ single-muon $+$ \ptmiss (lower) object groups. Measured data are shown as black markers, contributions from SM processes are represented by coloured histograms, and the shaded region represents the uncertainty in the SM background. The numbers above the plot indicate the observed \pvalue for the agreement of data and simulation.}
        \label{fig:Simpsons-singleelectron_met-Simpsons-1Ele-1Muon-MET}
\end{figure*}

\begin{figure*}[h]
    \centering
        \includegraphics[width=\cmsFigWidthApp]{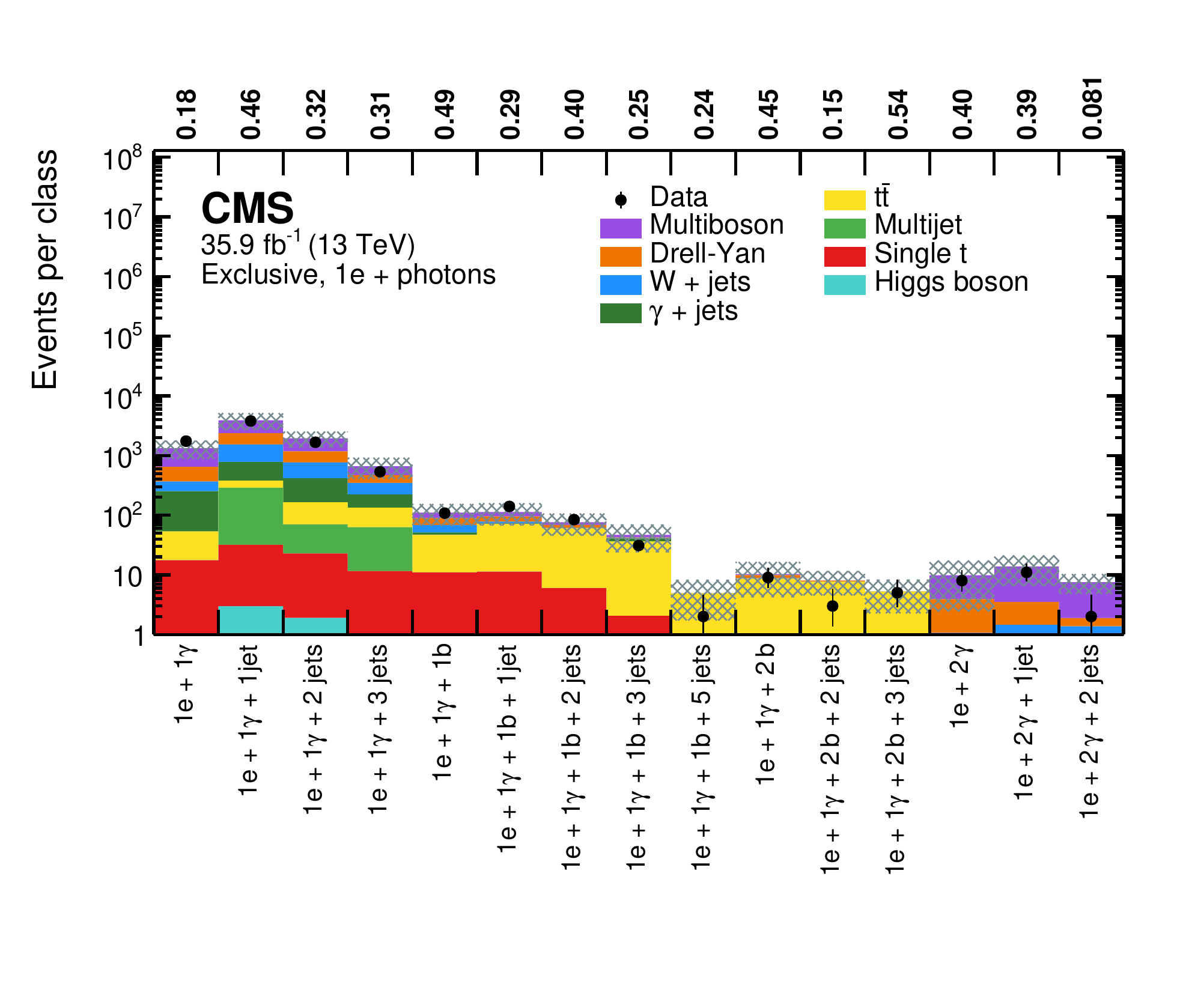}
        \includegraphics[width=\cmsFigWidthApp]{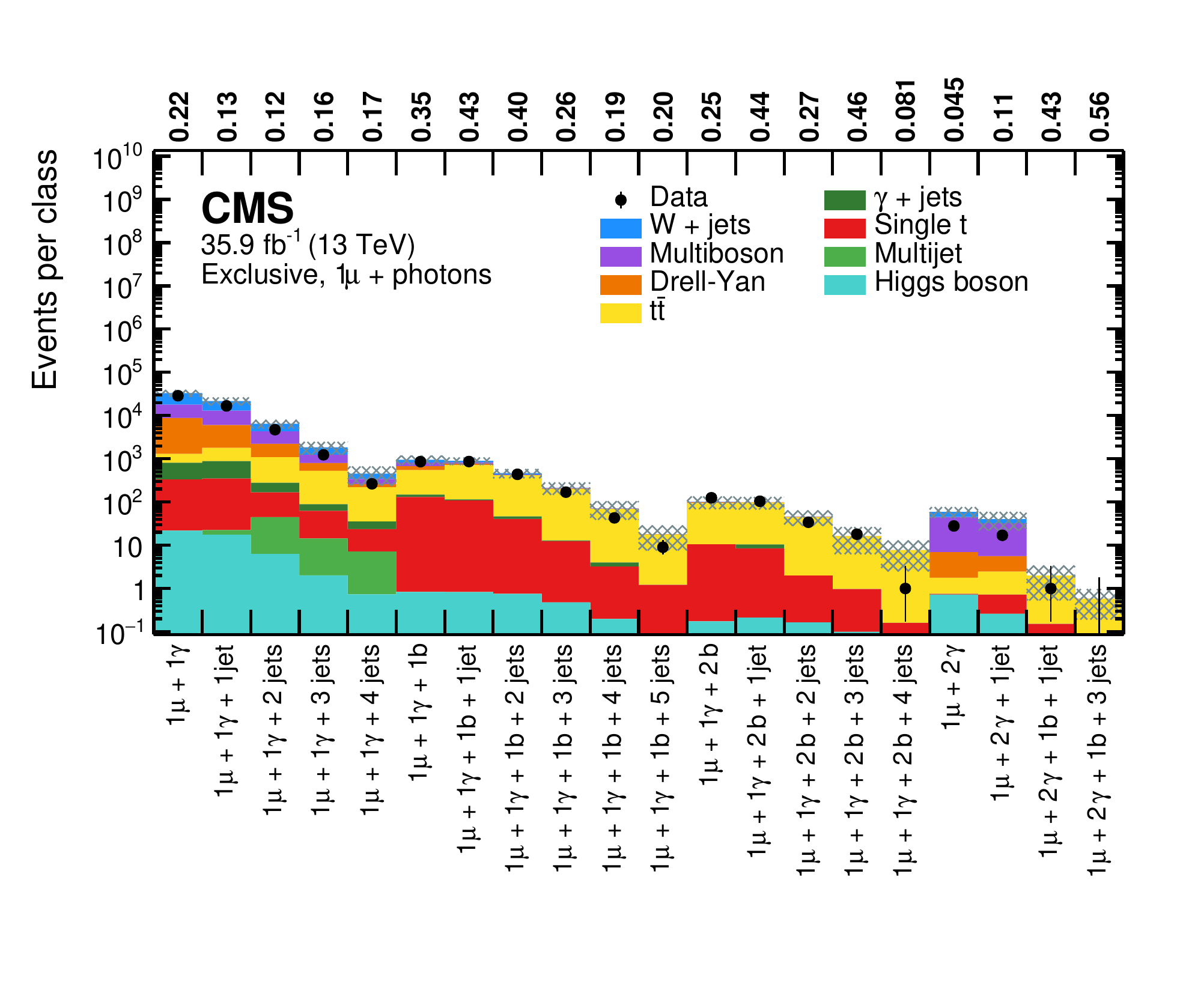}
        \caption{Overview of total event yields for the event classes of the single-electron $+$ photons (upper) and the single-muon $+$ photons (lower) object groups. Measured data are shown as black markers, contributions from SM processes are represented by coloured histograms, and the shaded region represents the uncertainty in the SM background. The numbers above the plot indicate the observed \pvalue for the agreement of data and simulation.}
        \label{fig:Simpsons-1Ele_photons-Simpsons-1Muon_photons}
\end{figure*}

\begin{figure*}[h]
    \centering
        \includegraphics[width=\cmsFigWidthApp]{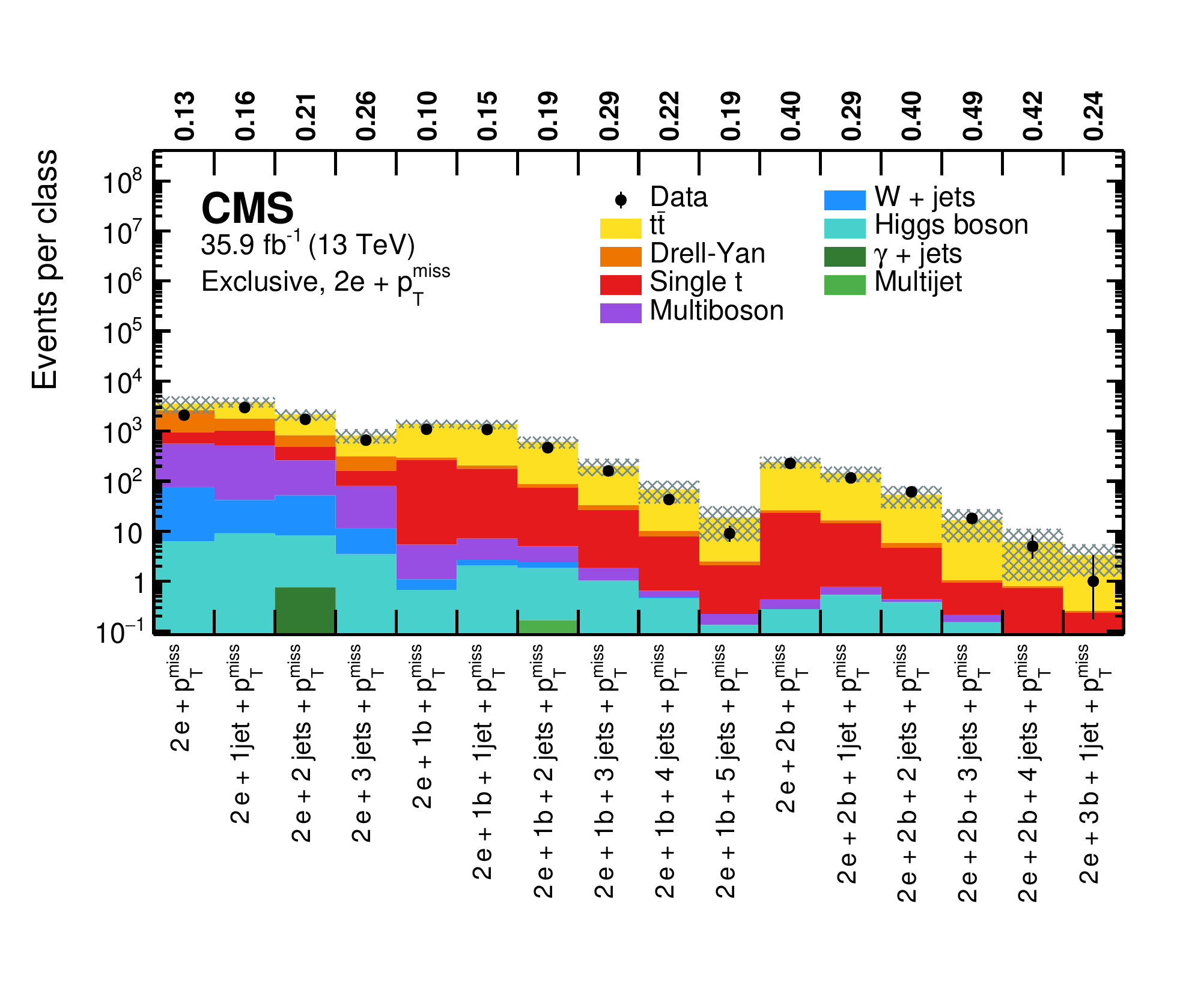}
        \includegraphics[width=\cmsFigWidthApp]{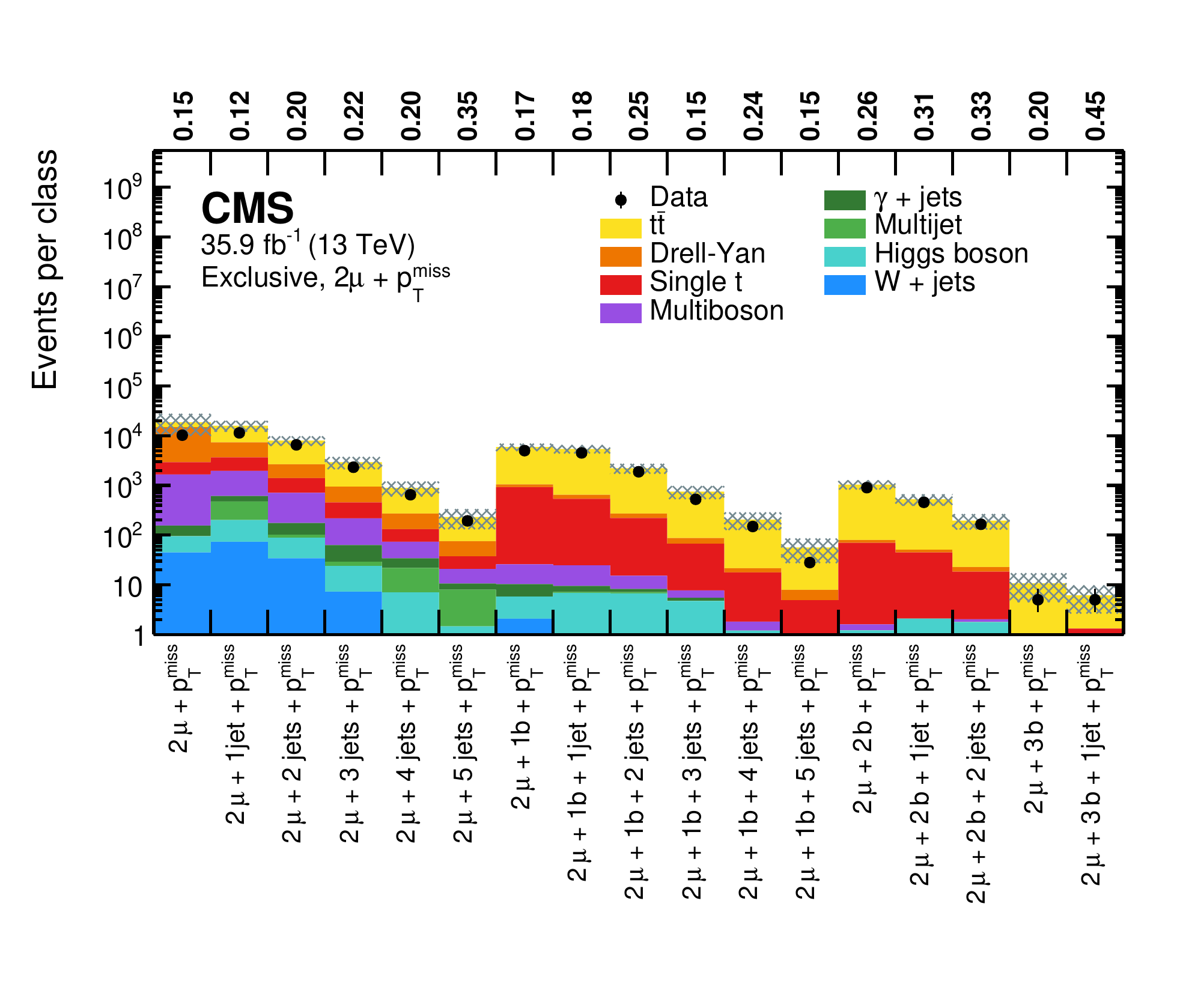}
        \caption{Overview of total event yields for the event classes of the double-electron $+$ \ptmiss (upper) and the double-muon $+$ \ptmiss (lower) object groups. Measured data are shown as black markers, contributions from SM processes are represented by coloured histograms, and the shaded region represents the uncertainty in the SM background. The numbers above the plot indicate the observed \pvalue for the agreement of data and simulation.}
        \label{fig:Simpsons-doublelepton_met}
\end{figure*}

\begin{figure*}[h]
    \centering
        \includegraphics[width=\cmsFigWidthApp]{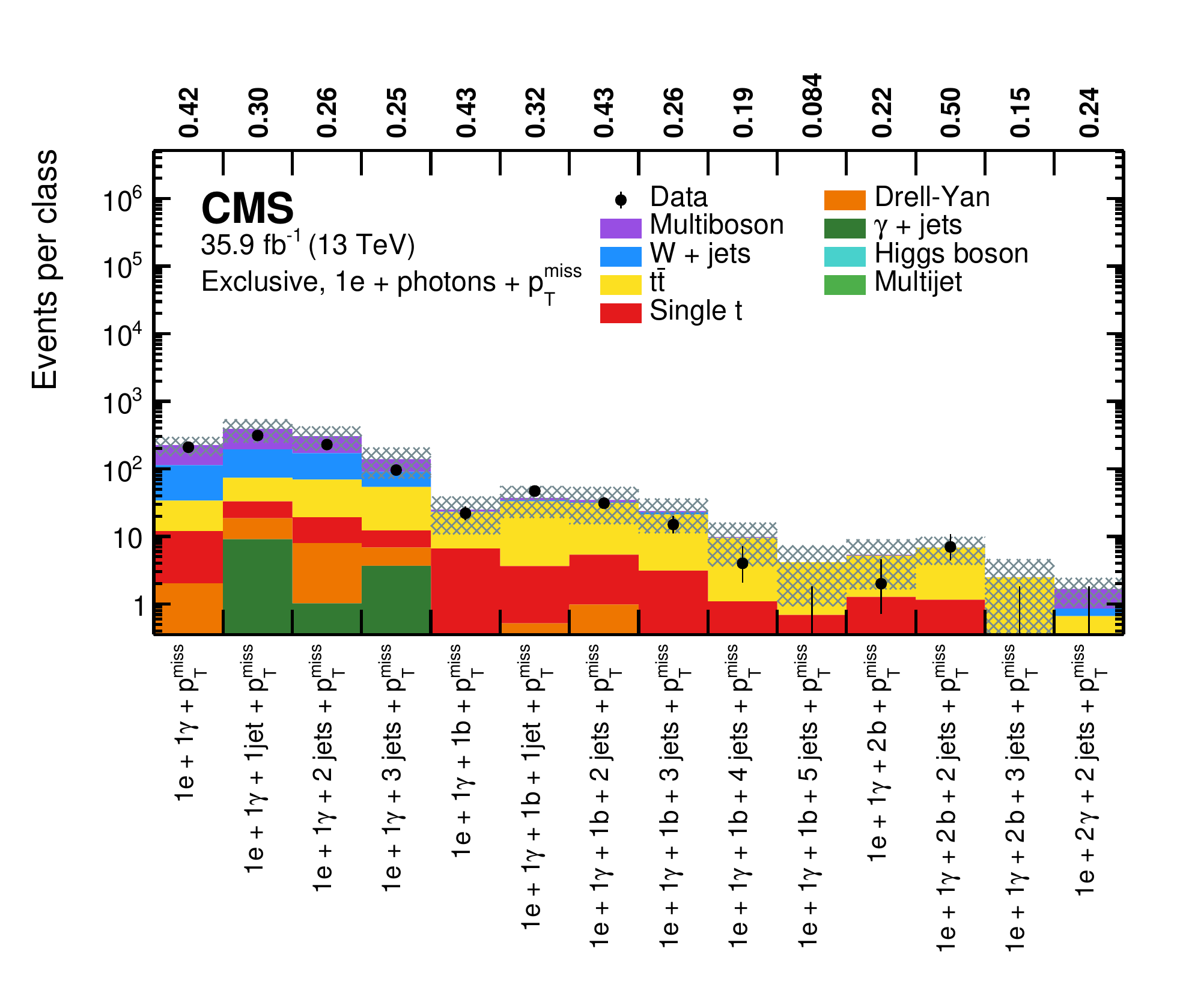}
        \includegraphics[width=\cmsFigWidthApp]{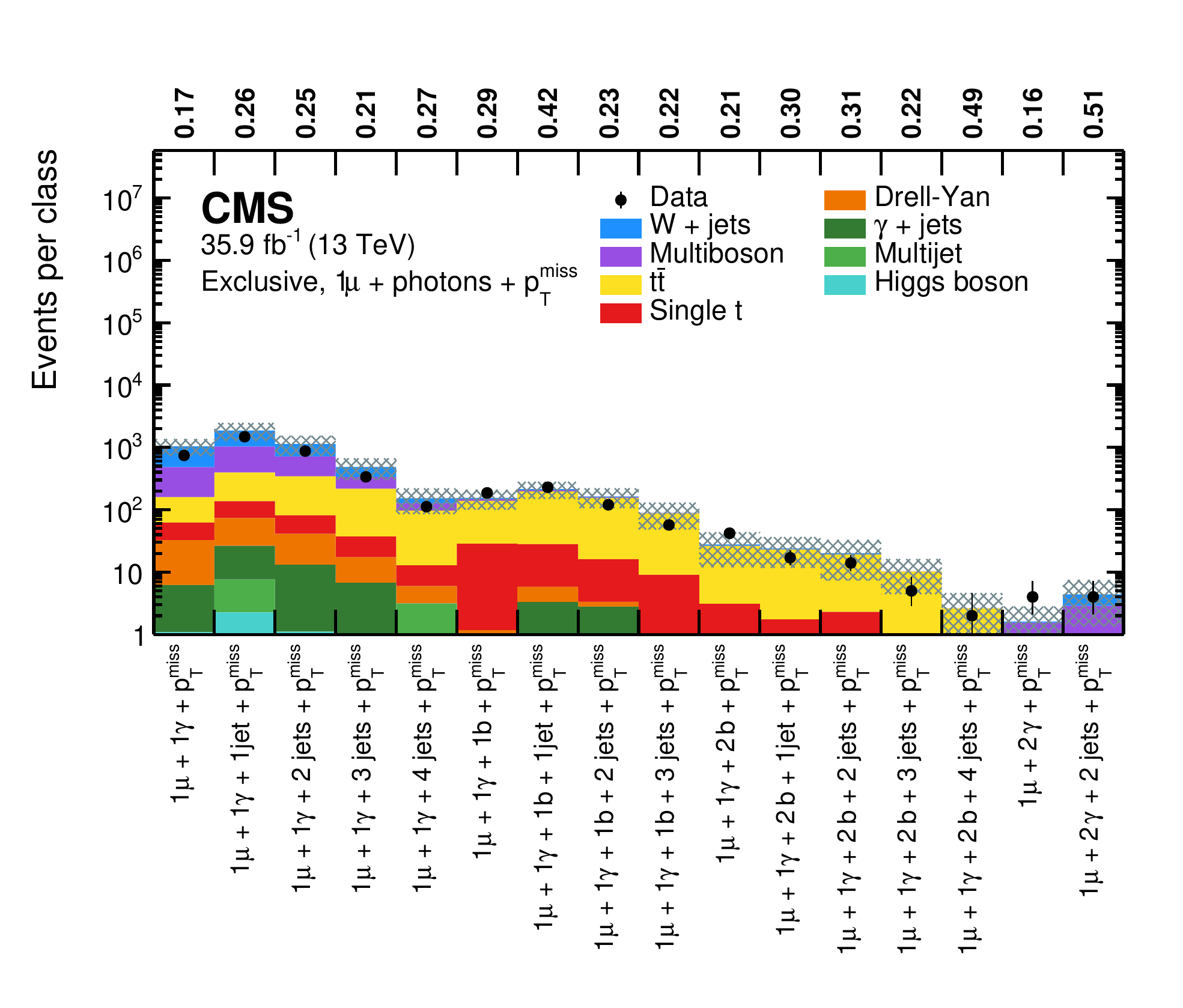}
        \caption{Overview of total event yields for the event classes of the single-electron $+$ photons $+$ \ptmiss (upper) and the single-muon $+$ photons $+$ \ptmiss (lower) object groups. Measured data are shown as black markers, contributions from SM processes are represented by coloured histograms, and the shaded region represents the uncertainty in the SM background. The numbers above the plot indicate the observed \pvalue for the agreement of data and simulation.}
        \label{fig:Simpsons-singlepton_photons_met}
\end{figure*}

\begin{figure*}[h]
    \centering
        \includegraphics[width=\cmsFigWidthApp]{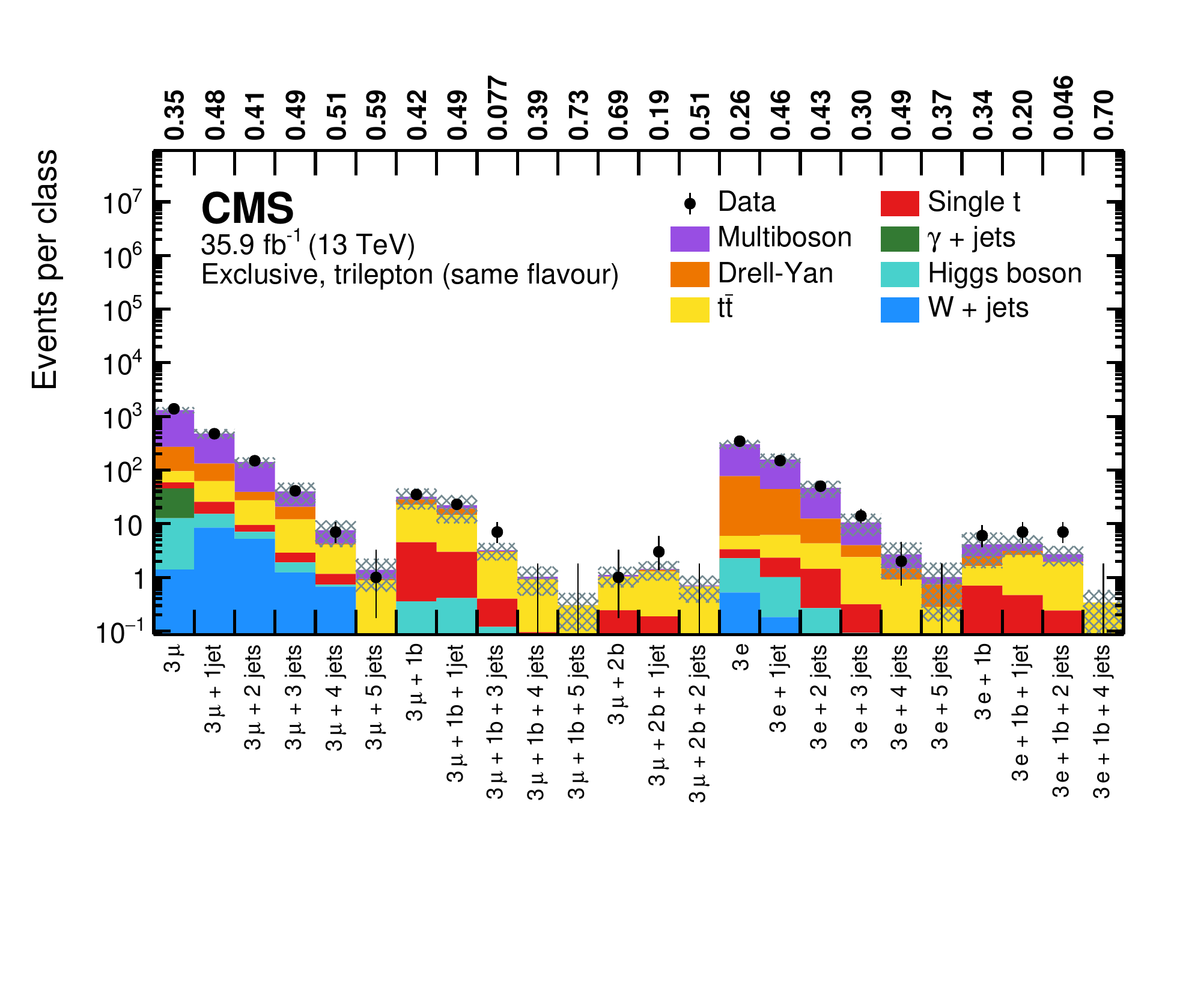}
        \includegraphics[width=\cmsFigWidthApp]{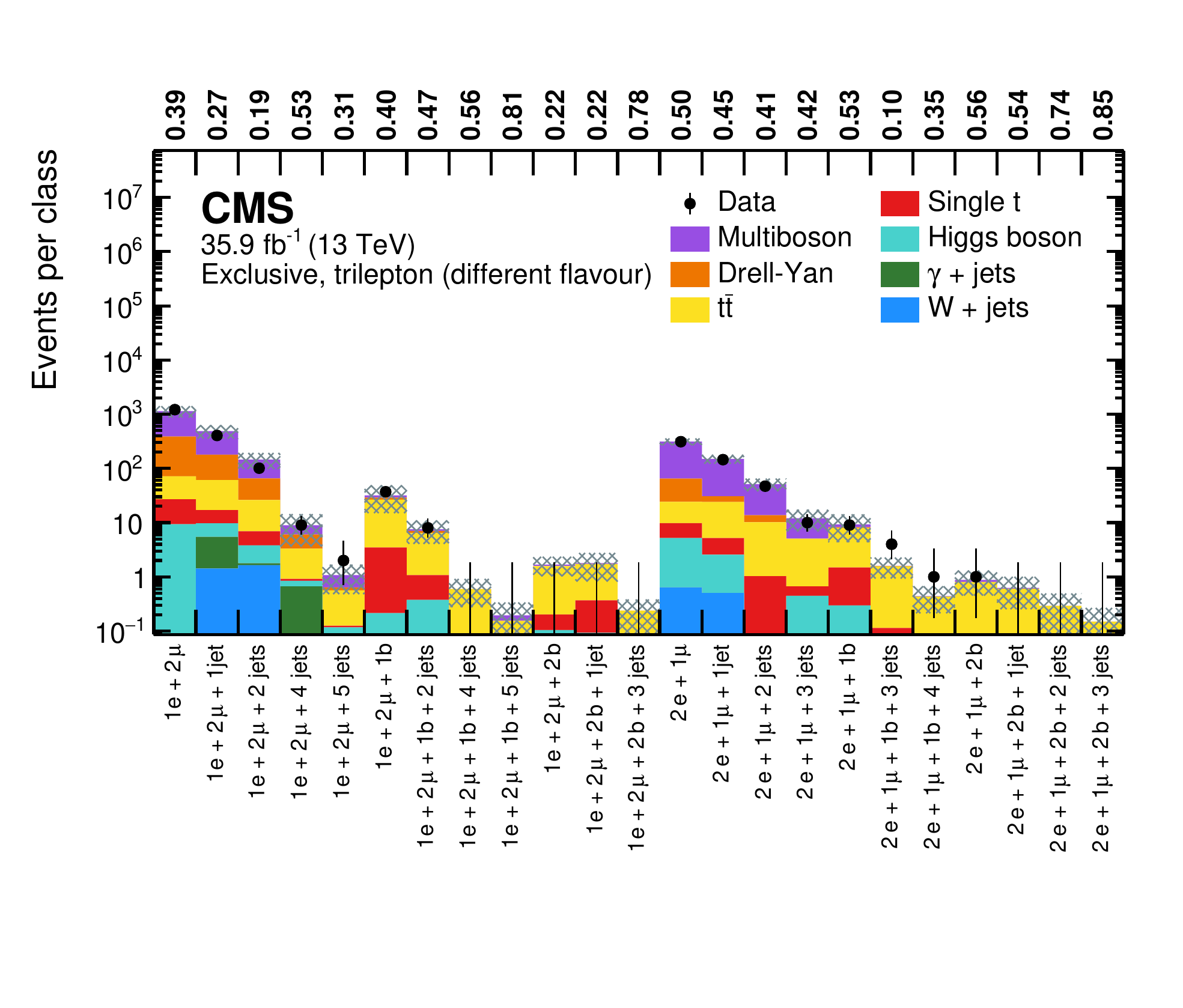}
        \caption{Overview of total event yields for the event classes of the three-lepton object groups with same flavour (upper) and different flavour (lower). Measured data are shown as black markers, contributions from SM processes are represented by coloured histograms, and the shaded region represents the uncertainty in the SM background. The numbers above the plot indicate the observed \pvalue for the agreement of data and simulation.}
        \label{fig:Simpsons-triplelepton}
\end{figure*}

\begin{figure*}[h]
    \centering
        \includegraphics[width=\cmsFigWidthApp]{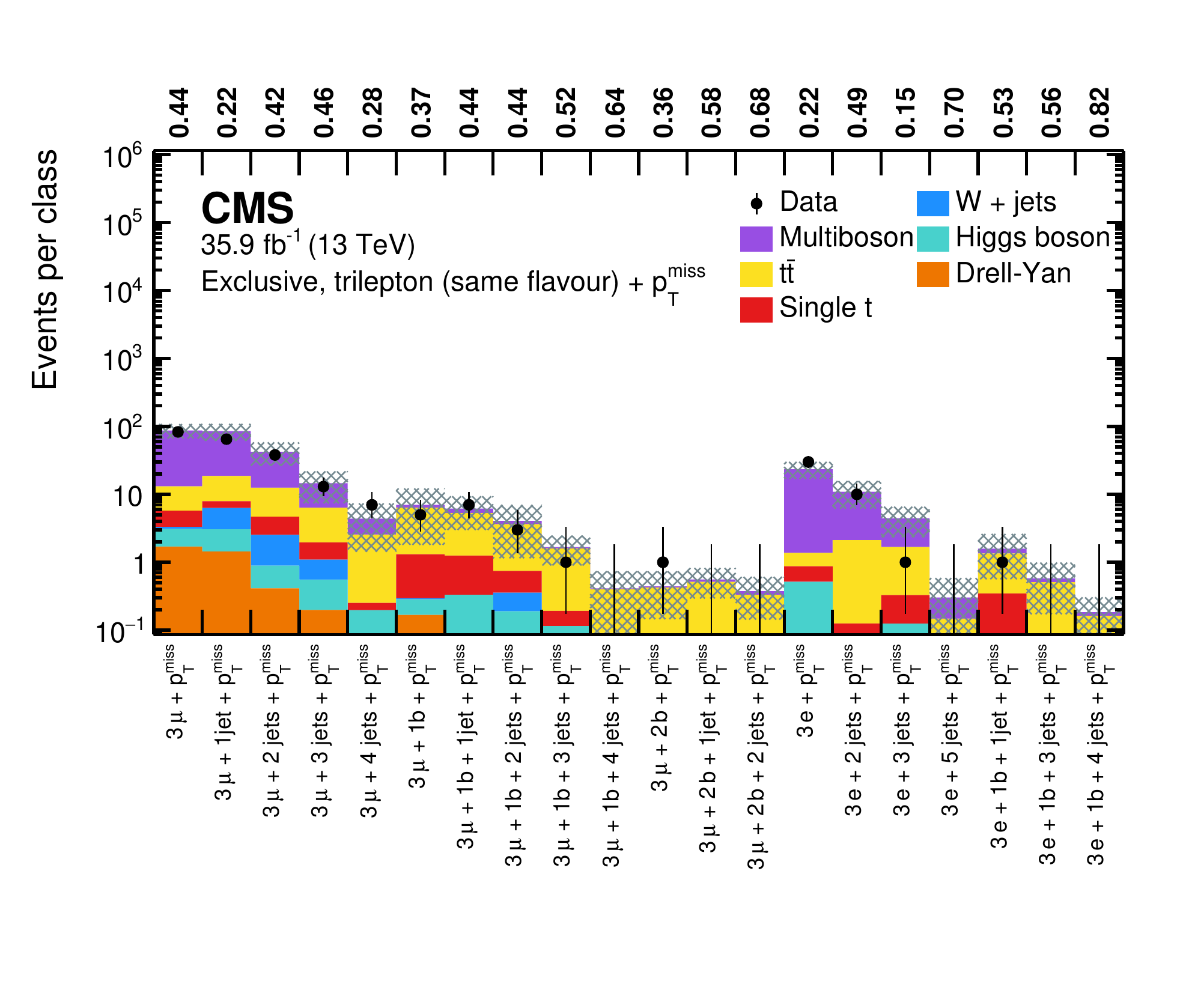}
        \includegraphics[width=\cmsFigWidthApp]{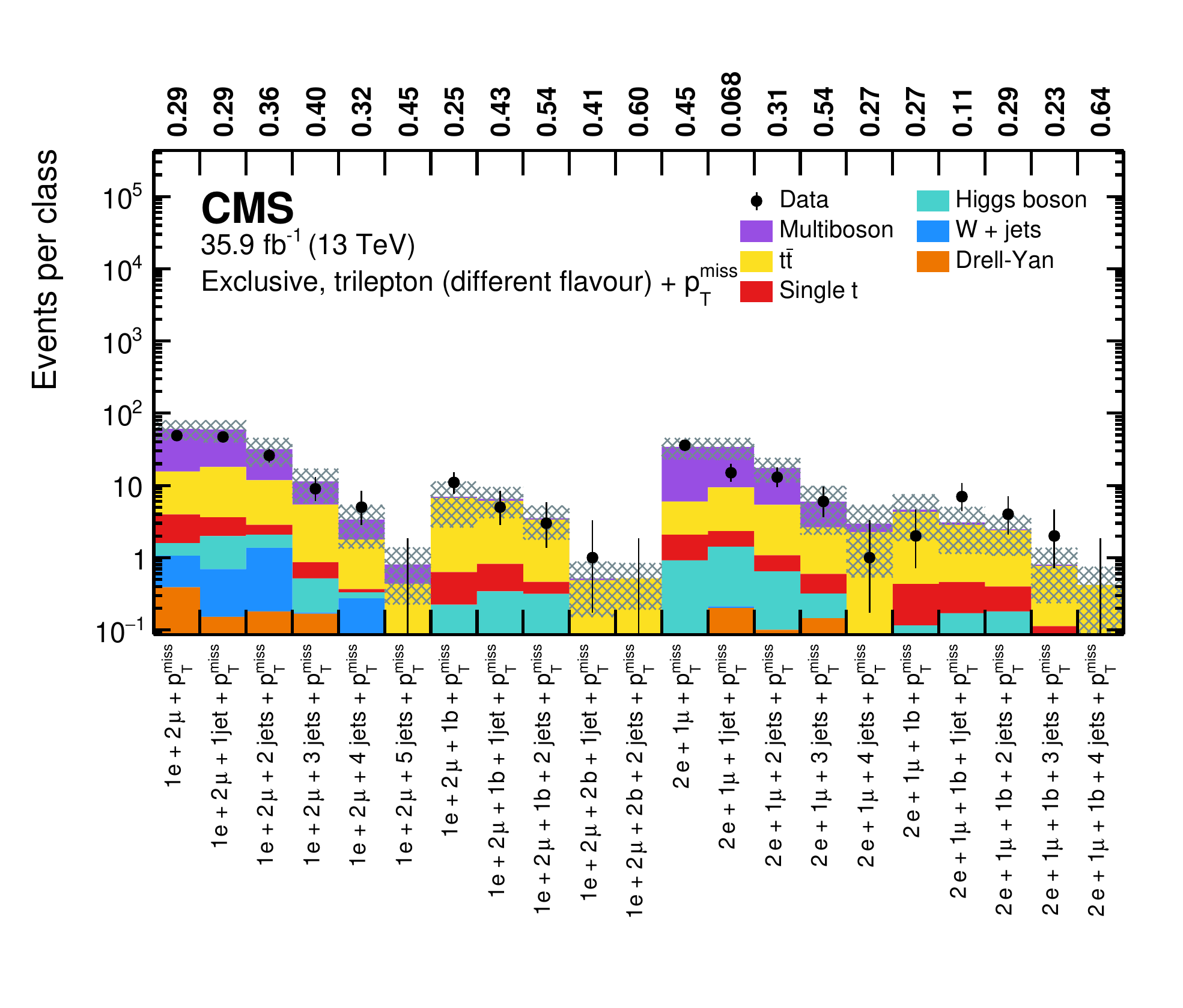}
        \caption{Overview of total event yields for the event classes of the three-lepton (same flavour) $+$ \ptmiss object group (upper), and the three-lepton (different flavour) $+$ \ptmiss object group (lower). Measured data are shown as black markers, contributions from SM processes are represented by coloured histograms, and the shaded region represents the uncertainty in the SM background. The numbers above the plot indicate the observed \pvalue for the agreement of data and simulation.}
        \label{fig:Simpsons-triplelepton_met}
\end{figure*}

\begin{figure*}[h]
    \centering
        \includegraphics[width=\cmsFigWidthApp]{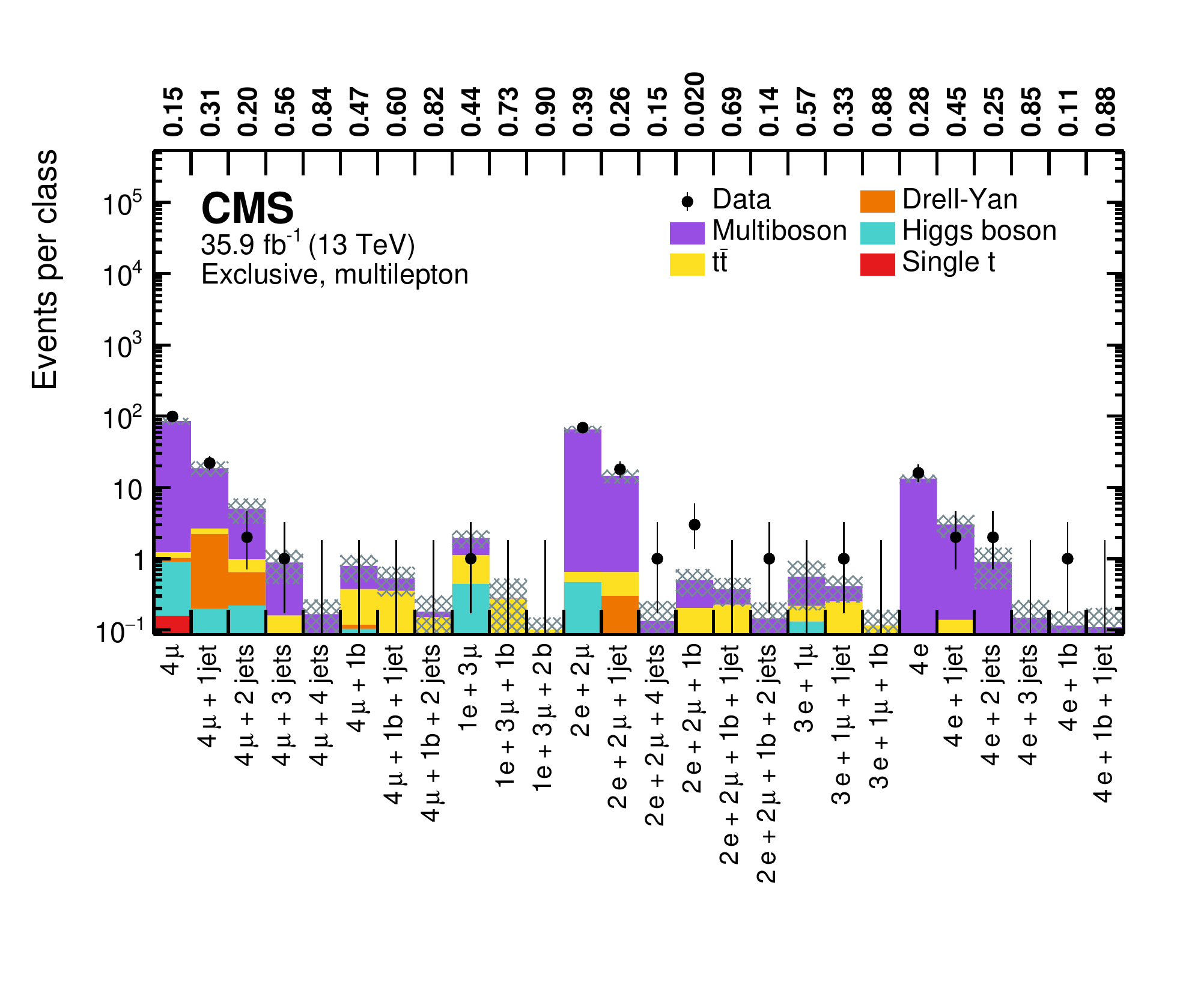}
        \includegraphics[width=\cmsFigWidthApp]{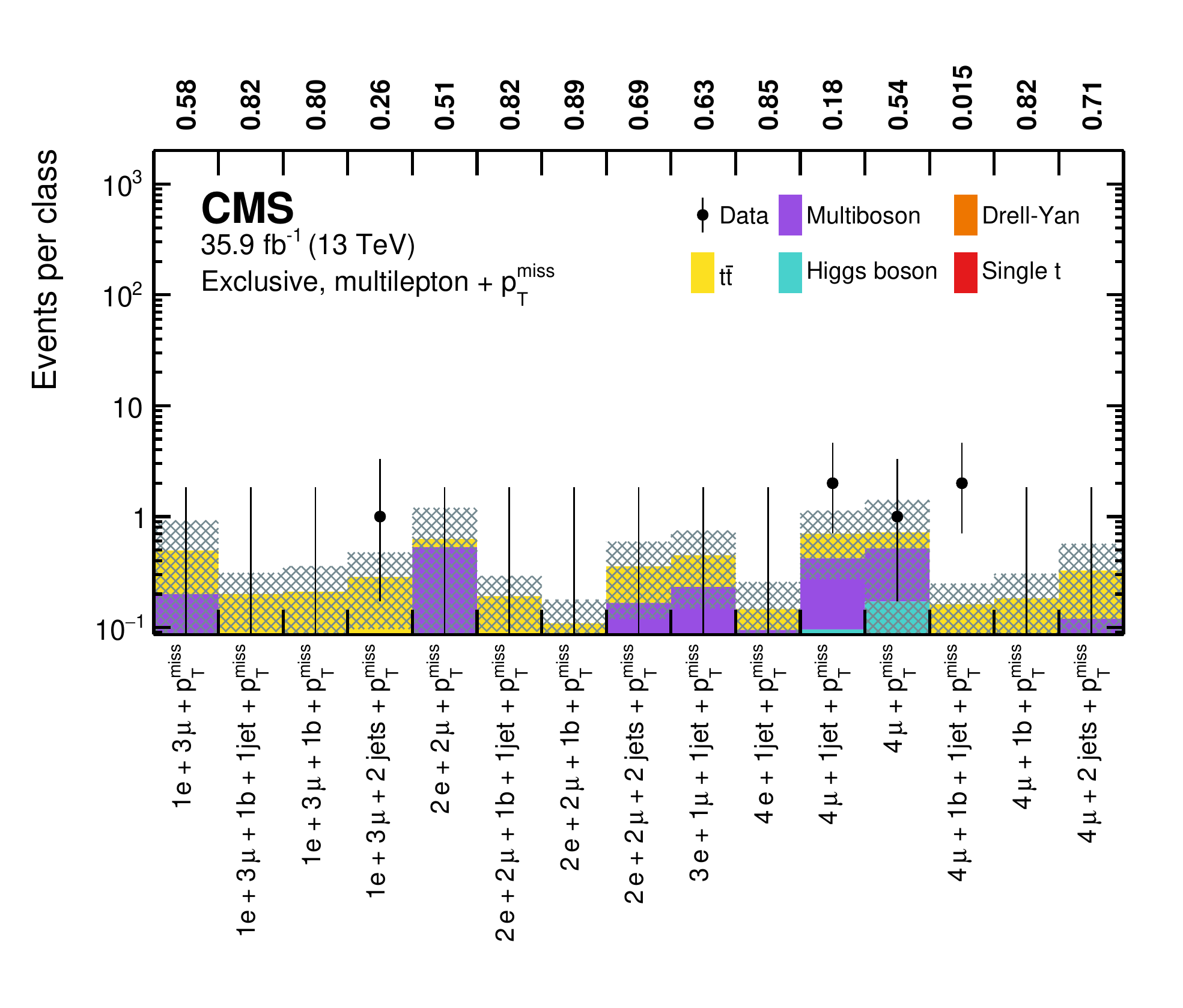}
        \caption{Overview of total event yields for the event classes of the ${\geq}4$ leptons object group (upper), and the ${\geq}4$ leptons $+$ \ptmiss object group (lower). Measured data are shown as black markers, contributions from SM processes are represented by coloured histograms, and the shaded region represents the uncertainty in the SM background. The numbers above the plot indicate the observed \pvalue for the agreement of data and simulation.}
        \label{fig:Simpsons-multilepton}
\end{figure*}

\begin{figure*}[h]
    \centering
        \includegraphics[width=\cmsFigWidthApp]{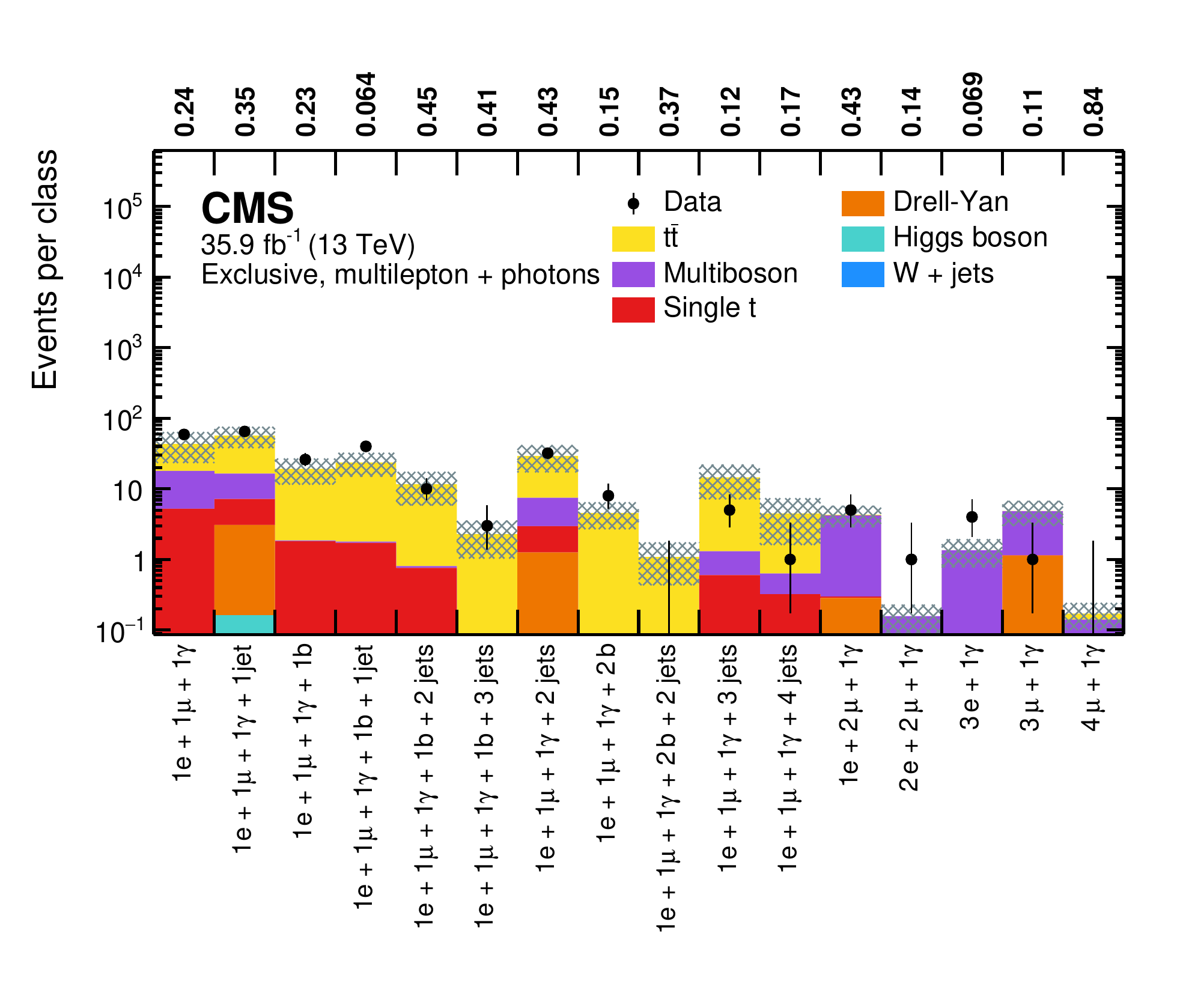}
        \includegraphics[width=\cmsFigWidthApp]{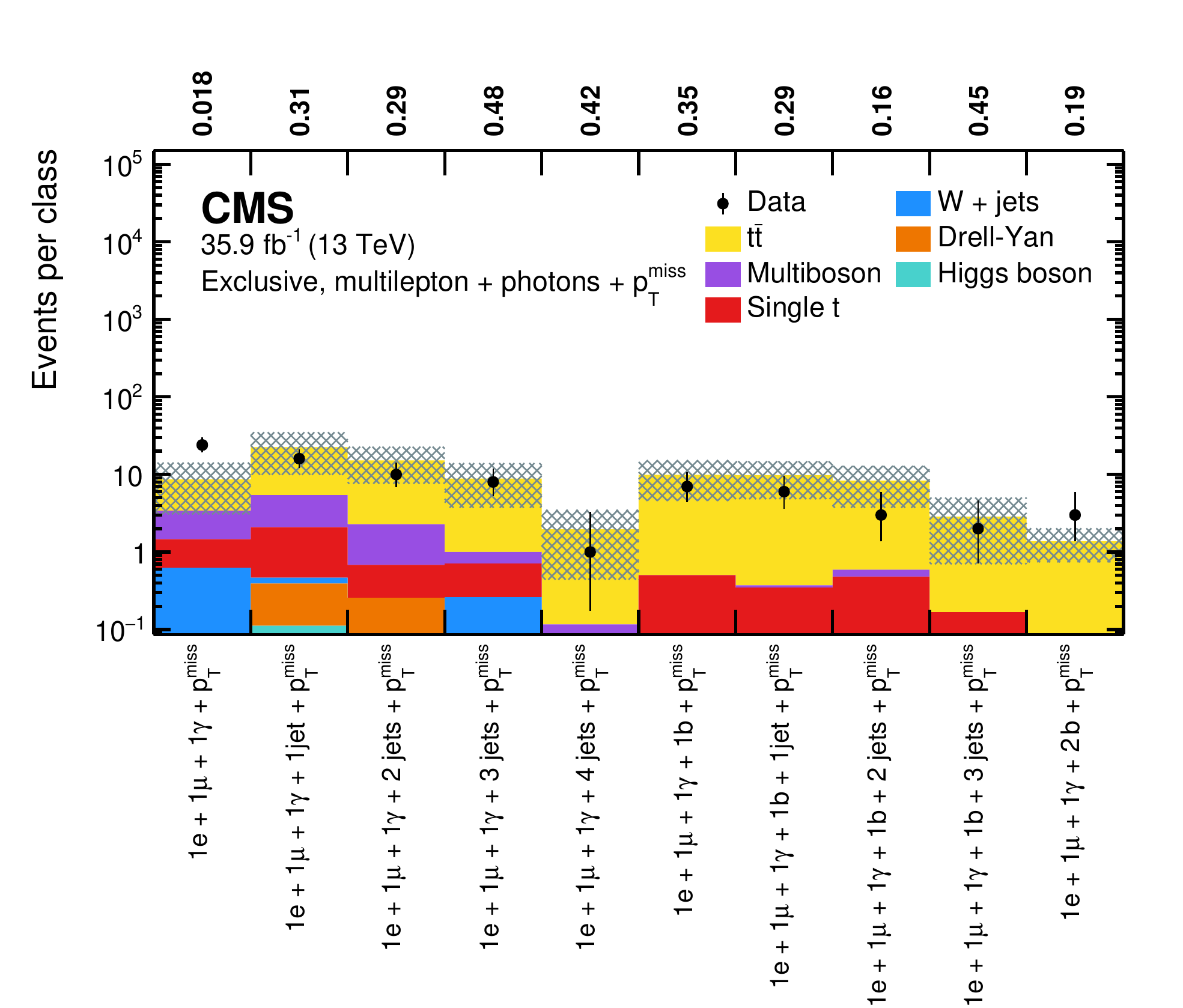}
        \caption{Overview of total event yields for the event classes of the $> 1$ lepton $+$ photons object group (upper), and the electron $+$ muon $+$ photons $+$ \ptmiss object group (lower). Measured data are shown as black markers, contributions from SM processes are represented by coloured histograms, and the shaded region represents the uncertainty in the SM background. The numbers above the plot indicate the observed \pvalue for the agreement of data and simulation.}
        \label{fig:Simpsons-multilepton_photons}
\end{figure*}

\cleardoublepage \section{The CMS Collaboration \label{app:collab}}\begin{sloppypar}\hyphenpenalty=5000\widowpenalty=500\clubpenalty=5000\vskip\cmsinstskip
\textbf{Yerevan Physics Institute, Yerevan, Armenia}\\*[0pt]
A.M.~Sirunyan$^{\textrm{\dag}}$, A.~Tumasyan
\vskip\cmsinstskip
\textbf{Institut f\"{u}r Hochenergiephysik, Wien, Austria}\\*[0pt]
W.~Adam, F.~Ambrogi, T.~Bergauer, M.~Dragicevic, J.~Er\"{o}, A.~Escalante~Del~Valle, R.~Fr\"{u}hwirth\cmsAuthorMark{1}, M.~Jeitler\cmsAuthorMark{1}, N.~Krammer, L.~Lechner, D.~Liko, T.~Madlener, I.~Mikulec, F.M.~Pitters, N.~Rad, J.~Schieck\cmsAuthorMark{1}, R.~Sch\"{o}fbeck, M.~Spanring, S.~Templ, W.~Waltenberger, C.-E.~Wulz\cmsAuthorMark{1}, M.~Zarucki
% ...
\vskip\cmsinstskip
\dag: Deceased\\
1:  Also at Vienna University of Technology, Vienna, Austria\\
% ...
\end{sloppypar}
\end{document}